\documentclass{cernyrep} 
\usepackage{texnames}
\usepackage[T1]{fontenc}
\usepackage{graphicx}
\usepackage{psfrag}
\usepackage{pstricks}
\usepackage{hyperref}
\usepackage{booktabs}
\renewcommand{\ie}{i.e.\ }
\newcommand{\fm}{\,\mathrm{fm}}
\newcommand{\MeV}{\,\mathrm{MeV}}
\newcommand{\TeV}{\,\mathrm{TeV}}
\newcommand{\GeV}{\,\mathrm{GeV}}
\newcommand{\as}{\alpha_\mathrm{s}}
\newcommand{\cL}{{\cal L}}
\newcommand{\cS}{{\cal S}}
\newcommand{\cA}{{\cal A}}
\newcommand{\cM}{{\cal M}}

\newcommand{\tw}{\textwidth}
\newcommand{\CA}{C_A}
\newcommand{\CF}{C_F}
\newcommand{\TR}{T_R}
\newcommand{\NC}{N_C}
\newcommand{\nf}{n_f}
\newcommand{\muR}{\mu_\text{\sc r}}
\newcommand{\muF}{\mu_\text{\sc f}}
\newcommand{\MSbar}{\ensuremath{\overline{\mbox{\scriptsize MS}}}\xspace}

\newcommand{\order}[1]{{\cal O}\left(#1\right)}
\DeclareMathOperator{\Tr}{Tr}

\def\slashi#1{\rlap{\sl/}#1}
\def\slashii#1{\setbox0=\hbox{$#1$}             
   \dimen0=\wd0                                 
   \setbox1=\hbox{\sl/} \dimen1=\wd1            
   \ifdim\dimen0>\dimen1                        
      \rlap{\hbox to \dimen0{\hfil\sl/\hfil}}   
      #1                                        
   \else                                        
      \rlap{\hbox to \dimen1{\hfil$#1$\hfil}}   
      \hbox{\sl/}                               
   \fi}                                         %
\def\slashiii#1{\setbox0=\hbox{$#1$}#1\hskip-\wd0\hbox to\wd0{\hss\sl/\/\hss}}
\newcommand{\slA}{\slashii{A}}
\newcommand{\slB}{\slashii{B}}
\newcommand{\slE}{\slashii{E}}
\newcommand{\slp}{\slashii{p}}
\newcommand{\slk}{\slashi{k}}
\newcommand{\sleps}{\slashii{\epsilon}}

\begin{document}
\title{Elements of QCD for hadron colliders\footnote{Lectures given at
    the 2009 European School of High Energy Physics, Bautzen, Germany,
    June 2009. The original version of this writeup was completed in
    April 2010 and published in CERN Yellow Report CERN-2010-002,
    pp.~45--100. This arXiv version contains a few small updates.}}
 
\author{Gavin~P.~Salam}

\institute{LPTHE, CNRS UMR 7589, UPMC Univ.\ Paris 6, Paris,
  France\thanks{Current address: CERN, Department of Physics, Theory
    Unit, CH-1211 Geneva 23, Switzerland and Department of Physics,
    Princeton University, Princeton, NJ 08544, USA.}}

\maketitle 

\begin{abstract}
  The aim of these lectures is to provide students with an
  introduction to some of the core concepts and methods of QCD that
  are relevant in an LHC context.
\end{abstract}

 
\tableofcontents

\section{Introduction}
 
Quantum Chromodynamics, QCD, is the theory of quarks, gluons and their
interactions. It is central to all modern colliders. And, for the most
part, it is what we are made of.

QCD bears a number of similarities to Quantum Electrodynamics (QED). 
Just as electrons carry the QED charge, i.e., electric charge, quarks
carry the QCD charge, known as colour charge.
Whereas there is only one kind of electric charge, colour charge comes
in three varieties, sometimes labelled red, green and
blue. Anti-quarks have corresponding anti-colour.
The gluons in QCD are a bit like the photons of QED. But while photons
are electrically neutral, gluons are not colour neutral. They can be
thought of as carrying both colour charge and anti-colour
charge. There are eight possible different combinations of (anti)colour for
gluons.
Another difference between QCD and QED lies in its coupling $\as$. In
QCD it is only moderately small, it tends to zero at high momentum
scales (asymptotic freedom, QED does the opposite), it blows up at
small scales, and in between its evolution with scale is quite fast:
at the LHC its value will range from $\as = 0.08$ at a scale of
$5\TeV$, to $\as \sim 1$ at a scale of $0.5\GeV$.
These differences between QCD and QED contribute to making QCD a much
richer theory.

In these lectures I will attempt to give you a feel for how QCD works
at high momentum scales, and for the variety of techniques used by
theorists in order to handle QCD at today's high-energy colliders. The
hope is that these basics will come in useful for day-to-day work with
the QCD facets of hadron collider physics.
In the fifty or so pages of these lectures, it will be impossible to give
full treatment of any of the topics we will encounter.
For that the reader is referred to any of the classic textbooks about
QCD at colliders \cite{Ellis:1991qj,Dissertori:2003pj,Brock:1993sz}.

\subsection{The Lagrangian and colour}

Let us start with a brief reminder of the components of the QCD
Lagrangian. This section will be rather dense, but we will return to
some of the points in more detail later.
As already mentioned, quarks come in three colours. So rather than
representing them with a single spinor $\psi$, we will need the spinor
to carry also a colour index $a$, which runs from $1\ldots3$,
\begin{equation}
  \label{eq:psi_a}
  \psi_a = 
  \begin{pmatrix}
    \psi_1 
    \cr 
    \psi_2 
    \cr 
    \psi_3
  \end{pmatrix}.
\end{equation}
The quark part of the Lagrangian (for a single flavour) can be written
\begin{equation}
  \label{eq:Lquark}
  \cL_q =  {\bar\psi}_a (i \gamma^\mu\partial_\mu \delta_{ab} -
    {g_s \gamma^\mu t^C_{ab} \cA^C_\mu} - m)\psi_b\,,
\end{equation}
where the $\gamma^\mu$ are the usual Dirac matrices; the $\cA^C_\mu$
are gluon fields, with a Lorentz index $\mu$ and a colour index $C$
that goes from $1\ldots8$. Quarks are in the fundamental representation of
the SU(3) (colour) group, while gluons are in the adjoint
representation. Each of the eight gluon fields acts on the quark
colour through one of the `generator' matrices of the SU(3) group,
the $t_{ab}^C$ factor in Eq.~(\ref{eq:Lquark}). One convention for
writing the matrices is $t^A \equiv \frac12 \lambda^A$ with
\begin{equation*}
  \lambda^1 = \left(\!
    \begin{array}{ccc}
      0 & 1 & 0 \\
      1 & 0 & 0 \\
      0 & 0 & 0 
    \end{array}
    \!\right),\;
  \lambda^2 = \left(\!
    \begin{array}{ccc}
      0 & -i & 0 \\
      i & 0 & 0 \\
      0 & 0 & 0 
    \end{array}
    \!\right),\;
  \lambda^3 = \left(\!
    \begin{array}{ccc}
      1 & 0 & 0 \\
      0 & -1 & 0 \\
      0 & 0 & 0 
    \end{array}
    \!\right),\;
  \lambda^4 = \left(\!
    \begin{array}{ccc}
      0 & 0 & 1 \\
      0 & 0 & 0 \\
      1 & 0 & 0 
    \end{array}
    \!\right),\;
\end{equation*}
\begin{equation*}
  \lambda^5 = \left(\!
    \begin{array}{ccc}
      0 & 0 & -i \\
      0 & 0 & 0 \\
      i & 0 & 0 
    \end{array}
    \!\right),\;
  \lambda^6 = \left(\!
    \begin{array}{ccc}
      0 & 0 & 0 \\
      0 & 0 & 1 \\
      0 & 1 & 0 
    \end{array}
    \!\right),\;
  \lambda^7 = \left(\!
    \begin{array}{ccc}
      0 & 0 & 0 \\
      0 & 0 & -i \\
      0 & i & 0 
    \end{array}
    \!\right),\;
  \lambda^8 =\left(\!
    \begin{array}{ccc}
      \!\!\!\frac{1}{\sqrt{3}}\!\!\! & 0 & 0 \\
      0 & \!\!\!\frac{1}{\sqrt{3}}\!\!\! & 0 \\
      0 & 0 & \!\!\!\frac{-2}{\sqrt{3}}\!\!\!
    \end{array}
    \!\right)\, .
\end{equation*}
By looking at the first of these, together with the $t_{ab}^C
\cA^C_\mu \psi_b$ term of $\cL_Q$, one can immediately get a feel for
what gluons do: a gluon with (adjoint) colour index $C=1$ acts on
quarks through the matrix $t^1 = \frac12\lambda^1$. That matrix takes
green quarks ($b=2$) and turns them into red quarks ($a=1$), and vice
versa. In other words, when a gluon interacts with a quark it
\emph{repaints} the colour of the quark, taking away one colour and
replacing it with another. The likelihood with which this happens is
governed by the strong coupling constant $g_s$.
Note that the repainting analogy is less evident for some of the other
colour matrices, but it still remains essentially correct.

The second part of the QCD Lagrangian is purely gluonic
\begin{equation}
  \label{eq:Lg}
  \cL_G = -\frac14 F^{\mu \nu}_A F^{A\,\mu \nu}
\end{equation}
where the gluon field tensor $F^A_{\mu\nu}$ is given by
\begin{equation}
  \label{eq:Fmunu}
  F^A_{\mu\nu} = \partial_\mu \cA_{\nu}^A - \partial_\nu \cA_{\nu}^A
  - {g_s\, f_{ABC} \cA^B_\mu \cA^C_\nu\,\qquad [t^A,t^B] =
    i f_{ABC}t^C}\,,
\end{equation}
where the $f_{ABC}$ are the structure constants of SU(3) (defined
through the commutators of the $t^{A}$ matrices). 
Note the major difference with QED here, namely the presence of a term
$g_s\, f_{ABC} \cA^B_\mu \cA^C_\nu$ with two gluon fields. 
The presence of such a term is one of the major differences with QED,
and, as we will discuss in more detail below, it will be responsible for
the fact that gluons interact directly with gluons.
For now, note simply that it has to be there in order for the theory
to be gauge invariant under local SU(3) transformations:
\begin{align}
  \psi_a &\to\; e^{i \theta_C(x) t_{ab}^C} \psi_b\\
  \cA^C t^C &\to\; e^{i\theta^D(x) t^D} \left(\cA^C t^C - \frac1{g_s}
  \partial_\mu \theta^C(x) t^C\right) e^{-i\theta^E(x) t^E} 
\end{align}
where, in the second line, we have dropped the explicit subscript
${ab}$ indices, and the $\theta^C(x)$ are eight arbitrary real
functions of the space-time position $x$.

\subsection{`Solving QCD'}

There are two main first-principles approaches to solving QCD: lattice
QCD and perturbative QCD.\footnote{In addition, effective-theory
  methods provide ways of looking at QCD that make it easier to solve,
  given certain `inputs' that generally come from lattice or
  perturbative QCD (and sometimes also from experimental
  measurements). These lectures won't discuss effective theory
  methods, but for more details you may consult the lectures at this
  school by Martin Beneke.
  Another set of methods that has seen much development in recent
  years makes use of the `AdS/CFT'
  correspondence~\cite{Maldacena:1997re,Gubser:1998bc,Witten:1998qj},
  relating QCD-like models at strong coupling to gravitational models
  at weak coupling (e.g., ~\cite{Erlich:2005qh,DaRold:2005zs}).  }

\subsubsection{Lattice QCD}
\begin{figure}
  \centering
  \includegraphics[width=0.5\tw]{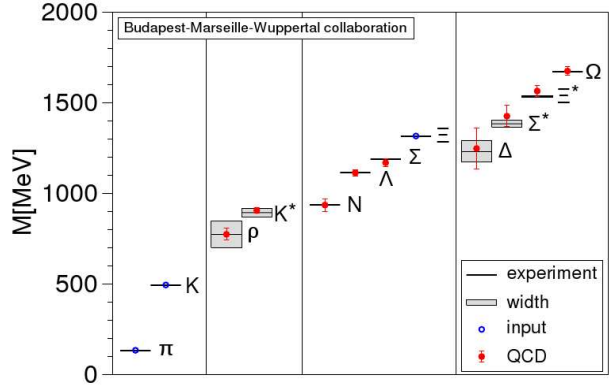}\\
  \caption{The measured spectrum of hadron masses, compared to a
    lattice calculation~\cite{Durr:2008zz}. The open blue circles
    are the hadron masses that have been used to fix the three
    parameters of the calculation: the value of the QCD coupling,
    the average of the up and down quark masses (taken equal) and the
    strange-quark mass.
    All other points are results of the calculation.
    \label{fig:lattice-spectrum}
  }
\end{figure}

The most complete approach is lattice QCD. It involves discretizing
space-time, and considering the values of the quark and gluon fields
at all the vertices/edges of the resulting 4-dimensional lattice (with
imaginary time).
Through a suitable Monte Carlo sampling over all possible field
configurations, one essentially determines the relative likelihood of
different field configurations, and this provides a solution to QCD.
This method is particularly suited to the calculation of static
quantities in QCD such as the hadron mass spectrum.
The results of such a lattice calculation are illustrated in
Fig.~\ref{fig:lattice-spectrum}, showing very good agreement.

Lattice methods have been successfully used in a range of contexts,
for example, in recent years, in helping extract fundamental
quantities such as the CKM matrix (and limits on new physics) from the
vast array of experimental results on hadron decays and oscillations
at flavour factories.
Unfortunately lattice calculations aren't suitable in all
contexts. Let us imagine, briefly, what would be required in order to
carry out lattice calculations for LHC physics: since the 
centre-of-mass energy is (will be) $14\TeV$, we need a lattice spacing of order
$1/(14\TeV) \sim 10^{-5} \fm$ to resolve everything that happens.
Non-perturbative dynamics for quarks/hadrons near rest takes place on
a timescale $t\sim \frac{1}{0.5\GeV} \sim 0.4\fm/c$.
But hadrons at LHC have a boost factor of up to $10^4$, so the extent
of the lattice should be about $4000\fm$.
That tells us that if we are to resolve  high-momentum transfer
interactions and at the same time follow the evolution of quark and
gluon fields up to the point where they form hadrons, we would need
about $4\times 10^8$ lattice units in each direction, of $\sim 3\times
10^{34}$ nodes. 
Not to mention the problem with high particle multiplicities (current
lattice calculations seldom involve more than two or three particles)
and all the issues that relate to the use of imaginary time in lattice
calculations.
Of course, that's not to say that it might not be possible, one day,
to find clever tricks that would enable lattice calculations to deal
with high-energy reactions.
However, with today's methods, any lattice calculation of the
properties of LHC proton--proton scattering seems highly unlikely.
For this reason, we will not give any further discussion of lattice
QCD here, but instead refer the curious reader to textbooks and
reviews for more
details~\cite{Smit:2002ug,DeGrand:2006zz,NaraProceedings,Onogi:2009eg}. 

\subsubsection{Perturbative QCD}

Perturbative QCD relies on the idea of an order-by-order expansion in
a small coupling $\as = \frac{g_s^2}{4\pi}\ll 1$. Some given
observable $f$ can then be predicted as
\begin{equation}
  f = f_1 \as + f_2 \as^2 + f_3 \as^3 + \ldots\,,
\end{equation}
where one might calculate just the first one or two terms of the
series, with the understanding that remaining ones should be small.

\begin{figure}
  \mbox{ }\hspace{-2em}
  \scalebox{1}{
  \begin{minipage}[t]{0.27\tw}
    \centering
    {\includegraphics[scale=0.8]{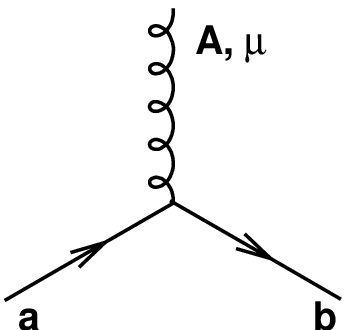}\\
      $-ig_s t^A_{ba} \gamma^\mu$}
  \end{minipage}\quad
  \begin{minipage}[t]{0.32\tw}
    \centering%
    {\includegraphics[scale=0.8]{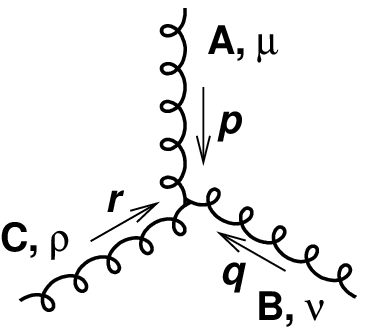}\\
      $-g_s f^{ABC}[(p-q)^\rho g^{\mu\nu}$\\
      $+(q-r)^\mu g^{\nu\rho}$\\$ + (r-p)^\nu g^{\rho\mu}]$}
  \end{minipage}\quad
  \begin{minipage}[t]{0.37\tw}
    \centering
    {\includegraphics[scale=0.8]{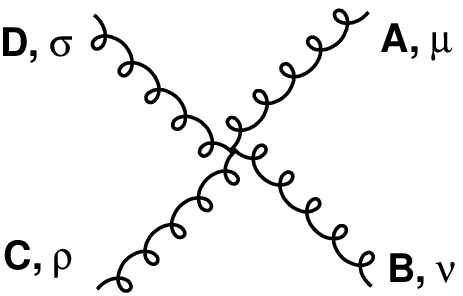}\\
      $-ig_s^2 f^{XAC}f^{XBD}[g^{\mu\nu}g^{\rho\sigma} -
      g^{\mu\sigma}g^{\nu\gamma}] + (C,\gamma)\leftrightarrow(D,\rho)
      + (B,\nu)\leftrightarrow(C,\gamma) $}
  \end{minipage}}
  \caption{The interaction vertices of the Feynman rules of QCD}
\label{fig:feynman-interactions}
\end{figure}

The principal technique to calculate the coefficients $f_i$ of the
above series is through the use of Feynman diagrammatic (or other
related) techniques. The interaction vertices of the QCD Feynman rules
are shown in Fig.~\ref{fig:feynman-interactions} (in some gauges one
also needs to consider ghosts, but they will be irrelevant for our
discussions here).

\begin{figure}
  \centering
  \includegraphics[scale=0.8]{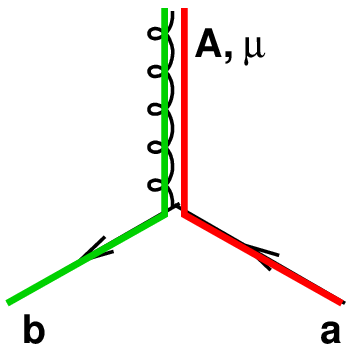}\qquad
  \includegraphics[scale=0.8]{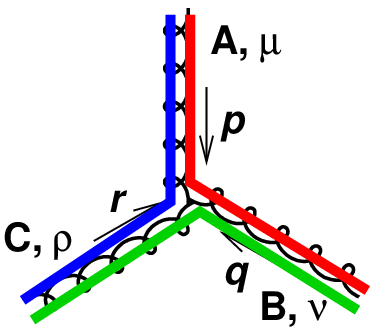}
  \caption{Schematic colour flow interpretation of the
    quark--quark--gluon ($t^A_{ab}$, left) and triple-gluon ($f_{ABC}$,
    right) vertices of QCD.
    These interpretations are only sensible insofar as one imagines
    that the number of colours in QCD, $N_c=3$, is large.  }
  \label{fig:colour-flow}
\end{figure}

The $qqg$ interaction in Fig.~\ref{fig:feynman-interactions} comes
from the ${\bar\psi}_a {g_s \gamma^\mu t^C_{ab} \cA^C_\mu}\psi_b$ term
of the Lagrangian. We have already discussed how the $t^C_{ab}$ factor
affects the colour of the quark, and this is represented in
Fig.~\ref{fig:colour-flow}(left), with the gluon taking away one
colour and replacing it with another.

The triple-gluon vertex in Fig.~\ref{fig:feynman-interactions} comes
from the $-\frac14 F^{\mu \nu}_A F^{A\,\mu \nu}$ part of the
Lagrangian, via the product of a $\partial_\mu \cA_\nu$ term in one
$F^{\mu \nu}_A$ factor with the $g_s f_{ABC} \cA_\mu^B \cA_\nu^C$ term
in the other.
It is the fact that gluons carry colour charge that means that they
must interact with other gluons. 
In terms of colour flows, we have the repetition of the idea that the
emission of a gluon can be seen as taking away the colour from the
gluon (or anti-colour) and replacing it with a different one.
Because of the double colour/anti-colour charge of a gluon, one can
anticipate that it will interact with (or emit) other gluons twice as
strongly as does a quark.
Before coming to mathematical formulation of that statement, let's
comment also on the 4-gluon vertex of
Fig.~\ref{fig:feynman-interactions}. This comes from the product of
two $g_s f_{ABC} \cA_\mu^B \cA_\nu^C$ type terms in $-\frac14 F^{\mu
  \nu}_A F^{A\,\mu \nu}$ and is order $g_s^2$ whereas the two other
interactions are order $g_s$.

Though Fig.~\ref{fig:colour-flow} gives some idea of how the colour
factors $t^C_{ab}$ and $f_{ABC}$ in the Feynman rules are to be
understood, it is useful to see also how they arise in calculations. After
squaring an amplitude and summing over colours of incoming and
outgoing particles, they often appear in one or other of the following
combinations:
\begin{subequations}
  \label{eq:colour}
  \begin{align}
    \label{eq:colour:g2qq}
    \Tr (t^A t^B) = T_R \delta^{AB}\,,\quad {T_R = \frac12} & \qquad
    \begin{minipage}[c]{0.3\linewidth}
      \includegraphics[scale=0.7]{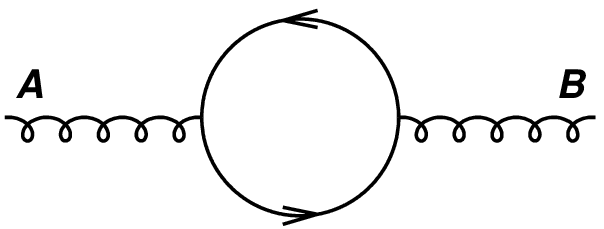}
    \end{minipage}
    \\
    \label{eq:colour:q2qg}
    \sum_A t^A_{ab} t^A_{bc} = C_F \delta_{ac}\,,\quad {C_F =
      \displaystyle\frac{\NC^2-1}{2\NC} = \frac43} & \qquad
    \begin{minipage}[b]{0.3\linewidth}
      \includegraphics[scale=0.7]{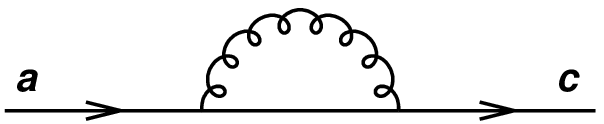}
    \end{minipage}
    \\
    \label{eq:colour:g2gg}
    \sum_{C,D} f^{ACD}f^{BCD} = \CA \delta^{AB}\,,\quad {\CA = \NC =
      3} & \qquad
    \begin{minipage}[c]{0.3\linewidth}
      \includegraphics[scale=0.7]{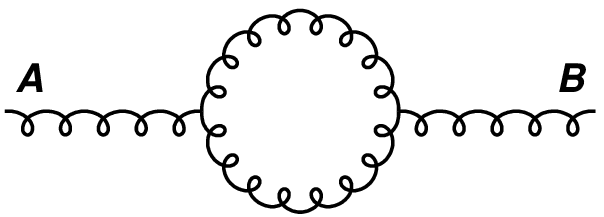}
    \end{minipage}
    \\
    \label{eq:colour:fierz}
    t^A_{ab} t^A_{cd} = \displaystyle\frac12 \delta_{bc} \delta_{ad} -
    \frac1{2\NC} \delta_{ab}\delta_{cd}\, \text{ (Fierz)} & \qquad
    \begin{minipage}[c]{0.3\linewidth}
      \includegraphics[scale=0.7]{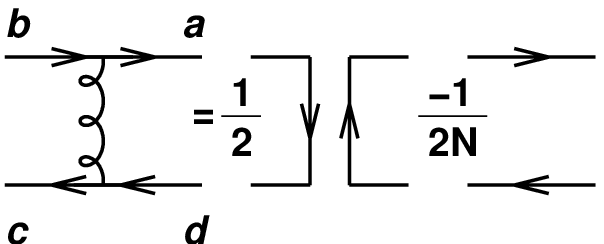}
    \end{minipage}
  \end{align}
\end{subequations}
where $N \equiv \NC = 3$ is the number of colours in QCD and it is useful to
express the results for general numbers of colours (because it is
sometimes useful to consider how results depend on $\NC$, especially
in the limit $\NC \to \infty)$.
Each mathematical combination of colour factors has a diagrammatic
interpretation. Equation~(\ref{eq:colour:g2qq}) corresponds to a gluon
splitting into $q\bar q$ which then join back into a gluon; or, the
sum over colours in the squared amplitude for $g\to q\bar
q$. Equation~(\ref{eq:colour:q2qg}) corresponds to the square of gluon
emission from a quark. Equation~(\ref{eq:colour:g2gg}) arises as the square
of gluon emission from a gluon. One sees that there is almost a factor
of 2 between Eqs.~(\ref{eq:colour:q2qg}) and (\ref{eq:colour:g2gg})
(modulo corrections terms $\sim 1/\NC$), which is the mathematical
counterpart of our statement above that gluons emit twice as strongly
as quarks.
Finally the approximate colour-flow interpretation that we had in
Fig.~\ref{fig:colour-flow}(left) can be stated exactly in terms of
the Fierz identity, Eq.~(\ref{eq:colour:fierz}).

\subsubsection{The running coupling}

Most higher-order QCD calculations are carried out with dimensional
regularization (the use of $4-\epsilon$ dimensions) in order to handle
the ultraviolet divergences that appear in loop diagrams.
In the process of going from $4$ to $4-\epsilon$ dimensions, one needs
to introduce an arbitrary `renormalization' scale, generally called
$\mu$, in order to keep consistent dimensions (units) for all
quantities.\footnote{The renormalization procedure itself, i.e., the
  removal of the $1/\epsilon$ divergences, is usually carried out in
  the modified minimal subtraction ($\MSbar$) scheme (see, e.g.,
  Section~11.4 of Ref.~\cite{Peskin:1995ev}), by far the most widespread
  scheme in QCD. }
The value of the QCD coupling, $\as = \frac{g_s^2}{4\pi}$, depends on
the scale $\mu$ at which it is evaluated.
That dependence can be expressed in terms of a renormalization group
equation
\begin{equation}
  \label{eq:renorm-group}
  \frac{d \as(\mu^2)}{d \ln \mu^2} = \beta(\as(\mu^2))\,,\qquad
  \beta(\as) = -\as^2 (b_0 + b_1 \as + b_2 \as^2 + \ldots)\,,
\end{equation}
where 
\begin{equation}
  \label{eq:beta-coeffs}
  b_0 = \frac{11\CA - 2\nf}{12\pi}\,,\qquad 
  { b_1 = \frac{17\CA^2 - 5\CA\nf -3\CF\nf}{24\pi^2} =
    \frac{153-19\nf}{24\pi^2}}\, ,
\end{equation}
with $\nf$ being the number of `light' quark flavours, those whose
mass is lower than $\mu$.
The negative sign in Eq.~(\ref{eq:renorm-group}) is the origin of
asymptotic freedom, the fact that the coupling becomes weaker at high
momentum scales, i.e., the theory almost becomes a free theory, in
which quarks and gluons don't interact. Conversely at low momentum
scales the coupling grows strong, causing quarks and gluons to be
tightly bound into hadrons.
The importance of the discovery of these features was recognized in
the 2004 Nobel prize to Gross, Politzer and Wilczek.
Why does the QCD $\beta$-function have the opposite sign of that in
QED? The fact that the vector particles (gluons) of the theory carry
colour charge is central to the result. 
However, while there have been various attempts to give simple but
accurate explanations for the negative sign~\cite{RunningA,RunningB},
in practice they all end up being quite involved.\footnote{You might
  still want to check the sign for yourself: if so, pick up a copy of
  Peskin and Schroeder~\cite{Peskin:1995ev}, arrange to have an
  afternoon free of interruptions, and work through the derivation.}
So, for the purpose of these lectures, let us just accept the results.

If we ignore all terms on the right of Eq.~(\ref{eq:renorm-group})
other than $b_0$, and also ignore the subtlety that the number of
`light' flavours $\nf$ depends on $\mu$, then there is a simple
solution for $\as(\mu^2)$:
\begin{equation}
  \label{eq:renorm-group-1loop-soln}
  \as(\mu^2) = 
  \frac{\as(\mu_0^2)}{1 + b_0 \as(\mu_0^2) \ln \frac{\mu^2}{\mu_0^2}} 
  = \frac{1}{b_0 \ln \frac{\mu^2}{\Lambda^2}}\,,
\end{equation}
where one can either express the result in terms of the value of the
coupling at a reference scale $\mu_0$, or in terms of a
non-perturbative constant $\Lambda$ (also called $\Lambda_{QCD}$), the
scale at which the coupling diverges. Only for scales $\mu \gg
\Lambda$, corresponding to $\as(\mu^2)\ll 1$, is perturbation theory
valid.
Note that $\Lambda$, since it is essentially a non-perturbative
quantity, is not too well defined: for a given $\as(\mu_0)$, its
value depends on whether we used just $b_0$ in
Eq.~(\ref{eq:renorm-group}) or also $b_1$, etc.
However, its order of magnitude, $200\MeV$, is physically meaningful
insofar as it is closely connected with the scale of hadron masses.

\begin{figure}
  \centering
  \includegraphics[width=0.45\tw]{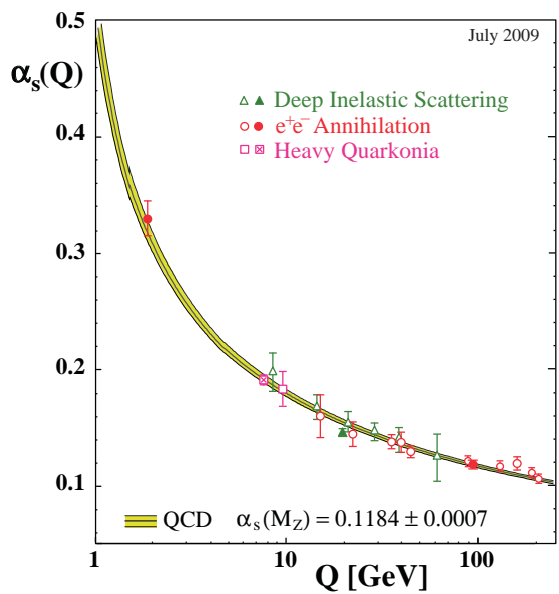}
  \caption{The QCD coupling as measured in physics processes at
    different scales $Q$, together with the band obtained by running
    the world average for $\as$ within its uncertainties. Figure taken
    from Ref.~\cite{Bethke:2009jm}.}
  \label{fig:running-coupling}
\end{figure}

One question that often arises is how $\mu$, the renormalization
scale, should relate to the physical scale of the process. We will
discuss this in detail later (Section~\ref{sec:fixed-order}), but for
now the following simple statement is good enough: the strength of the
QCD interaction for a process involving a momentum transfer $Q$ is
given by $\as(\mu)$ with $\mu\sim Q$.
One can measure the strength of that interaction in a range of
processes, at various scales, and 
Fig.~\ref{fig:running-coupling}~\cite{Bethke:2009jm} shows a compilation of such measurements,
together with the running of an average over many measurements,
$\as(M_Z) = 0.1184\pm 0.0007$, illustrating the good consistency of
the measurements with the expected running.

\subsubsection{QCD predictions and colliders}

Colliders like the Tevatron and the LHC are mainly geared to
investigating phenomena involving high-momentum transfers (more
precisely large transverse-momenta), say in the range $50\GeV$ to
$5\TeV$. There, the QCD coupling is certainly small and we would hope
to be able to apply perturbation theory.
Yet, the initial state involves protons, at whose mass scale, $m_p
\simeq 0.94\GeV$, the coupling is certainly not weak.
And the final states of collider events consist of lots of
hadrons. Those aren't perturbative either. 
And there are lots of them --- tens to hundreds. Even if we wanted to
try, somehow, to treat them perturbatively, we would be faced with
calculations to some very high order in $\as$, at least as high as the
particle multiplicity, which is far beyond what we can
calculate exactly: depending on how you count, at hadron colliders,
the best available complete calculation (i.e., all diagrams at a given
order), doesn't go beyond $\as^2$ or $\as^3$. Certain subsets of
diagrams (e.g., those without loops) can be calculated up $\as^{10}$
roughly.

So we are faced with a problem. Exact lattice methods can't deal with
the high momentum scales that matter, exact perturbative methods can't
deal with low momentum scales that inevitably enter the problem, nor
the high multiplicities that events have in practice.
Yet, it turns out that we are reasonably successful in making
predictions for collider events.
These lectures will try to give you an understanding of the methods
and approximations that are used.
%

\section{Considering $e^+e^- \to \text{hadrons}$}
\label{sec:ee2hadrons}

One simple context in which QCD has been extensively studied over the
past 30 years is that of $e^+e^-$ annihilation to hadrons.
This process has the theoretical advantage that only the final state
involves QCD.
Additionally, huge quantities of data have been collected at quite a
number of colliders, including millions of events at the $Z$ mass at
LEP and SLC.
We therefore start our investigation of the properties of QCD by
considering this process.

\subsection{Soft and collinear limits}
\label{sec:soft-coll}

There is one QCD approximation that we will repeatedly make use of,
and that is the soft and collinear approximation. 
`Soft' implies that an emitted gluon has very little energy compared
to the parton (quark or gluon) that emitted it.
`Collinear' means that it is emitted very close in angle to another
parton in the event.
By considering gluons that are soft and/or collinear one can
drastically simplify certain QCD calculations, while still retaining
much of the physics.

The soft and collinear approximation is sufficiently important that
it's worth, at least once, carrying out a calculation with it, and
we'll do that in the context of the emission of a gluon from $e^+e^- \to
q\bar q$ events.
Though there are quite a few equations in the page that follows, the
manipulations are all quite simple!
We're interested in the hadronic side of the $e^+e^- \to q\bar q$
amplitude, so let's first write the QED matrix element for a virtual photon
$\gamma^* \to q\bar q$ (we can always put back the $e^+e^- \to
\gamma^*$ and the photon propagator parts later if we need to ---
which we won't):
\begin{equation*}
  \cM_{q\bar q} = \bar u_a(p_1) i e_q \gamma_\mu \delta_{ab} v_b(p_2)\,\qquad
  \begin{minipage}[c]{0.15\tw}
    \includegraphics[width=\tw]{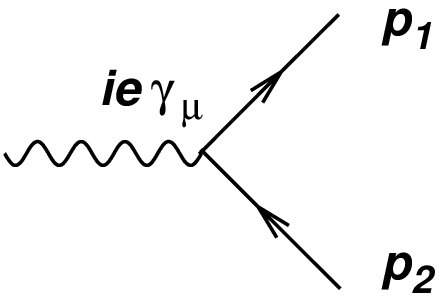}
  \end{minipage}\,,
\end{equation*}
where the diagram illustrates the momentum labelling. Here $\bar
u(p_1)$ and $v(p_2)$ are the spinors for the outgoing quark and
anti-quark (taken massless), $e_q$ is the quark's electric charge and
the $\gamma_\mu$ are the Dirac matrices. In what follows we shall drop
the $a,b$ quark colour indices for compactness and reintroduce them
only at the end.

The corresponding amplitude including the emission of a gluon with
momentum $k$ and polarization vector $\epsilon$ is
\begin{subequations}
  \label{eq:Mqqg-start}
  \begin{align}
    \cM_{q\bar qg} &=\;
    \begin{minipage}[c]{0.15\linewidth}
      \includegraphics[width=\tw]{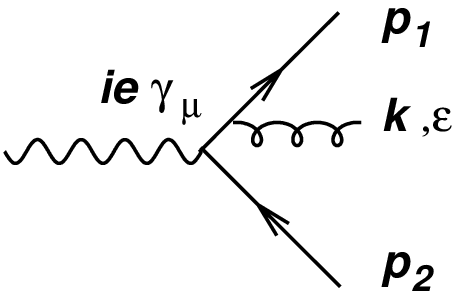}
    \end{minipage}
    \quad+ \quad
    \begin{minipage}[c]{0.15\linewidth}
      \includegraphics[width=\tw]{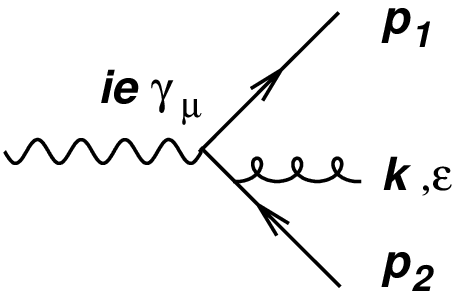}
    \end{minipage}
    \\
    & = -\bar u(p_1) ig_s \sleps t^A \frac{i(\slp_1+\slk)}{(p_1+k)^2}
    i e_q \gamma_\mu v(p_2) + \bar u(p_1) i e_q \gamma_\mu
    \frac{i(\slp_2+\slk)}{(p_2+k)^2} ig_s \sleps t^A v(p_2)\,,
  \end{align}
\end{subequations}
with one term for emission from the quark and the other for emission
from the anti-quark and use of the notation $\slashii{p} = p_\mu
\gamma_\mu$.
Let's concentrate on the first term, collecting the factors of $i$,
and using the anti-commutation relation of the $\gamma$-matrices,
$\slA \slB = 2A.B - \slB\slA$, to write 
\begin{subequations}
  \begin{align}
    i \bar u(p_1) g_s \sleps t^A \frac{(\slp_1+\slk)}{(p_1+k)^2} e_q
    \gamma_\mu v(p_2) &= i g_s \bar u(p_1) \frac{[ 2\epsilon.(p_1+k)
      - (\slp_1+\slk) \sleps]}{(p_1+k)^2} e_q \gamma_\mu t^A v(p_2)\,,
    \\ %
    &\simeq i g_s\frac{ p_1.\epsilon }{p_1.k}\, \bar u(p_1) e_q
    \gamma_\mu t^A v(p_2)\,,
  \end{align}
\end{subequations}
where to obtain the second line we have made use of the fact that
$\bar u(p_1) \slp_1 = 0$, $p_1^2 = k^2 = 0$, and taken the soft
approximation $k_\mu \ll p_\mu$, which allows us to neglect the terms
in the numerator that are proportional to $k$ rather than $p$.
The answer including both terms in Eq.~(\ref{eq:Mqqg-start}) is
\begin{equation}
  \label{eq:Mqqg-result}
   \cM_{q\bar qg} \simeq 
   \bar u(p_1) i e_q
    \gamma_\mu t^A v(p_2)
    \cdot g_s \left(\frac{ p_1.\epsilon }{p_1.k} - \frac{ p_2.\epsilon }{p_2.k}\right),
\end{equation}
where the first factor has the Lorentz structure of the $\cM_{q\bar
  q}$ amplitude, i.e., apart from the colour matrix $t^A$, $\cM_{q\bar
  q}$ is simply proportional to the $\cM_{q\bar q}$ result.
We actually need the squared amplitude, summed over polarizations and
colour states, 
\begin{multline}
  \label{eq:Mqqg_squared}
  |\cM_{q\bar qg}|^2 \simeq \sum_{A,a,b,\text{pol}} 
    \left|\bar u_a(p_1) i e_q
      \gamma_\mu {t^A} v_b(p_2) \;g_s 
      { \left(\frac{p_1.\epsilon}{p_1.k} -
          \frac{p_2.\epsilon}{p_2.k}\right)} \right|^2
  \\       = -|M_{q\bar q}^2| {\CF} g_s^2
    {\left(\frac{p_1}{p_1.k}-\frac{p_2}{p_2.k}\right)^2}
    = |M_{q\bar q}^2| 
    {\CF} g_s^2
    {\frac{2p_1.p_2}{(p_1.k) (p_2.k) }}\,.
\end{multline}
We have now explicitly written the quark colour indices $a,b$ again.
To obtain the second line we have made use of the result that
$\sum_{A,a,b} t^A_{ab} t^{A}_{ba} = 
\CF \NC$ [cf.\ Eq.~(\ref{eq:colour:q2qg})], whereas for $|M_{q\bar
  q}^2|$ we have $\sum_{A,a,b} \delta_{ab} t^{A}_{ba}= \NC$.
To carry out the sum over gluon polarizations we have exploited the fact that
$\sum_{\text{pol}} \epsilon_\mu(k) \epsilon_\nu^*(k) = -g_{\mu\nu}$,
plus terms proportional to $k_\mu$ and $k_\nu$ that disappear when
dotted with the amplitude and its complex conjugate.

One main point of the result here is that in the soft limit, the
$|\cM_{q\bar qg}|^2$ 
squared matrix element \emph{factorizes} into two terms: the
$|\cM_{q\bar q}|^2$ matrix element and a piece with a rather simple
dependence on the gluon momentum.

The next ingredient that we need is the the phase space for the $q\bar q
g$ system, $d\Phi_{q\bar q g}$. In the soft approximation, we can
write this 
\begin{equation}
  d\Phi_{q\bar q g} \simeq d\Phi_{q\bar q} \frac{d^3\vec k}{2E (2\pi)^3}\,,
\end{equation}
where $E\equiv E_k$ is the energy of the gluon $k$. We see that the
phase space also factorizes. Thus we can write the full differential
cross section for $q\bar q$ production plus soft gluon emission as the
$q\bar q$ production matrix element and phase space, $|\cM_{q\bar
  q}|^2 d\Phi_{q\bar q}$, multiplied by a soft gluon emission
probability $d\cS$,
\begin{equation}
  |\cM_{q\bar qg}|^2 d\Phi_{q\bar q g} \simeq |\cM_{q\bar q}|^2
  d\Phi_{q\bar q} d\cS\,,
\end{equation}
with
\begin{equation}
  \label{eq:dS}
  d\cS =  E d E\, d\!\cos\theta
        \,\frac{d\phi}{2\pi} \cdot \frac{2\as\CF}{\pi}
        \frac{2p_1.p_2}{(2p_1.k) (2p_2.k) }\,,
\end{equation}
where we have used $d^3 k = E^2 dE d\cos \theta d\phi$, expressing the
result in terms of the polar ($\theta$) and azimuthal ($\phi$) angles
of the gluon with respect to the quark (which itself is back-to-back
with the antiquark, since we work in the centre-of-mass frame and
there is negligible recoil from the soft gluon).
With a little more algebra, we get our final result for the
probability of soft gluon emission from the $q\bar q$ system
\begin{equation}
  \label{eq:dS-final}
  d\cS = \frac{2\as\CF}{\pi} \,\frac{d E}{ E}
        \frac{d\theta}{\sin\theta} \, \frac{d\phi}{2\pi}\,.
\end{equation}
This result has two types of non-integrable divergence: one, called
the soft (or infrared) divergence when $E \to 0$ and the other, a
collinear divergence, when $\theta \to 0$ (or $\pi$), i.e., the gluon
becomes collinear with the quark (or antiquark) direction.
Though derived here in the specific context of $e^+e^-\to q\bar q$
production, these soft and collinear divergences are a very general
property of QCD and appear whenever a gluon is emitted from a quark,
regardless of the process.

\subsection{The total cross section}
\label{sec:total-xsct}

If we want to calculate the $\order{\as}$ corrections to the total
cross section, the diagrams included in Eq.~(\ref{eq:Mqqg-start}) are
not sufficient. We also need to include a one-loop correction
(`virtual'), specifically, the interference between one-loop
$\gamma^* \to q\bar q$ diagrams and the tree-level $\gamma^* \to q\bar
q$ amplitude, for example a contribution such as
\begin{equation*}
  \includegraphics[scale=0.6]{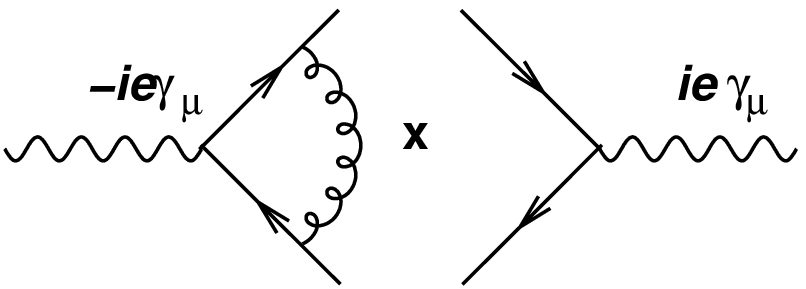}
\end{equation*}
which has the same perturbative order (number of $g_s$ factors) as
the square of Eq.~(\ref{eq:Mqqg-start}). 

The total cross section for the production of hadrons must be
finite. 
The integral over the gluon-emission correction has two
non-integrable, logarithmic divergences.
These divergences must therefore somehow be cancelled by corresponding
divergences in the virtual term. 
This is the requirement of \emph{unitarity}, which is basically the
statement that probability of anything happening must add up to $1$.
The most straightforward way of doing the full calculation for the
total cross section is to use dimensional regularization in the phase
space integral for the real emission diagram and for the integration
over the loop momentum in the virtual diagram. 
However, in order just to visualize what is happening one can also
write
\begin{equation}
      \sigma_{tot} = \sigma_{q\bar q}\left(
      1 + \frac{2\as\CF}{\pi}
      \int\frac{d{E}}{{E}}
      \int\frac{d\theta}{\sin\theta}
      \, { R\!\left(\frac{E}{Q},\theta\right)} 
      - \frac{2\as\CF}{\pi}
      \int\frac{d{E}}{{E}}
      \int\frac{d\theta}{\sin\theta}
      \, { V\!\left(\frac{E}{Q},\theta\right)} 
    \right)\,,
\end{equation}
where the first term, $1$, is the `Born' term, i.e., the production
of just $q\bar q$, the second term is the real emission term, and the
third term is the loop correction.
Since we need to integrate gluon emission beyond the soft and
collinear region, we have introduced a function $R(E/Q,\theta)$, which
parametrizes the deviation of the matrix-element from its soft limit
when $E \sim Q$, with $Q$ the centre-of-mass energy of the
process. $R$ has the property $\lim_{E\to 0} R(E/Q,\theta)
= 1$.
We have written the virtual term in a similar form, using
$V(E/Q,\theta)$ to parametrize its structure (we cheat a bit, since
the loop momentum integral includes regions of phase space where the
gluon is offshell; this won't matter though for us here).
The statement that real and virtual divergences cancel means that $V$
should be identical to $R$ in the soft or collinear limits
\begin{equation}
  \label{eq:VR-properties}
   \lim_{{E}\to 0}
  (R-V) = 0\,,\qquad\quad \lim_{\theta\to 0,\pi} (R-V) = 0   \, .  
\end{equation}
Thus the corrections to the total cross section come from the region
of hard ($E\sim Q$), large-angle gluons (for which perturbation theory
is valid).
There's a good reason for this: soft and collinear emission takes
place on a time-scale $\sim 1/(E\theta^2)$ that is long compared to that,
$\sim 1/Q$, for the production of the $q\bar q$ pair from the virtual
photon.  
Anything that happens long after the production of the $q\bar q$ pair
cannot change the fact that there will be a QCD final state (though it
can change the properties of that final state), and so it does not
affect the total cross section.
Similarly, whatever dynamics is involved in effecting the transition
between partons and hadrons is also expected to occur on a long
time-scale ($\sim 1/\Lambda$) and so should not modify the total cross
section. 
This is important because it allows us to neglect the issue that we cannot
directly compute the properties of hadron production.

The fact that the corrections to the total cross section are dominated
by a region of hard gluon emission is reflected in a reasonable
behaviour for the perturbative series 
\begin{equation}
  \label{eq:tot-xsct-series}
  \sigma_{tot} = \sigma_{q\bar q}\left( 1 + 1.045 \frac{\as(Q)}{\pi} + 
    0.94 \left(\frac{\as(Q)}{\pi}\right)^2 - 15
    \left(\frac{\as(Q)}{\pi}\right)^3 + \cdots\right)\,,
\end{equation}
where we have expressed the result in terms of $\as$ evaluated at a
renormalization $\mu=Q$ and the coefficients correspond to $Q=M_Z$
(for all known terms in the series, including electroweak effects, see 
Refs.~\cite{Gorishnii:1990vf,Surguladze:1990tg,Hebbeker:1994ih,Chetyrkin:1996ia,Baikov:2008jh,Baikov:2009uw}
as well as references therein).

\subsection{The final state}
\label{sec:ee-final-state}

As a first step towards asking questions about the final state, our
next exercise is to attempt to determine the mean number of gluons 
that are emitted from a quark with energy $\sim Q$. If the emission
probability is small ($\propto \as$) then to first order in the
coupling the mean number of emitted gluons is equal to the probability
of emitting one gluon
\begin{equation}
  \label{eq:Ng-integral}
  \langle N_g \rangle \simeq \frac{2\as \CF}{\pi} \int^Q \frac{dE}{E}
  \int^{\pi/2} \Theta(E \theta > Q_0)\,.
\end{equation}
The integral diverges for $E\to 0$ and $\theta \to 0$, however, we can
reasonably argue that the divergent structure only holds as long as
perturbation theory is valid. This motivates us to cut the divergences
off at a scale $Q_0 \sim \Lambda$, because below that scale the
language of quarks and gluons loses its meaning.
That immediately tells us that we should have $E \gtrsim Q_0$, but
it's not so immediately clear how the $\theta$ integral will be cut
off.
It turns out, for reasons to do with invariance of the small-angle
emission pattern as one boosts the quark in the (longitudinal)
direction of its motion, that the correct variable to cut on is
\emph{transverse momentum}, $k_t \sim E\theta$.
We therefore find, to first order in the coupling,
\begin{equation}
  \label{eq:Ng-result}
  \langle N_g \rangle \simeq \frac{\as \CF}{\pi} \ln^2 \frac{Q}{Q_0} +
  \order{\as \ln Q}\,,
\end{equation}
where we have explicitly kept track only of the term with the largest
number of logarithms.
If we take $Q_0 = \Lambda$, how big is this result? We have to decide
on the scale for $\as$. Being hopelessly optimistic, i.e., taking $\as
= \as(Q) = (2b \ln Q/\Lambda)^{-1}$ gives us 
\begin{equation}
  \label{eq:Ng-result-running}
  \langle N_g \rangle \simeq \frac{\CF}{2b\pi} \ln \frac{Q}{\Lambda}
  \simeq \frac{\CF}{4b^2\pi \as}\,,
\end{equation}
which, numerically, corresponds to $\langle N_g \rangle \simeq 2$.
This is neither small numerically, nor parametrically ($\sim 1/\as$).
Does this render perturbation completely useless for anything other
than total cross sections?

We can follow two possible avenues to help answer this question. One
approach is to calculate the next order, and see what structure it
has. Alternatively we can ask whether there are final-state observables that
have a better-behaved perturbative series than `the mean number of
gluons'.

\subsubsection{Gluon (and hadron) multiplicity}
\label{sec:gluon-multiplicity}

Once one gluon has been emitted, it can itself emit further gluons. 
To understand what the gluon multiplicity might look like to higher
orders, it's useful to write down the general pattern of emission of a
soft gluon both from a quark and from a gluon, which is essentially
independent of the process that produced the `emitter':
\begin{subequations}
  \label{eq:fs-universal-splitting}
  \begin{align}
    \label{eq:fs-quark-splitting}
    \begin{minipage}[c]{0.34\linewidth}
      \includegraphics[width=0.8\tw]{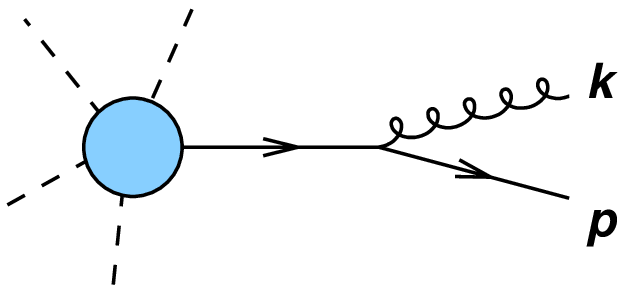}
    \end{minipage}
    &\simeq\quad \frac{2\as \CF}{\pi} \frac{dE}{E}
    \frac{d\theta}{\theta}\,,
    \\
    \label{eq:fs-gluon-splitting}
    \begin{minipage}[c]{0.34\linewidth}
      \includegraphics[width=0.8\tw]{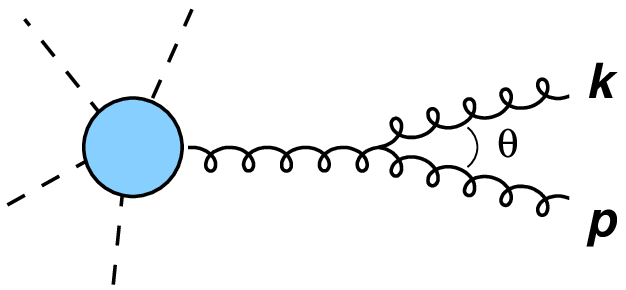}
    \end{minipage}
    &\simeq\quad \frac{2\as \CA}{\pi} \frac{dE}{E}
    \frac{d\theta}{\theta}\,.
  \end{align}
\end{subequations}
These expressions are valid when the emitted gluon is much lower in
energy than the emitter, $k \ll p$, and when the emission angle
$\theta$ is much smaller than the angle between the emitter and any
other parton in the event (this is known as the condition of angular
ordering~\cite{Coherence}).
The structure of emission of a soft gluon is almost identical from a
quark and from a gluon, except for the substitution of the $\CF=4/3$
colour factor in the quark case with the $\CA=3$ colour factor in the
gluon case.

\begin{figure}
  \centering
  a)\!\!\includegraphics[height=0.22\tw]{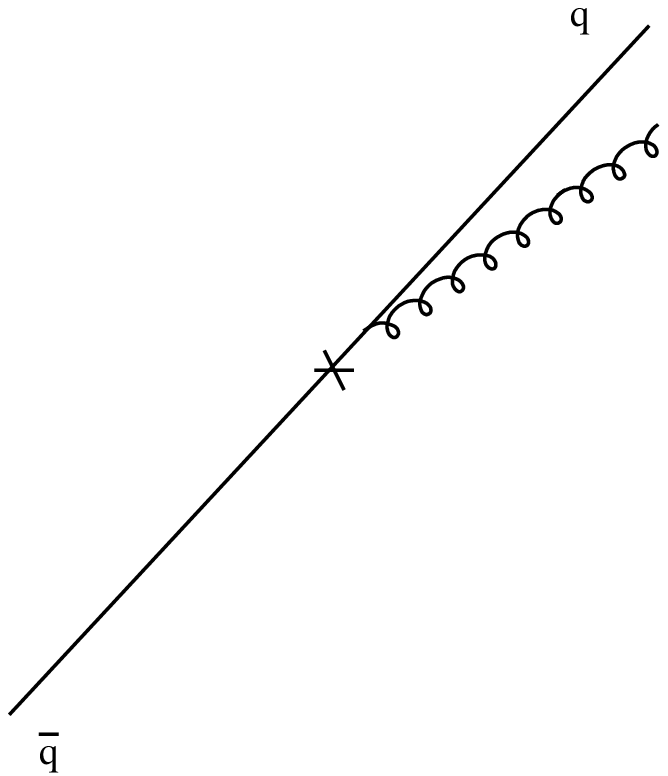}
  b)\!\!\includegraphics[height=0.22\tw]{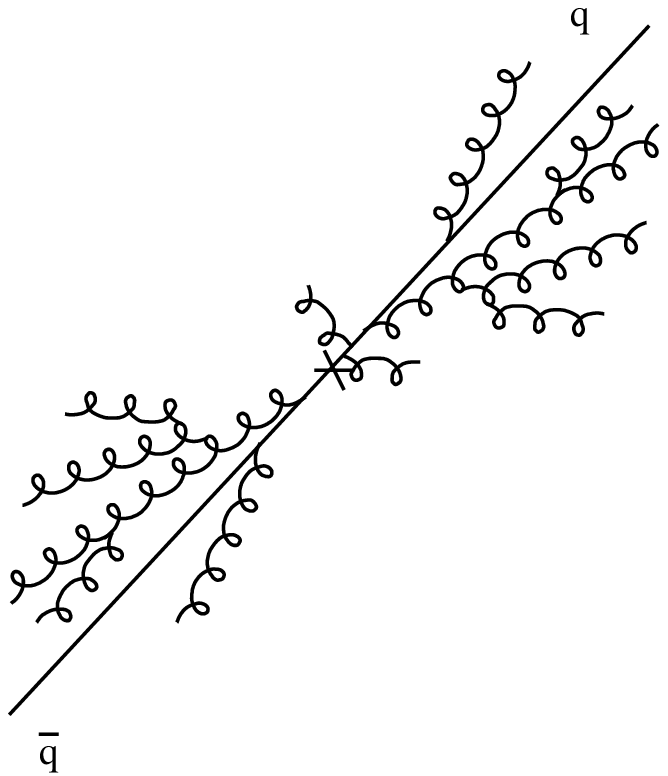}
  c)\!\!\includegraphics[height=0.22\tw]{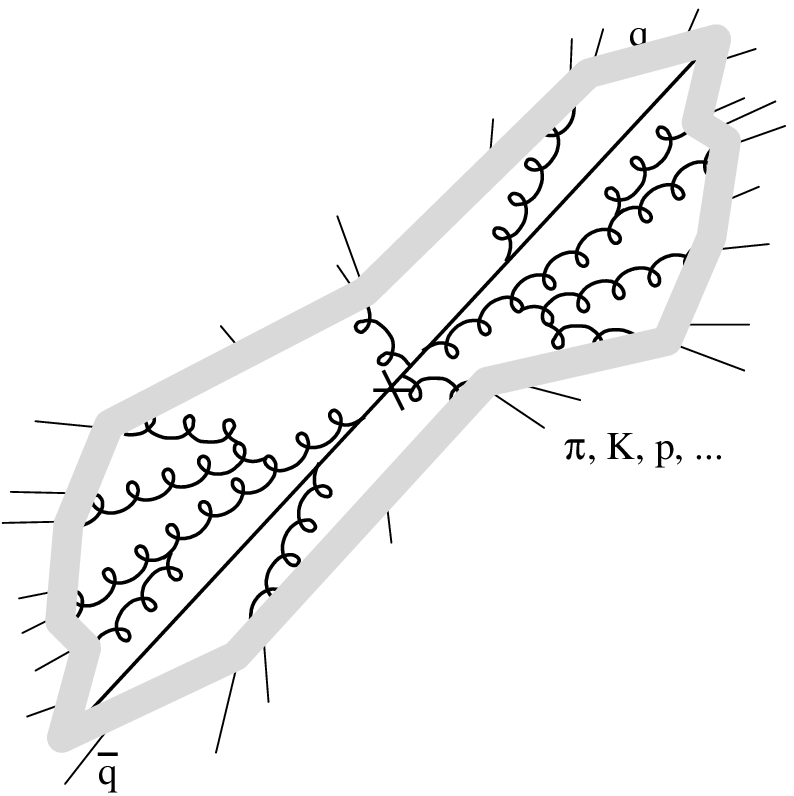}\qquad
  d)\includegraphics[height=0.22\tw]{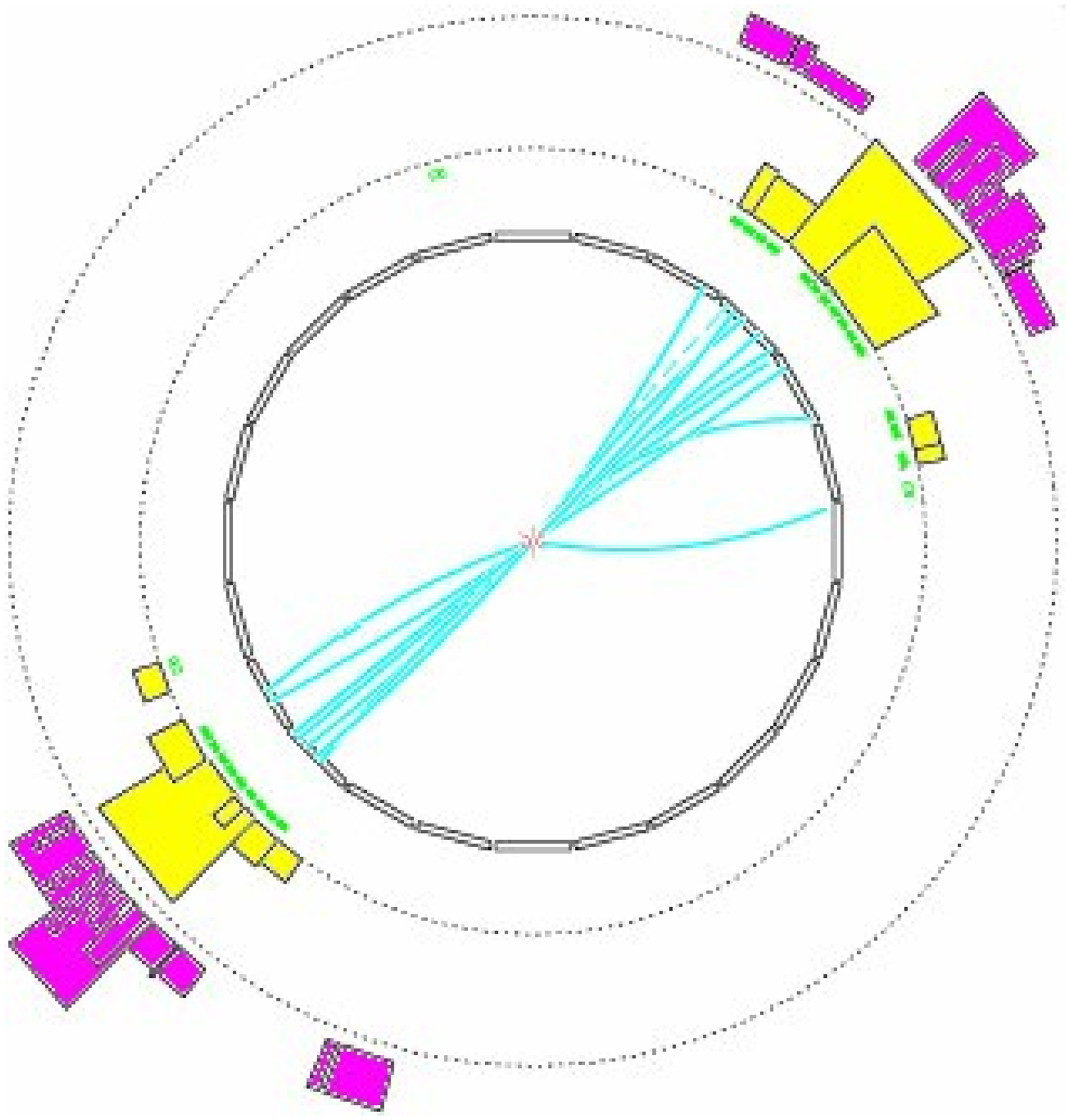}
  \caption{Emission pattern from a $q\bar q$ event, with first a
    single gluon (a), then multiple emissions of gluons both from the
    $q\bar q$ pair and from the previously emitted gluons (b),
    followed by some process, `hadronization', that causes hadrons
    to be produced from the gluons, giving an event (c), that
    structurally resembles a real event (d) ($e^+e^- \to Z \to
    $hadrons at LEP in the OPAL detector)}
  \label{fig:soft-coll-emission-v-event}
\end{figure}

Since quarks and gluons emit in similar ways, every gluon that is
emitted from the quark can itself emit further gluons, and so
forth. Most of the emissions will either be in almost the same
direction as the original quark (due to the collinear divergence)
and/or be soft.
This is represented in Figs.~\ref{fig:soft-coll-emission-v-event}(a) 
and~(b) (for simplicity we've not shown gluons splitting to $q\bar q$ pairs,
which also occurs, with just a collinear divergence).
This still only gives a description of events in terms of quarks and
gluons, whereas real events consist of hadrons. 
Though hadronization, the transition from quarks and gluon to hadrons
is not something that we know how to calculate from first principles,
one idea that has had some success is Local Parton Hadron Duality
(LPHD) (see, e.g., Ref.~\cite{Dokshitzer:1991wu}).
It states that after accounting for all gluon and quark production down
to scales $\sim \Lambda$, the transition from partons to hadrons is
essentially local in phase space.
Thus the hadron directions and momenta will be closely related to the
partons', and the hadron multiplicity will reflect the parton
multiplicity too. This is illustrated in
Fig.~\ref{fig:soft-coll-emission-v-event}(c), comparing it also to the
picture of a real event,
Fig.~\ref{fig:soft-coll-emission-v-event}(d). The latter illustrates how
the hadrons do tend to have the same collimated angular distribution
as is predicted for gluons, with the small number of exceptions having
low energy (i.e., soft) as can be seen from the larger curvature in
the experiment's magnetic field.

This comparison with a single event is suggestive that our picture of
gluon emission and hadronization might be reasonable.
A more quantitative test can be obtained by calculating the number of
emitted gluons. 
This requires the extension of
Eqs.~(\ref{eq:Ng-integral})--(\ref{eq:Ng-result-running}) to multiple gluon
emission. 
The full calculation doesn't fit into the space available for these
lectures (see instead textbook discussions in
Refs.~\cite{Ellis:1991qj,Dokshitzer:1991wu}), but the basic idea is that
there are terms $(\as \ln^2 Q/Q_0)^n$ for all 
orders $n$ and that one can calculate their coefficients
analytically. The structure of the result is
\begin{equation}
  \label{eq:Ng-all-orders}
  \langle N_g \rangle \sim \frac{\CF}{\CA} \sum_{n=1}^\infty \frac{1}{(n!)^2}
  \left(\frac{\CA}{2\pi b^2 \as}\right)^n 
  \sim \frac{\CF}{\CA} \exp \left(\sqrt{\frac{2\CA}{\pi b^2 \as(Q)} }\right)\,,
\end{equation}
where we've neglected to write the prefactor in front of the
exponential, and we've also not given the subleading
terms~\cite{Mueller:1982cq}.

\begin{figure}
  \centering
  \includegraphics[width=0.5\tw]{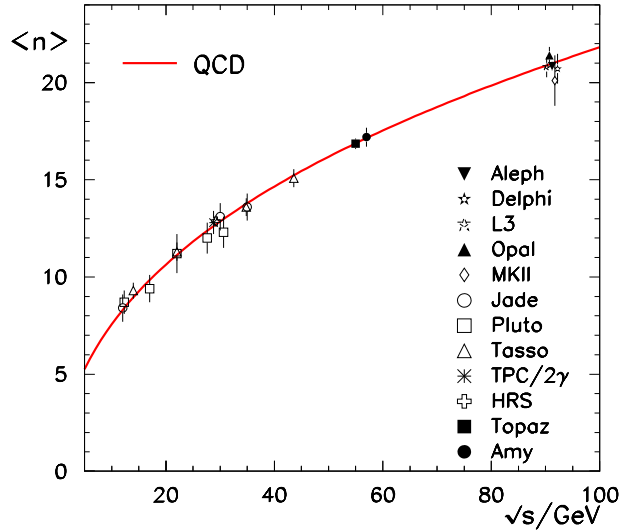}\\
  \caption{Multiplicity of charged hadrons in $e^+e^-\to$hadrons
   events, comparing the experimental data at a range of
   centre-of-mass energies $Q$, with the QCD prediction using a fitted
   normalisation and non-perturbative scale $\Lambda$. Figure adapted
   from Ref.~\cite{Ellis:1991qj}.}
  \label{fig:hadron-multiplicity}
\end{figure}%

How is Eq.~(\ref{eq:Ng-all-orders}) to be related to the hadron
multiplicity? The simplest assumption is that each final parton gives
some (unknown) fixed number of hadrons which must be fitted to data.
Equation~(\ref{eq:Ng-all-orders}) then predicts not the total hadron
multiplicity but its energy dependence. 
That prediction is illustrated in Fig.~\ref{fig:hadron-multiplicity}
and shows remarkable agreement with data over a range of energies,
providing strong evidence that the picture outlined above is a fair
reflection of `reality'.

The above approach can be extended to calculate other properties of
events such as the energy spectrum of hadrons, the fluctuations in the
number of hadrons, and even correlations between hadrons, generally
with reasonable success.
However, as one probes more detailed features of events, the
analytical calculations become significantly more complicated and one
also becomes increasingly sensitive to the oversimplicity of the LPHD
concept.
Having said that, the same ideas that we are using, i.e., the
importance of multiple soft and collinear splitting together with a
transition from partons to hadrons, are, in numerical form, at the
base of widely used Monte Carlo parton-shower event generators like
{\sc Pythia}, {\sc Herwig} and {\sc Sherpa}.
We will discuss them in more detail in Section~\ref{sec:MC}.

\subsubsection{Infrared safe observables}
\label{sec:ir-safe-obs}

It is heartening that the above soft-collinear discussion gave such a
good description of the data. However, it did involve the application
of perturbation theory to kinematic regions where its validity is
questionable, the need to calculate dominant contributions at all
orders in $\as$, and the introduction of a free parameter to `fudge'
the fact that we don't understand the non-perturbative physics. 
A natural question is therefore whether one can formulate final-state
observables for which these problems are not present.

The answer is that one can. For an observable to be calculate based on
just one or two orders of perturbation theory it should be infrared
and collinear (IRC) safe. In the words of Ref.~\cite{Ellis:1991qj}:
\begin{quote}
  For an observable's distribution to be calculable in [fixed-order]
  perturbation theory, the observable should be infra-red safe, \ie
  insensitive to the emission of soft or collinear gluons.  In
  particular if ${\vec p}_i$ is any momentum occurring in its
  definition, it must be invariant under the branching
  \begin{equation*}
    \label{eq:ESW3.39}
    {\vec p}_i \to {\vec p}_j + {\vec p}_k
  \end{equation*}
  whenever ${\vec p}_j$ and ${\vec p}_k$ are parallel [collinear] or
  one of them is small [infrared].
\end{quote}
For example, the multiplicity of gluons is not IRC safe, because it is
modified by soft and collinear splitting. 
The energy of the hardest particle in an event is not IRC safe,
because it is modified by collinear splitting.
However, the total energy flowing into a given cone is IRC safe,
because soft emissions don't modify the energy flow, and collinear
emissions don't modify its direction.

This last example comes from Sterman and Weinberg
\cite{Sterman:1977wj}, who defined an $e^+e^-$ event as having 2
`jets' if at least a fraction $(1-\epsilon)$ of the event's energy
is contained in two cones of half-angle $\delta$. 
If we take $\delta$ to be $30^\degree$ and $\epsilon = 0.1$, then
Fig.~\ref{fig:soft-coll-emission-v-event}(d) is an example of such a
2-jet event.
We can adapt our expression for the total cross section,
Eq.~(\ref{eq:tot-xsct-series}), to give us the 2-jet cross section as
follows
\begin{equation}
      \sigma_{\text{2-jet}} = \sigma_{q\bar q}\left(
      1 + \frac{2\as\CF}{\pi}
      \int\frac{d{E}}{{E}}
      \int\frac{d\theta}{\sin\theta} \left[
         R\!\left(\frac{E}{Q},\theta\right)
         \left(1 -
           \Theta\left(\frac{E}{Q}-\epsilon\right)\Theta(\theta-\delta)\right) 
      -  V\!\left(\frac{E}{Q},\theta\right)  \right]
    \right)\,.
\end{equation}
For small $E$ or $\theta$ this is just like the total cross section,
with full cancellation of divergences between real and virtual terms
[cf.\ Eq.~(\ref{eq:VR-properties})].
For large $E$ and large $\theta$ a \emph{finite} piece of
real-emission cross section is cut out by the factor 
$(1 - \Theta(\frac{E}{Q}-\epsilon)\Theta(\theta-\delta))$
and it corresponds to scales
with $E\sim Q$ and large angles, for which perturbation theory is valid.
This then gives 
\begin{equation}
  \label{eq:2-jet-series}
  \sigma_{\text{2-jet}} = \sigma_{q\bar q} (1 - c_1 \as + c_2 \as^2 + \cdots)\,,
\end{equation}
where $c_1$, $c_2$, etc. are all of order $1$ (as long as $\epsilon$
and $\delta$ were not taken too small).
Similarly one could define a 3-jet cross section by requiring that it
not be a 2-jet event and that all but a fraction $\epsilon$ of the
energy be contained in 3 cones of half angle $\delta$. 
This would give a cross section of the form
\begin{equation}
  \label{eq:3-jet-series}
  \sigma_{\text{3-jet}} = \sigma_{q\bar q} (c_1' \as + c_2' \as^2 + \cdots)\,,
\end{equation}
where, again, the coefficients are all $\order{1}$.
So whereas the cross section for getting an extra gluon is divergent,
the cross section for an extra jet is finite and small, $\order{\as}$.
One difficulty with the extension of the Sterman--Weinberg definition
to multiple jets is to know how to place the cones. 
Since jet-finding is a well-developed subject in its own right, we
will return to in detail in Section~\ref{sec:jets}.

The Sterman--Weinberg jet cross section gives a discrete classification
of events: an event either has two jets, or more. 
An alternative class of infrared and collinear safe observables is
that of event shapes, which give a continuous classification of
events.
The most widely studied example is the thrust, $T$, 
\begin{equation}
  T = \max_{{\vec n}_T} \frac{\sum_i |{\vec p}_i.{\vec n}_T|}{ \sum_i |{\vec p}_i|}\,,
\end{equation}
where the sum runs over all particles, and one chooses the thrust axis
${\vec n}_T$ (a 3-dimensional unit vector) so as to maximize the
projection in the numerator.
For a perfectly collimated $2$-jet event, the thrust axis aligns with
the jet axes and the thrust is $1$.
For a `Mercedes' type event with three identical collimated jets, the
thrust axis will be aligned with any one of the three jets and the
thrust will be $2/3$.
Intermediate events will have intermediate values of the thrust.

\begin{figure}
  \centering
  \includegraphics[width=0.6\tw]{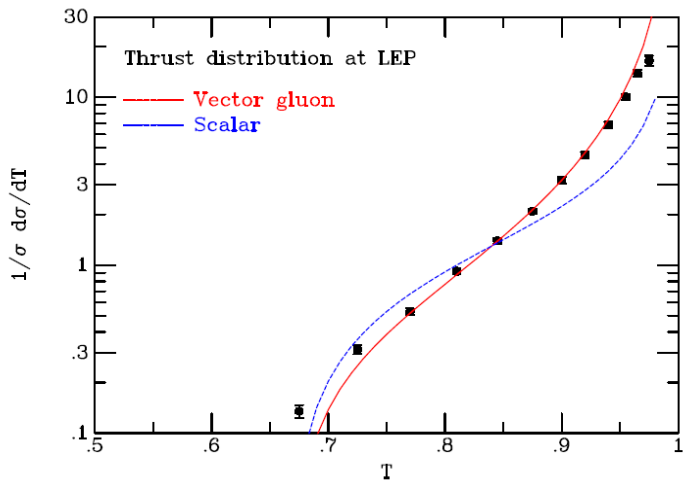}
  \caption{Measured thrust distribution at LEP compared to leading
    order predictions based on QCD (vector gluon, solid red line) and
    a modified version of QCD in which the gluon is a scalar (spin 0,
    dashed blue line) rather than a vector (spin-1) particle. Figure
    taken from CERN academic training lectures by B.~R.~Webber.}
  \label{fig:vector-scalar-thrust}
\end{figure}

One application of the thrust variable is given in
Fig.~\ref{fig:vector-scalar-thrust}. It shows data for the thrust
distribution from LEP, compared to $\order{\as}$ calculations of the
thrust distribution in QCD and in a variant of QCD in which the gluon
is a scalar particle rather than a vector particle.
The scalar gluon case does not have a divergence for soft emission
(the collinear divergence is still there), with the result is that the
distribution diverges less strongly in the 2-jet limit than for vector
gluons.
The data clearly prefer the vector-gluon case, though they do also
show the need for higher-order corrections at thrust values close to
$2/3$ and to $1$.

More generally, event shapes like the thrust have seen broad use in
measurements of the QCD coupling, tuning of Monte Carlo event
generators (see Section~\ref{sec:MC}), studies of the hadronization
process, and also as an event-topology discriminant in searches for
decays of particles beyond the Standard Model.

\subsection{Summary}

The $e^+e^-\to \text{hadrons}$ process has allowed us to examine many
of the basic ideas of perturbative QCD: soft and collinear
divergences, the question of which observables are perturbatively
calculable or not (related to infrared and collinear safety) and even
what happens if one takes perturbation theory seriously outside its
strict domain of applicability (one acquires a rough understanding of
the collimated, high-multiplicity structure of real events).

\section{Parton distribution functions}
\label{sec:PDFs}

Having considered processes that involve hadrons in the final state,
let us now examine what happens when they are present in the initial
state. 
The importance of understanding initial-state hadrons is obvious at
the LHC. Within the `parton model', we write, for example, the hadron
collider cross section to produce a $Z$ and a Higgs boson as
\begin{equation}
  \label{eq:parton-model-pp}
  \sigma = \int dx_1 { f_{q/p}(x_1)} \int dx_2 {
      f_{\bar q/p}(x_2)}
    \,{ \hat \sigma_{q\bar q \to ZH} (x_1 x_2 s)}\,, 
    \begin{minipage}[c]{0.35\linewidth}
      \includegraphics[width=\tw]{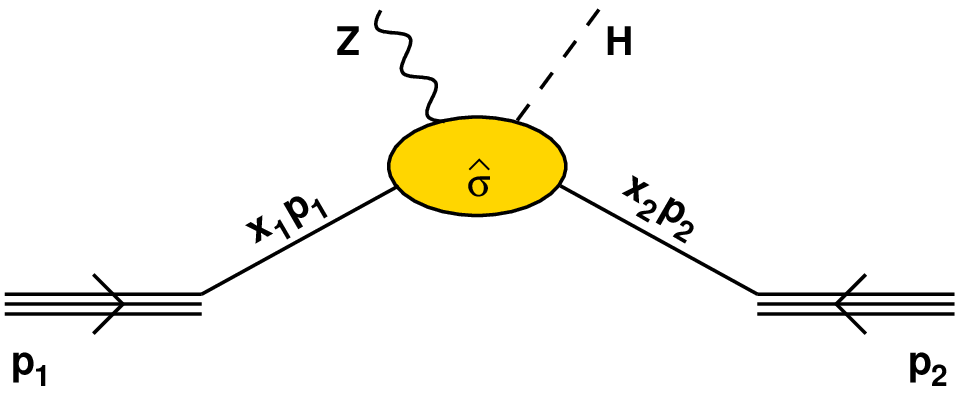}
    \end{minipage}
\end{equation}
where $s = (p_1+p_2)^2$ is the squared $pp$ centre-of-mass energy,
$f_{q/p}(x_1)$ is the number density of quarks of type $q$ 
carrying a fraction $x_1$ of the momentum of the first proton, and
similarly with $f_{\bar q/p}(x_2)$ for the other proton.
The $f_{q/p}(x)$ functions are known as `parton distribution
functions' (PDFs).
They multiply the `hard' (here, electroweak) cross section, $\hat
\sigma_{q\bar q \to ZH} (x_1  x_2 s)$ for the process $q\bar q
\to ZH$, a function of the squared partonic ($q\bar q$) centre-of-mass energy,
$\hat s = x_1 x_2 s$.
After integrating over $x_1$ and $x_2$ (and summing over quark
species), one obtains the total cross section for $p p \to ZH$.
The above form seems quite intuitive, but still leaves a number of
questions: for example, how do we determine the momentum distributions
of quarks and gluons inside the proton? 
How does the `factorization' into PDFs and a hard part stand up to
QCD corrections?
Understanding these questions is crucial if, one day, we are to take
measured cross sections for $ZH$ and interpret them, for example, in
terms of the coupling of the Higgs to the $Z$.
And they're just as crucial for practically any other physics analysis
we might want to carry out at the LHC.

The parton distribution functions are properties of the
(non-perturbative) proton. 
A natural first question is whether we can calculate the PDFs based on
lattice QCD. 
In principle, yes, and there is ongoing work in this direction (see,
e.g., Ref.~\cite{Zanotti:2007zz}), however, currently lattice QCD has
not reached an accuracy in these calculations that is competitive with
the combination of experimental measurements and perturbative QCD
analyses that we discuss below.

\subsection{Deep Inelastic Scattering}
\label{sec:DIS}

\begin{figure}
  \centering
  \begin{minipage}[c]{0.25\linewidth}
    \includegraphics[width=\tw]{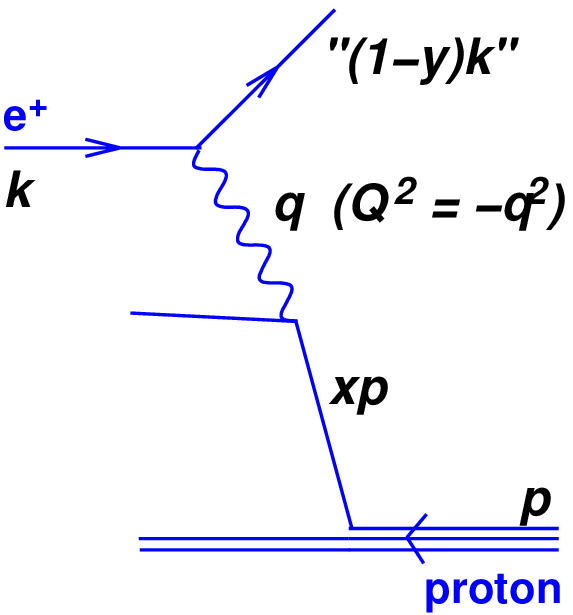}
  \end{minipage}\qquad\qquad
  \begin{minipage}[c]{0.50\linewidth}
    \includegraphics[width=\tw]{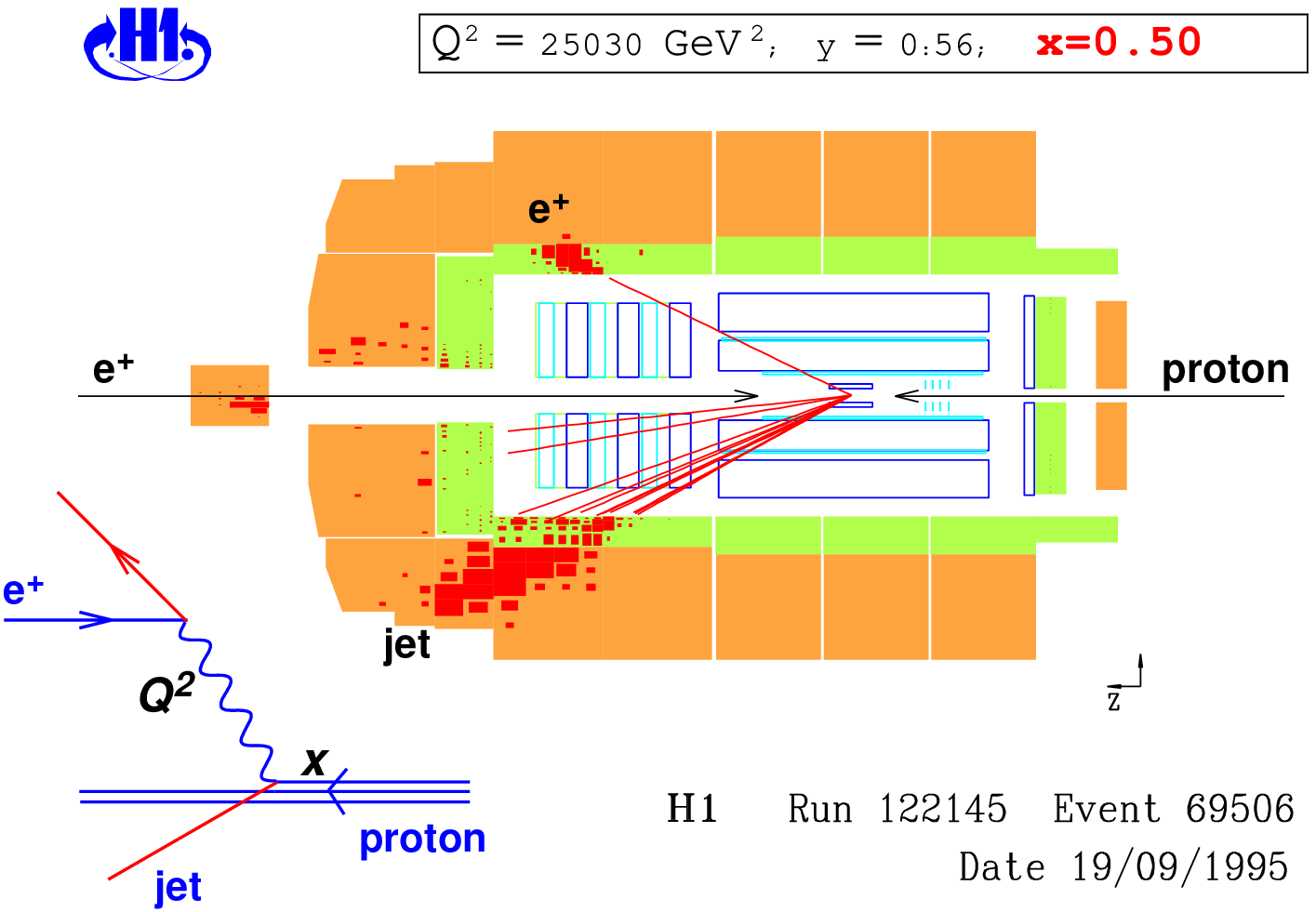}
  \end{minipage}
  \caption{Left: kinematic variables of DIS; right: illustration of an
    event as it appeared in practice in the H1 detector at HERA}
  \label{fig:DIS}
\end{figure}

The process where we have learnt the most about PDFs is Deep Inelastic
Scattering (DIS), i.e., lepton--proton scattering in which the photon
that is exchanged between lepton and proton has a large virtuality.
The kinematics of the ``quark-parton-model'' DIS process is
represented in Fig.~\ref{fig:DIS} (left) and an event from the H1
detector at HERA is shown on the right. Kinematic variables that are
usually defined are
\begin{equation}
  \label{eq:DIS-kin-var}
  Q^2 = -q^2\,,\qquad 
  x = \frac{Q^2}{2p.q}\,,\qquad 
  y = \frac{p.q}{p.k}\,,
\end{equation}
where $Q^2$ is the photon virtuality, $x$ is the longitudinal momentum
fraction of the struck quark in the proton, and $y$ is the momentum
fraction lost by the electron (in the proton rest frame).

To zeroth order in $\as$ (the `quark parton model'), the DIS cross
section can be written as
\begin{equation}
  \label{eq:DIS-xsct}
  \frac{d^2 \sigma^{em}}{dx dQ^2} = \frac{4\pi \alpha^2}{xQ^4} \left(
    \frac{1+(1-y)^2}{2}F_2^{em} + \order{\as}\right)\,,
\end{equation}
written in terms of the $F_2$ structure function, which, to zeroth
order in $\as$ is given by 
\begin{equation}
  \label{eq:F2em}
  F_2^{em} = x\sum_{i=q,\bar q} e_i^2 f_{i/p}(x) + \order{\as}\,,
\end{equation}
(the ``$em$'' superscript is a reminder that we're only considering
the electromagnetic part here)\,.

Given the sum over all flavours in Eq.~(\ref{eq:F2em}), disentangling
the information about individual flavours might seem like quite a
daunting task.
We can attempt to see where the information comes from by starting off
with the assumption that the proton consists just of up and down
quarks, in which case
\begin{equation}
  \label{eq:F2ud}
  F_2^{\text{proton}} = x(e_u^2 u_p(x) + e_d^2 d_p(x)) =
  x\left(\frac49 u_p(x) + \frac19 d_p(x)\right)\,,
\end{equation}
where we have introduced the shorthand $f_{u/p}(x) = u_p(x)$,
etc. (later we will drop the ``$p$'' subscript altogether).
In Eq.~(\ref{eq:F2ud}) we now have a linear combination of just two
functions.
The next step is to use isospin symmetry, the fact that the neutron is
essentially just a proton with $u \leftrightarrow d$, i.e., $u_n(x)
\simeq d_p(x)$ (ignoring small electromagnetic effects), so that
\begin{equation}
  \label{eq:F2udneutron}
  \frac1xF_2^{\,\text{neutron}} = \frac{4}{9} u_n(x) + \frac{1}{9} u_n(x)
  \simeq \frac{4}{9}d_p(x) + \frac{1}{9}u_p(x)\, .
\end{equation}
Appropriate linear combination of $F_2^{\text{proton}}$ and
$F_2^{\text{neutron}}$ (in practice one uses deuterons, or nuclei as a
source of neutrons) therefore provides separate information on $u_p(x)$
and $d_p(x)$.

\begin{figure}
  \centering
  \begin{minipage}[c]{0.4\linewidth}
    \psfrag{F2ndToxU}[Br][Br]{\footnotesize$3F_2^p - \frac65 F_2^d \to $``${x}u(x)$''}
    \psfrag{F2ndToxD}[Br][Br]{\footnotesize$-3F_2^p + \frac{24}5 F_2^d \to $``${x}d(x)$''}
    \includegraphics[width=\tw]{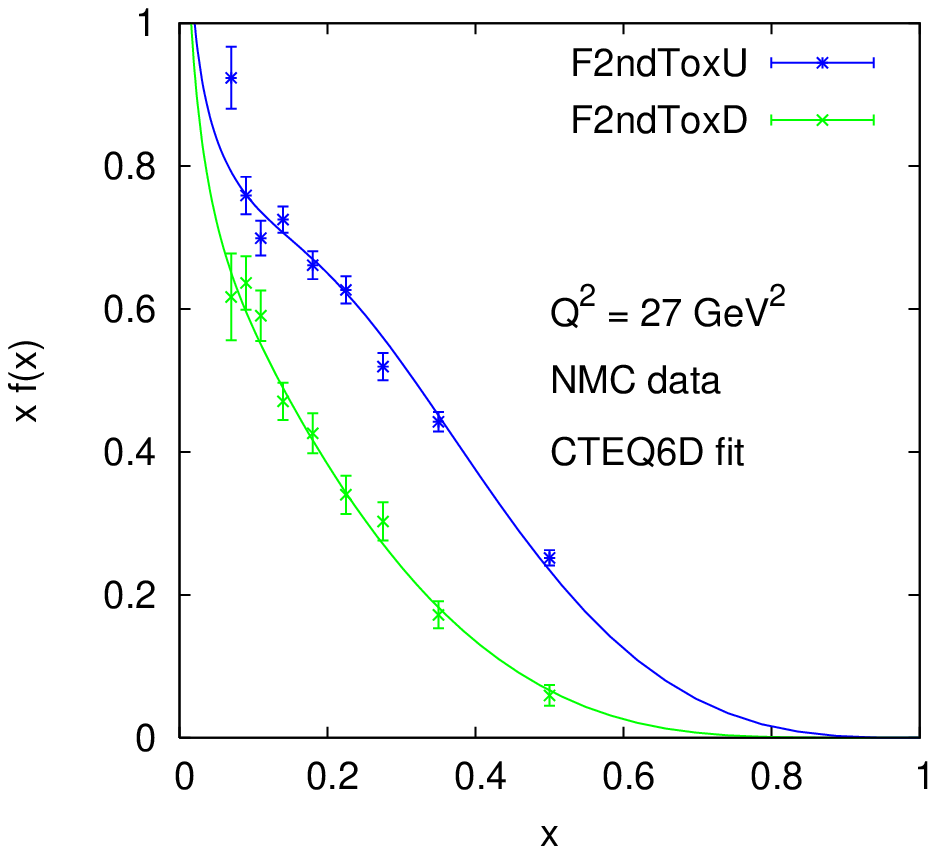}
  \end{minipage}\qquad\quad
  \begin{minipage}[c]{0.33\linewidth}
    \includegraphics[width=\tw]{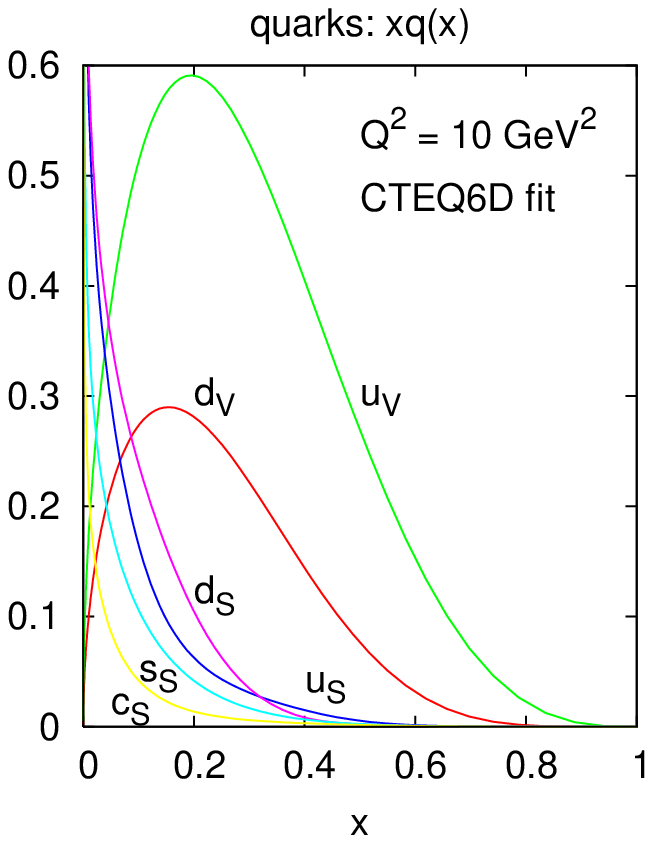}
  \end{minipage}
  \caption{Left: linear combinations of NMC $F_2$ data \cite{Arneodo:1996qe}
    for protons and deuterons so as to obtain $x u(x)$ and $x d(x)$,
    assuming only $u$ and $d$ quarks inside the proton, together with
    the expectations from the same linear combination based on the
    CTEQ6D PDF parametrizations \cite{Pumplin:2002vw}.
    Right: results for different valence and sea quark distributions
    from CTEQ6D at $Q^2 = 10\GeV^2$.
  }
  \label{fig:nmc-data}
\end{figure}

The results of this exercise are shown in Fig.~\ref{fig:nmc-data}
(left).  As expected we see more up quarks than down quarks.
But there's also a problem: let's try to extract the total number of
up quarks, $U = \int_0^1 dx u(x)$. We see that the data increase
towards small $x$, and a parametrization (CTEQ6D
\cite{Pumplin:2002vw}) of these and other data even seems to diverge
as $x\to 0$, $xu(x) \sim xd(x) \sim x^{-0.25}$.\footnote{The
  particular parametrization shown here is not very widely used
  nowadays, but is particularly convenient for comparisons to DIS data
  (it corresponds to a definition of PDFs known as the ``DIS
  scheme,'' for which higher-order corrections to many DIS cross
  sections are particularly simple). The behaviour at small $x$ is
  common to essentially all parametrizations.} %
Given that the plot is supposed to be for $xu(x)$ and that we need to
integrate $u(x)$ to get the total number of up quarks in the proton,
it looks like we'll obtain an infinite number of up quarks, which is
hardly consistent with expectations from the picture of a proton as
being a $uud$ state.

One thing we've `neglected' is that there can also be anti-up and
anti-down quarks in the proton, because the proton wavefunction can
fluctuate, creating $u\bar u$ and $d\bar d$ pairs, `sea quarks', and
so give rise to $\bar u(x)$ and $\bar d(x)$ distributions. Therefore
instead of Eq.~(\ref{eq:F2ud}), we should have written
\begin{equation}
  \label{eq:F2udubardbar}
  F_2^{\text{proton}} = 
  \frac49 (xu_p(x) + x\bar u_p(x)) + \frac19 (d_p(x) + \bar d_p(x))\,,
\end{equation}
since quarks and antiquarks have identical squared charges. 
So what we called ``$xu(x)$'' in Fig.~\ref{fig:nmc-data}(left) was
actually $xu(x) + x\bar u(x)$ (with some admixture of strange and
charm quarks too).
The infinite number of quarks and antiquarks can then just be
interpreted as saying that fluctuations within the proton create
infinite numbers of $q\bar q$ pairs, mostly carrying a small fraction
of the proton's momentum.

Returning to the statement that the proton has 2 up and 1 down quark,
what we mean is that the net number of up minus anti-up quarks is 2, 
\begin{equation}
  \label{eq:number-sum-rules}
  \int_0^1 dx (u(x) - \bar u(x)) = 2\,,\qquad
  \int_0^1 dx (d(x) - \bar d(x)) = 1\,,
\end{equation}
where $u(x) - \bar u(x)$ is also called the valence quark distribution
$u_V(x)$. How can we measure the difference between quarks and
antiquarks? The answer is through charged-current processes (e.g.,
neutrino scattering), since a $W^+$ interacts only with $d$ and $\bar
u$, but not with $\bar d$ or $u$.

We could imagine forming linear combinations of data sets with proton
and nuclear targets, with photon and $W^\pm$ exchange, etc., in order
to obtain the different quark-flavour PDFs. 
In practice it is simpler to introduce parametrizations for each
flavour, deduce predictions for DIS, neutrino-scattering and other
cross sections, and then fit the parameters so as to obtain agreement
with the experimental data. 
This is known as a `global fit'. We will return to such fits below,
but for now let's just look at the results for different quark
distributions, shown in Fig.~\ref{fig:nmc-data} (right).

We see that the valence quark distributions are mainly situated at
moderate $x$ values. For $x\to 1$ they fall off as a moderate power of
$(1-x)$, roughly as $(1-x)^3$. This fall-off comes about because to
have one $u$ quark carrying most of the proton momentum implies that
the other $u$ and the $d$ quark must both carry only a small fraction,
and the phase space for that to occur vanishes as $x\to1$ (the proper
treatment for this is through `quark-counting rules').
For $x\to0$, $xq_V(x) \sim x^{0.5}$, which is related to something
called Regge physics.

The sea-quark distributions ($u_S(x) \equiv u(x)+\bar u(x) - u_V(x)\equiv 2\bar u(x)$, etc.) are
concentrated at small $x$ values. They fall off much more steeply at
large $x$: for a $\bar u$ quark to have a large momentum fraction, 3
$u$-quarks and one $d$-quark must have correspondingly small-$x$, so
the phase space is even more restricted than in the case of valence
quarks. 
For $x\to 0$, the exact behaviour is not entirely simple, but as a
rough rule of thumb $xq_S(x) \sim x^{-\omega}$ with $\omega \sim$
0.2--0.4.

Of course we're still missing the PDFs for one kind of parton:
gluons. 
It's a well known fact that if we evaluate the fraction of the
proton's momentum carried by all the quarks,
\begin{equation}
  \label{eq:momentum-sum}
  \sum_{q} \int_0^1 dx\, xq(x)\,,
\end{equation}
the result is about $0.5$. 
It's fair to suspect that the gluon is responsible for the other half,
but how are we to establish its shape given that it's not directly
probed by photons or $W^\pm$?
To answer that question we must go beyond the `naive' leading-order
partonic picture of the proton's quarks interacting with probes, and
bring in QCD splitting.

\subsection{Initial-state parton splitting, DGLAP evolution}
\label{sec:dglap}

\subsubsection{Final and initial-state divergences}

In Eq.~(\ref{eq:fs-quark-splitting}) we wrote the universal form for
the final-state `splitting' of a quark into a quark and a soft
gluon.
Let's rewrite it with different kinematic variables, considering a
hard process $h$ with cross section $\sigma_h$, and examining the
cross section for $h$ with an extra gluon in the final state,
$\sigma_{h+g}$. We have
\begin{equation}
  \begin{minipage}[c]{0.21\tw}
    \includegraphics[width=\tw]{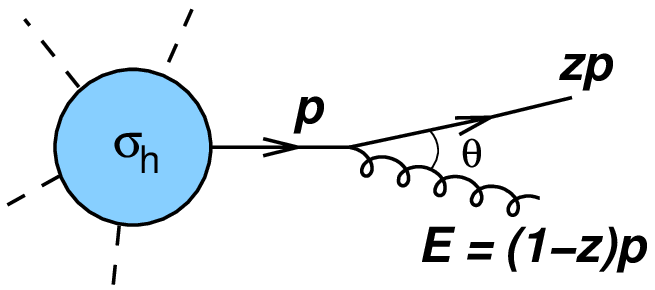}
  \end{minipage}
  \qquad\qquad
  \sigma_{h+g} \simeq \sigma_{h} \frac{\as\CF}{\pi}
  \frac{dz}{1-z} \frac{dk_t^2}{k_t^2}\,,
\end{equation}
where $E$ in Eq.~(\ref{eq:fs-quark-splitting}) corresponds to $E =
(1-z)p$ and we've introduced $k_t = E \sin \theta\simeq E\theta$.
If we avoid distinguishing a collinear $q+g$ pair from a plain quark
(measurements with IRC safe observables) then, as we argued before,
the divergent part of the
gluon emission contribution always cancels with a related virtual
correction
\begin{equation}
  \begin{minipage}[c]{0.21\tw}
    \includegraphics[width=\tw]{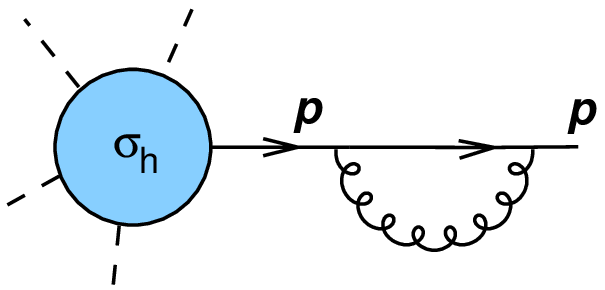}
  \end{minipage}
  \qquad\qquad
  \sigma_{h+V} \simeq -\sigma_{h} \frac{\as\CF}{\pi}
  \frac{dz}{1-z} \frac{dk_t^2}{k_t^2}\,.
\end{equation}
Now let us examine what happens for initial-state splitting, where the
hard process occurs \emph{after} the splitting and the momentum
entering the hard process is modified $p \to zp$:
\begin{equation}
  \begin{minipage}[c]{0.21\tw}
    \includegraphics[width=\tw]{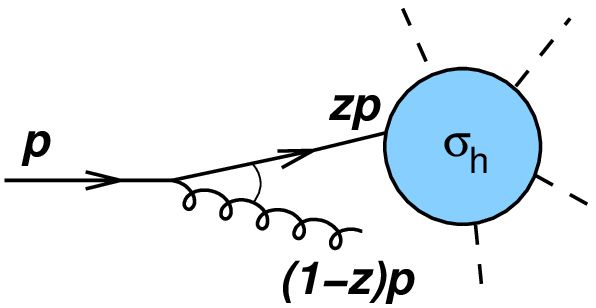}
  \end{minipage}
  \qquad\qquad
  \sigma_{g+h}(p) \simeq \sigma_{h}(zp) \frac{\as\CF}{\pi}
  \frac{dz}{1-z} \frac{dk_t^2}{k_t^2}\,,
\end{equation}
where we have made explicit the hard process's dependence on the
incoming momentum, 
and we assume that $\sigma_h$ involves momentum transfers $\sim Q \gg
k_t$, so that we can ignore the extra transverse momentum entering
$\sigma_h$.
For virtual terms, the momentum entering the process is unchanged, so
we have
\begin{equation}
  \begin{minipage}[c]{0.21\tw}
    \includegraphics[width=\tw]{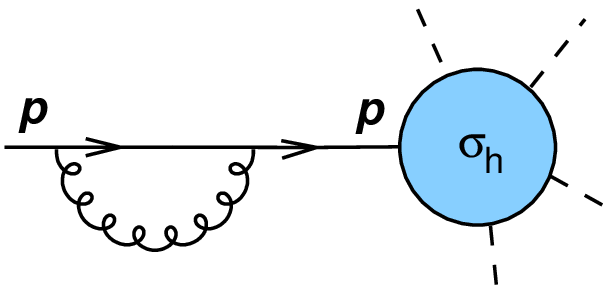}
  \end{minipage}
  \qquad\qquad
  \sigma_{g+h}(p) \simeq -\sigma_{h}(p) \frac{\as\CF}{\pi}
  \frac{dz}{1-z} \frac{dk_t^2}{k_t^2}\,,
\end{equation}
The total cross section then gets contributions with two different
hard cross sections:
\begin{equation}
  \label{eq:total-two-different-hard}
   \sigma_{g+h} + \sigma_{V+h} \simeq 
  \frac{\as\CF}{\pi}
  \underbrace{\int_0^{Q^2} \frac{dk_t^2}{k_t^2}}_{\text{infinite}}\;
  \underbrace{\int_0^1 \frac{dz}{1-z} [\sigma_{h}({ zp}) - \sigma_{h}({ p})]}_{\text{finite}}\,.
\end{equation}
Note the limits on the integrals, in particular the $Q^2$ upper limit
on the transverse-momentum integration: the approximations we're using
are valid as long as the transverse momentum emitted in the initial
state is much smaller than the momentum transfers $Q$ that are present
in the hard process.
Of the two integrations in Eq.~(\ref{eq:total-two-different-hard}),
the one over $z$ is finite, because in the region of the soft
divergence, $z\to1$, the difference of hard cross sections,
$[\sigma_{h}({ zp}) - \sigma_{h}({ p})]$, tends to zero.
In contrast, the $k_t$ integral diverges in the collinear limit: the
cross section with an incoming parton (and virtual corrections)
appears not to be collinear safe. 
This is a general feature of processes with incoming partons: so how
are we then to carry out calculations with initial-state hadrons?

In Section~\ref{sec:gluon-multiplicity}, when trying to make sense of
final-state divergences, we introduced a (non-perturbative)
cutoff. 
Let's do something similar here, with a cutoff, $\muF$, called a
factorization scale (which will usually be taken at perturbative
scales).
The main idea in using this cutoff is that any emissions that occur
with $k_t \lesssim \muF$ are absorbed (`factorized') into the PDF
itself.
Thus the PDFs become a function of $\muF$.
We can write the lowest order term for the cross section, $\sigma_0$,
and the correction with one initial-state emission or virtual loop,
$\sigma_1$, as follows
\\
\begin{minipage}[c]{0.24\linewidth}\hfill
  \includegraphics[width=\tw]{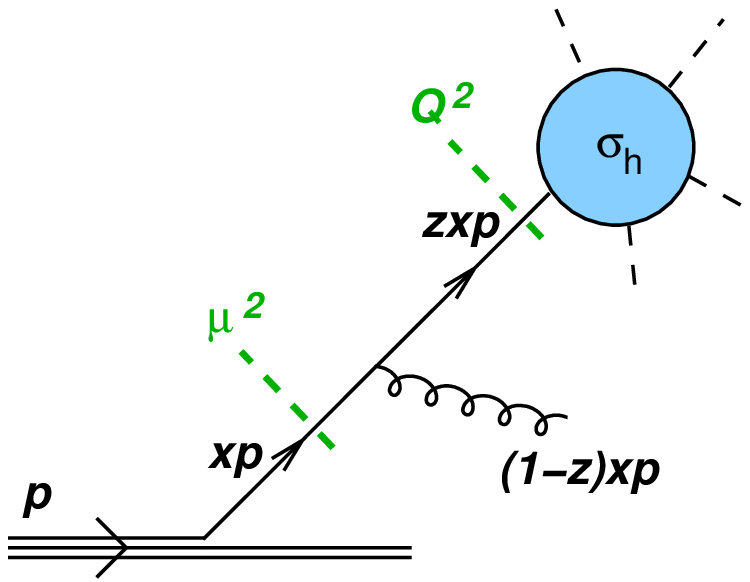}%
\end{minipage}
\hfill
\begin{minipage}[c]{0.7\linewidth}
  \begin{subequations}
    \label{eq:sigma-with-factorization}
    \begin{align}
      \sigma_{0} &= \int dx \; \sigma_h(xp)\; q( x, \muF^2)\,,
      \\
      \sigma_{1} &\simeq \frac{\as\CF}{\pi} \underbrace{
        \int_{\muF^2}^{{Q^2}} \frac{dk_t^2}{k_t^2}}_{\mathsf{finite\;
          (large?)}} \; \underbrace{\int \frac{dx\,dz
        }{1-z}\;[\sigma_{h}({ z}{ x}p) - \sigma_{h}({ x}p)]\; q({x}, {
          \muF^2})}_{\mathsf{finite}}\,,
    \end{align}
  \end{subequations}
\end{minipage}
\\[0.5em]
where we have now included also the integral over the longitudinal
momentum fraction $x$ of the parton extracted from the proton.
The emissions and virtual corrections with $k_t \lesssim \muF$ are
now implicitly included inside the $q( x, \muF^2)$ PDF factor that
appears in the $\sigma_0$ contribution, and only those with $k_t
\gtrsim \muF$ arise explicitly in the $\order{\as}$ term.
This term (whose real-emission part is represented in the diagram to
the left) is now finite, albeit potentially large if $\muF \ll Q$.

This situation of having a non-integrable divergence that somehow
needs to be regularized and absorbed with a scale choice into some
`constant' of the theory (here the PDFs), is reminiscent of
renormalization for the coupling constant.
The differences are that here we are faced with a divergence in the
(collinear) infrared rather than one in the ultraviolet. And that
unlike the coupling, the PDFs are not fundamental parameters of the
theory, but rather quantities that we could calculate if only we had
sufficiently sophisticated theoretical `technology'.
Nevertheless, as for the coupling, the freedom in choosing the scale
that enters the regularization, called the factorization scale,
implies the presence of a renormalization group equation for the
PDFs, the Dokshitzer--Gribov--Lipatov--Altarelli--Parisi (DGLAP) equation.

\subsubsection{The DGLAP equation}

To see what form the DGLAP equation takes, let us fix the longitudinal
momentum of the quark entering the hard process to be $xp$ (whereas
above we'd fixed the momentum for the quark extracted from the
proton).
Next we examine the effect on the PDFs of integrating over a small
region of transverse momentum $\muF^2 < k_t^2 < (1+\epsilon) \muF^2$, 
\begin{subequations}
  \label{eq:dglap-picture+eq}
  \begin{align}
    \frac{d q(x,\muF^2)}{d\ln \muF^2} &= \frac{1}{\epsilon}\left(
      \begin{minipage}[c]{0.45\linewidth}
        \includegraphics[width=\tw]{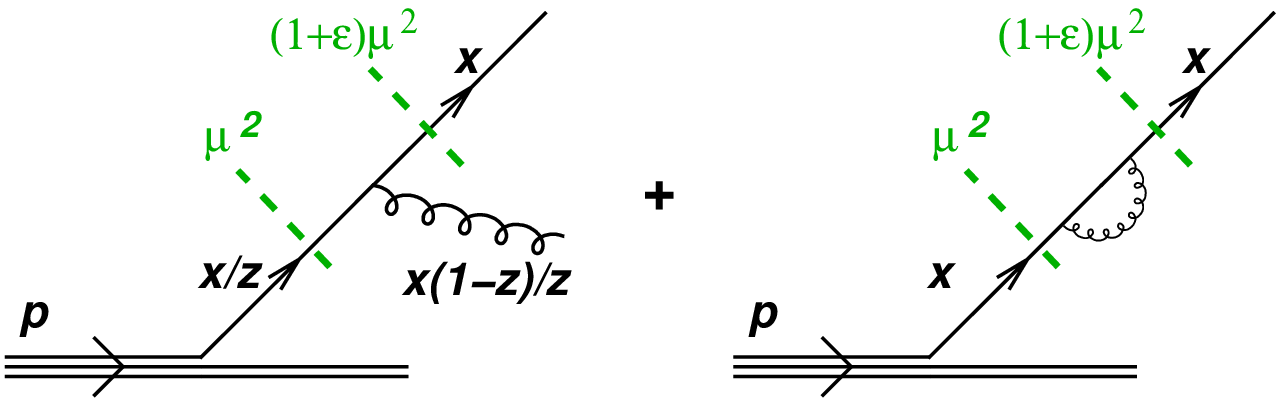}
      \end{minipage}\right)
    \\
    & = \frac{\as}{2\pi} \int_x^1 dz \,{ p_{qq}(z)}\,
    \frac{q(x/z,\muF^2)}{z} \;-\; \frac{\as}{2\pi} \int_0^1 dz \,{
      p_{qq}(z)}\, q(x,\muF^2)
  \end{align}
\end{subequations}
where $p_{qq}(z)$ is the real part of the `splitting kernel' for
a quark splitting to a quark plus a gluon,
\begin{equation}
  \label{eq:pqq}
  p_{qq}(z) = \CF \frac{1+z^2}{1-z}\, .
\end{equation}
Until now, we had concentrated on the soft limit, which was equivalent to
approximating $p_{qq}(z) \simeq \frac{2\CF}{1-z}$.
What Eq.~(\ref{eq:dglap-picture+eq}) tells us is that as we increase
the factorization scale, we get extra partons with longitudinal
momentum fraction $x$ that come from the branching of partons in the
proton at lower factorization scales but larger momentum fractions
$x/z$ ($x<z<1$).
There are also loop contributions (second term on the RHS) to the
parton density at a fixed $x$ value, which are negative contributions
to the evolution.
The way to think about these is that when a parton with momentum
fraction $x$ branches to partons with lower momentum fractions, the
original parton is lost and the loop diagram accounts for that.

It's a bit awkward to write the real and virtual parts separately in
Eq.~(\ref{eq:dglap-picture+eq}), especially if one wants to explicitly
see the cancellation of the divergences for $z\to 1$. It's therefore
standard to use the more compact notation 
\begin{equation}
  \label{eq:dglap-qq-compact}
  \frac{d q(x,\muF^2)}{d\ln \muF^2} = \frac{\as}{2\pi}
  \underbrace{\int_x^1 dz \,{ P_{qq}(z)}\,
    \frac{q(x/z,\muF^2)}{z}}_{P_{qq} \otimes q},\qquad
  P_{qq} = \CF\left(\frac{1+z^2}{1-z}\right)_+\,,
\end{equation}
where the subscript plus, known as the `plus' prescription, is to be
understood as follows:
\begin{align}
  \label{eq:plus-def}
  \int_x^1 dz \,{ [g(z)]_+} \,f(z) &= \int_x^1 dz \, {g(z)}
  \,{ f(z)} - \int_0^1 dz \, {g(z)}\, { f(1)}
  \\
  &= \int_x^1 dz \, {g(z)}\, (f(z) - f(1)) - \int_0^x dz \, {g(z)}\, { f(1)}
\end{align}
so that the factor $(f(z)-f(1))$, which goes to zero at $z=1$, kills
the divergence due the singular behaviour of $g(z)$ for $z\to 1$.

Equation~(\ref{eq:dglap-qq-compact}) involves just quarks, but the proton
contains both quarks and gluons, so the full DGLAP equations are
actually coupled evolution equations. Schematically, for just a single
quark flavour, they read
\begin{equation}
  \label{eq:matrix-dglap}
  \frac{d}{d \ln \muF^2}
  \left( 
    \begin{array}{c}
      q\\g
    \end{array}
  \right)
  = 
  \frac{\as(\muF^2)}{2\pi}
  \left(
    \begin{array}{cc}
      P_{q\leftarrow q} & P_{q\leftarrow g} \\
      P_{g\leftarrow q} & P_{g\leftarrow g} \\
    \end{array}
  \right)
  \otimes 
  \left( 
    \begin{array}{c}
      q\\g
    \end{array}
  \right)
\end{equation}
and more generally they span all quark flavours and anti-flavours.
In labelling the different flavour entries, we've included arrows
(usually not shown), e.g.\ $q\leftarrow g$, so as to emphasize that we
have evolution from the right-hand parton type to the left-hand
parton type.
The splitting functions other than $P_{qq}$ are given by
\begin{subequations}
  \label{eq:splitting-fns}
  \begin{align*}
    P_{qg}(z) &= \TR \left[z^2 + (1-z)^2 \right],\qquad\quad P_{gq}(z)
    = \CF\left[ \frac{1+(1-z)^2}{z}
    \right]\,,\\
    P_{gg}(z) &= 2\CA\left[ \frac{z}{(1-z)_+}+
      \frac{1-z}{z}+z(1-z)\right]
    +\delta(1-z)\frac{(11\CA-4n_f\TR)}{6} \,.
  \end{align*}
\end{subequations}
Additionally, $P_{\bar qg} = P_{qg}$ and, to this first order in the coupling,
$P_{qq'}$ and $P_{q\bar q}$ are both zero.

Several features of the splitting functions are worth noting: $P_{qg}$
and $P_{gg}$ are both symmetric in $z\leftrightarrow 1-z$ (except for
the virtual part).
$P_{qq}$ and $P_{gg}$ diverge for $z\to 1$, which corresponds to
soft-gluon emission.
And $P_{gg}$ and $P_{gq}$ both diverge for $z\to 0$ (corresponding to a soft
gluon entering the `hard' process). 
This last point implies that PDFs $q(x)$ and $g(x)$ must grow at least
as fast as $1/x$ for $x\to 0$: even if such a divergence is absent in
the non-perturbative `initial conditions' for the quark and gluon
distributions at low scales $\muF$, DGLAP evolution inevitably
introduces it into the result for $q(x,\muF^2)$ and $g(x,\muF^2)$ at
higher scales.

\subsubsection{Results of DGLAP evolution}

\begin{figure}
  \centering
  \includegraphics[width=0.65\tw]{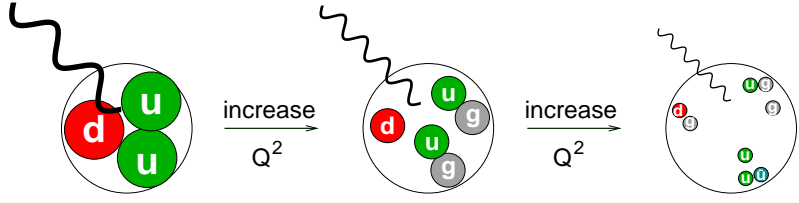}
  \caption{An illustration how with ever shorter wavelength photon
    probes, one resolves more and more structure inside the proton}
  \label{fig:dglap}
\end{figure}
\begin{figure}
  \centering
  \includegraphics[width=0.23\tw]{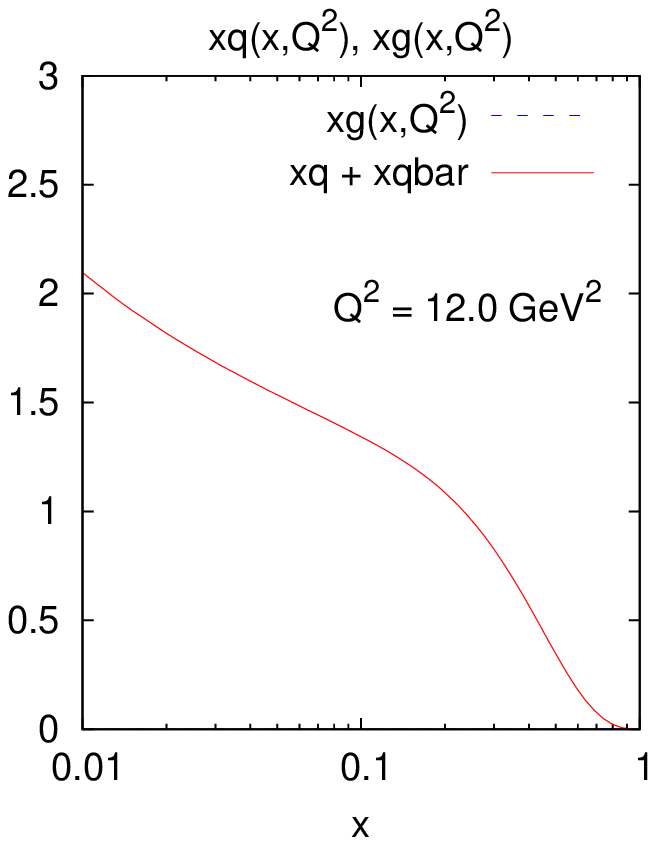}\;
  \includegraphics[width=0.23\tw]{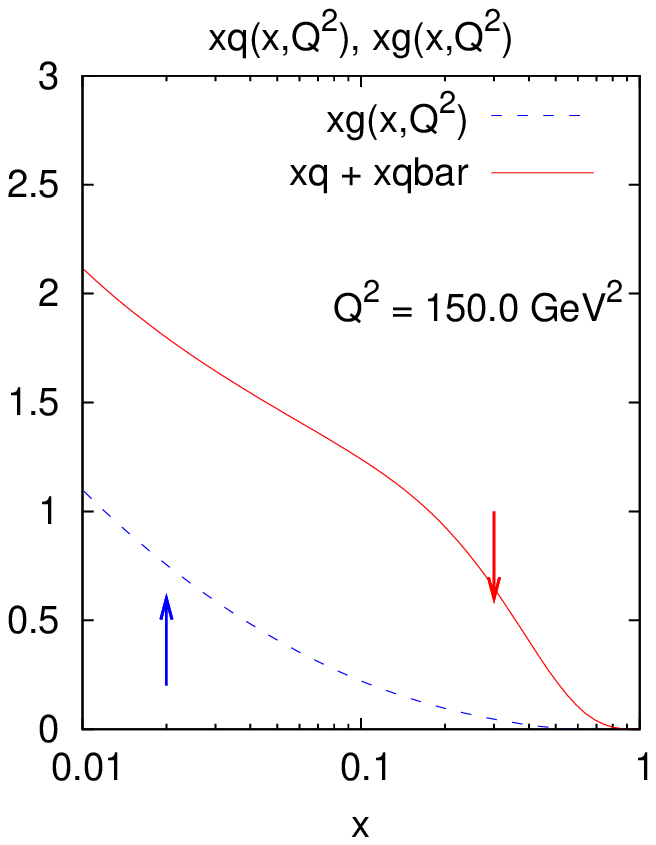}\;
  \includegraphics[width=0.23\tw]{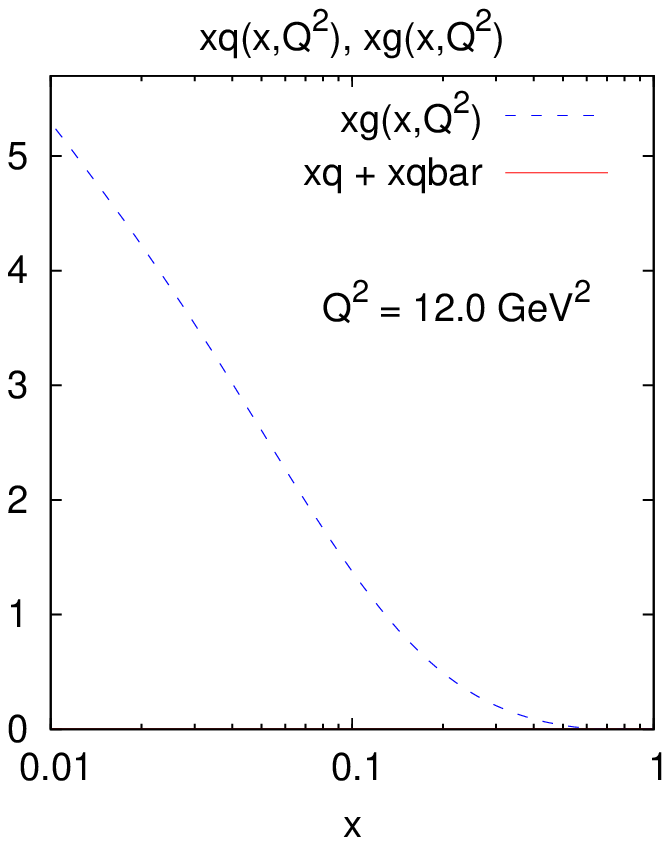}\;
  \includegraphics[width=0.23\tw]{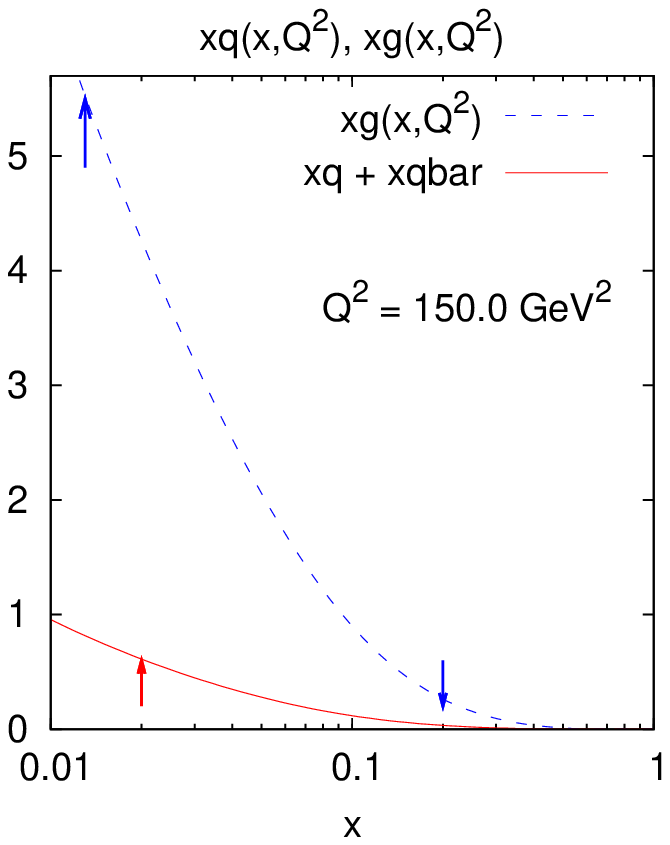}\;
  \caption{An illustration of the impact of DGLAP evolution. From left
    to right: (a) initial condition consisting just of quarks and
    anti-quarks at $\muF^2 \equiv Q^2 = 12\GeV^2$; (b) the result of
    evolution to $Q^2 = 150\GeV^2$; (c) a purely gluonic initial
    condition at $Q^2 = 12\GeV^2$; and (d) the result of its evolution
    to $Q^2 = 150\GeV^2$.}
  \label{fig:dglap-evolution}
\end{figure}

Pictorially, the effect of DGLAP evolution is illustrated in
Fig.~\ref{fig:dglap}.
A more quantitative view is given in Fig.~\ref{fig:dglap-evolution},
which shows the effect of DGLAP evolution with an initial condition
that is pure quark (two left plots) or pure gluon (two right plots).
In both cases one sees that evolution generates some amount of the
missing parton type; one also sees how it depletes the parton
distributions at large $x$, and increases them at small $x$
(especially in the case of the gluon).
The attentive reader may have observed that the figure labels the
scale of the PDFs as $Q^2$ rather than $\muF^2$: this is because it
is standard to take $\muF^2 = Q^2$ so as to minimize the size of the
the $\order{\as}$ term of Eq.~(\ref{eq:sigma-with-factorization})
which arises, roughly, from the integral over transverse momenta from
$\muF^2$ to $Q^2$. I.e., one usually chooses to factorize essentially
\emph{all} initial-state radiation into the PDFs and so into the LO
cross section.

\begin{figure}
  \centering
  \includegraphics[width=0.3\tw]{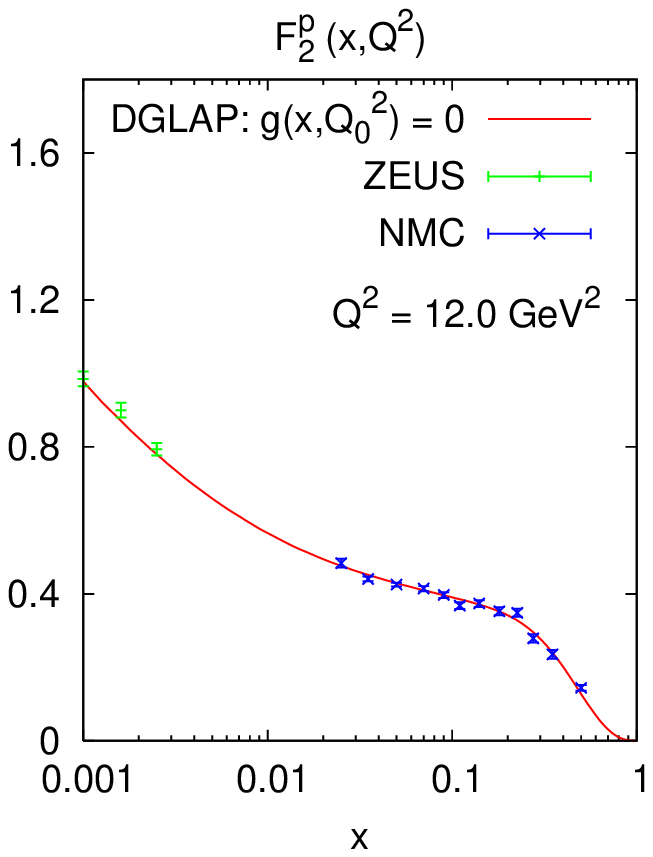}\;
  \includegraphics[width=0.3\tw]{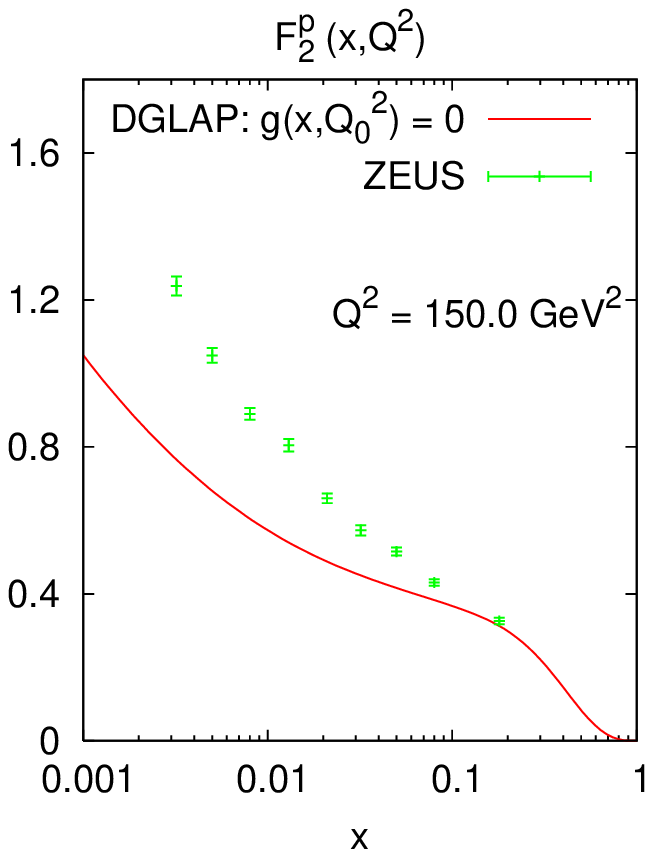}\;
  \includegraphics[width=0.3\tw]{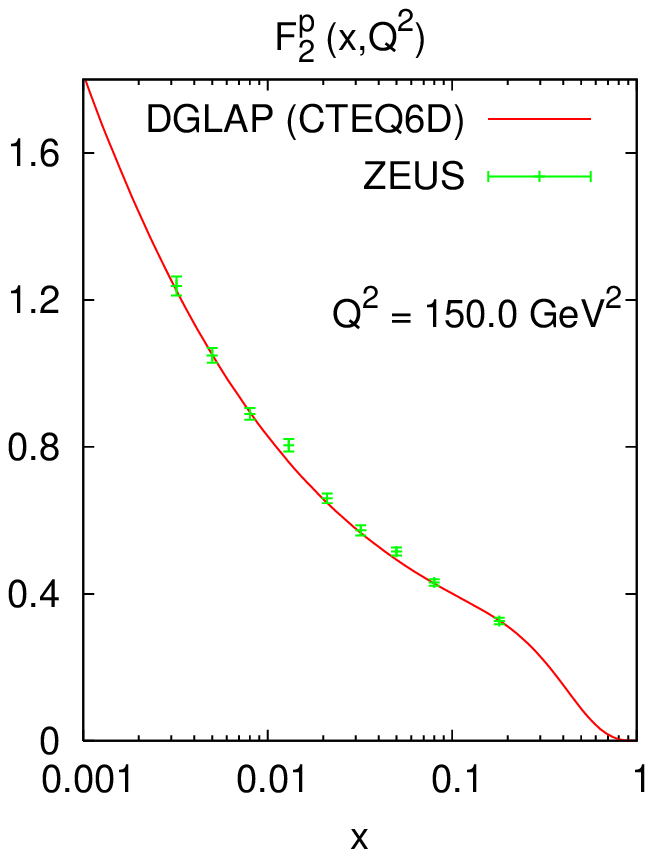}
  \caption{ZEUS and NMC data together with an initial condition that
    gives a good fit at low $Q^2$ (left-most plot). If one evolves
    that initial condition assuming a gluon distribution that is zero
    at low $Q^2$, then agreement with high-scale data is poor (central
    plot); whereas with a significant low-scale gluon component (taken
    from the CTEQ6D parametrization), agreement becomes good at high
    scales (right-most plot).
    \label{fig:gluon-dglap-impact}
  }
\end{figure}

Since, as we've see in Fig.~\ref{fig:dglap-evolution}, the presence of
a gluon distribution helps drive quark evolution, we can use the
experimentally observed pattern of quark evolution to help constrain
the gluon. The left-hand plot of Fig.~\ref{fig:gluon-dglap-impact}
shows data from ZEUS \cite{Chekanov:2001qu} and NMC \cite{Arneodo:1996qe} on
$F_2(x,Q^2)$ at some low but still perturbative scale $Q^2 = Q_0^2
\equiv 12\GeV^2$. The data are compared 
to the expectations based on the CTEQ6D PDFs' quark content at that
scale, illustrating the good agreement. Since these are data for
$F_2$, they have no direct sensitivity to the gluon distribution.
The middle plot shows data for $150\GeV^2$, together with the results
of DGLAP evolution from $Q_0^2=12\GeV^2$, assuming that the gluon
distribution was zero at $Q_0^2$. There's a clear discrepancy.
In the right-hand plot, the comparison is made with evolution whose
initial condition at $Q_0^2$ contained a quite large gluon component
(exactly that in the CTEQ6D distributions), causing the quark
distribution at small $x$ values to increase faster with $Q^2$ than
would otherwise be the case, bringing the evolution into agreement
with the data.
%

\subsection{Global fits}
\label{sec:global-fits}

\begin{figure}
  \centering
  \includegraphics[width=0.32\tw]{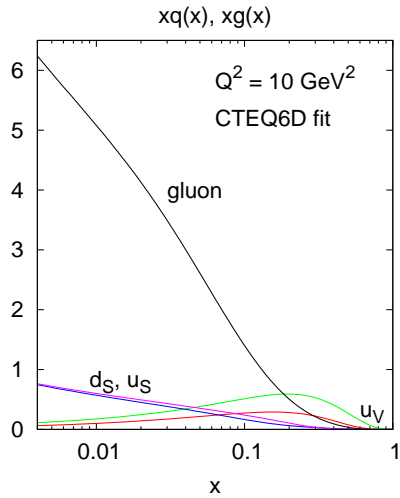}
  \caption{The distributions of different parton species in the CTEQ6D
    parametrization at a scale $Q^2= 10\GeV^2$}
  \label{fig:quark-gluon-dists-cteq6d}
\end{figure}

It's interesting to ask just how much of a gluon distribution is
needed in order to get the agreement shown in
Fig.~\ref{fig:gluon-dglap-impact}. 
The answer is given in Fig.~\ref{fig:quark-gluon-dists-cteq6d} and one
sees that the gluon distribution is \emph{enormous}, especially at
small values of $x$.
It is fair to ask whether we can trust a result such as
Fig.~\ref{fig:quark-gluon-dists-cteq6d}, so in this section we will
examine some of ingredients and issues that are relevant to the
`global fits' that inform our knowledge of PDFs.

\begin{figure}
  \centering
  \begin{minipage}[c]{0.52\tw}
    \includegraphics[width=\tw]{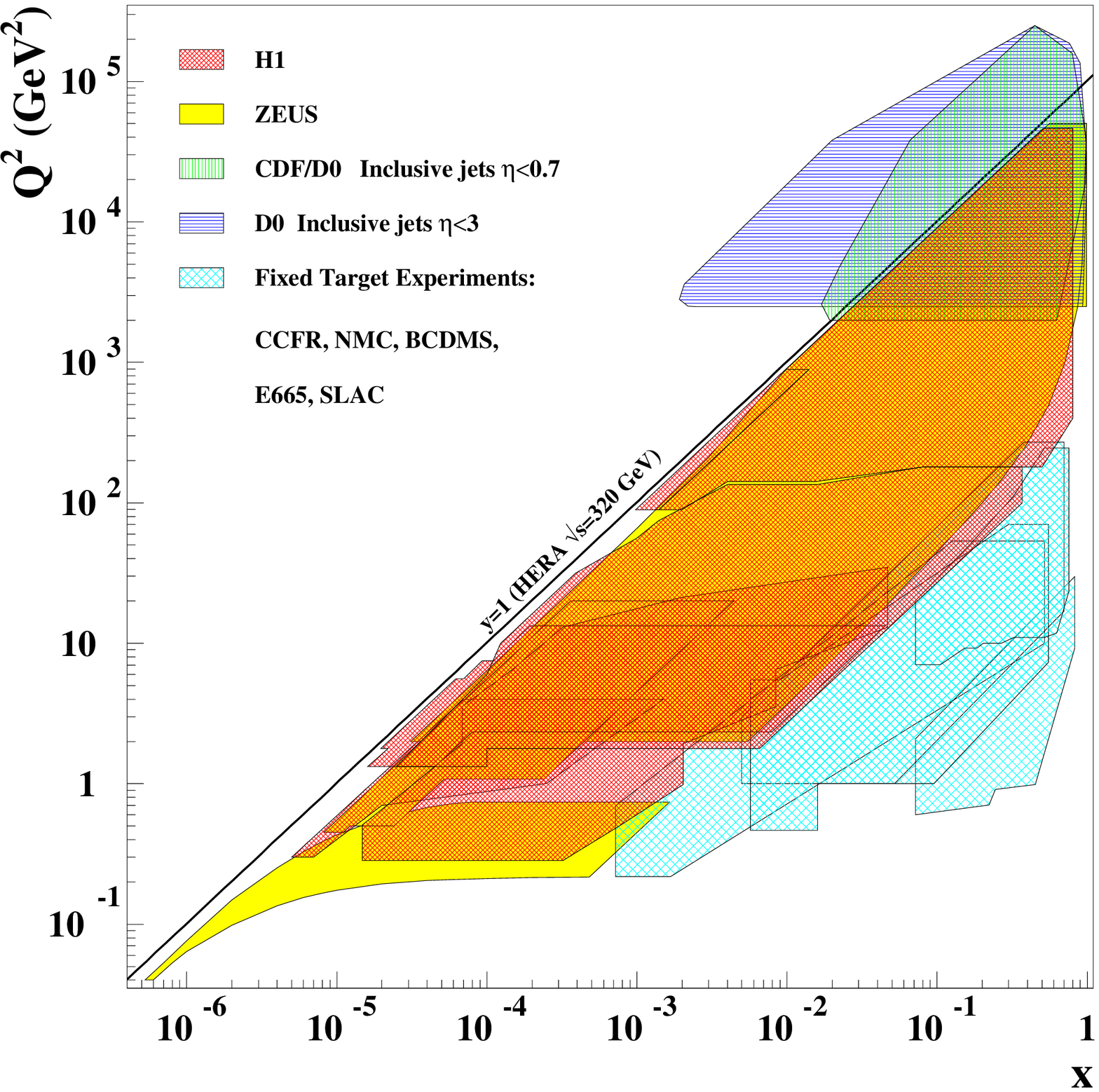}
  \end{minipage}\hfill
  \begin{minipage}[c]{0.44\tw}
    \includegraphics[width=\tw]{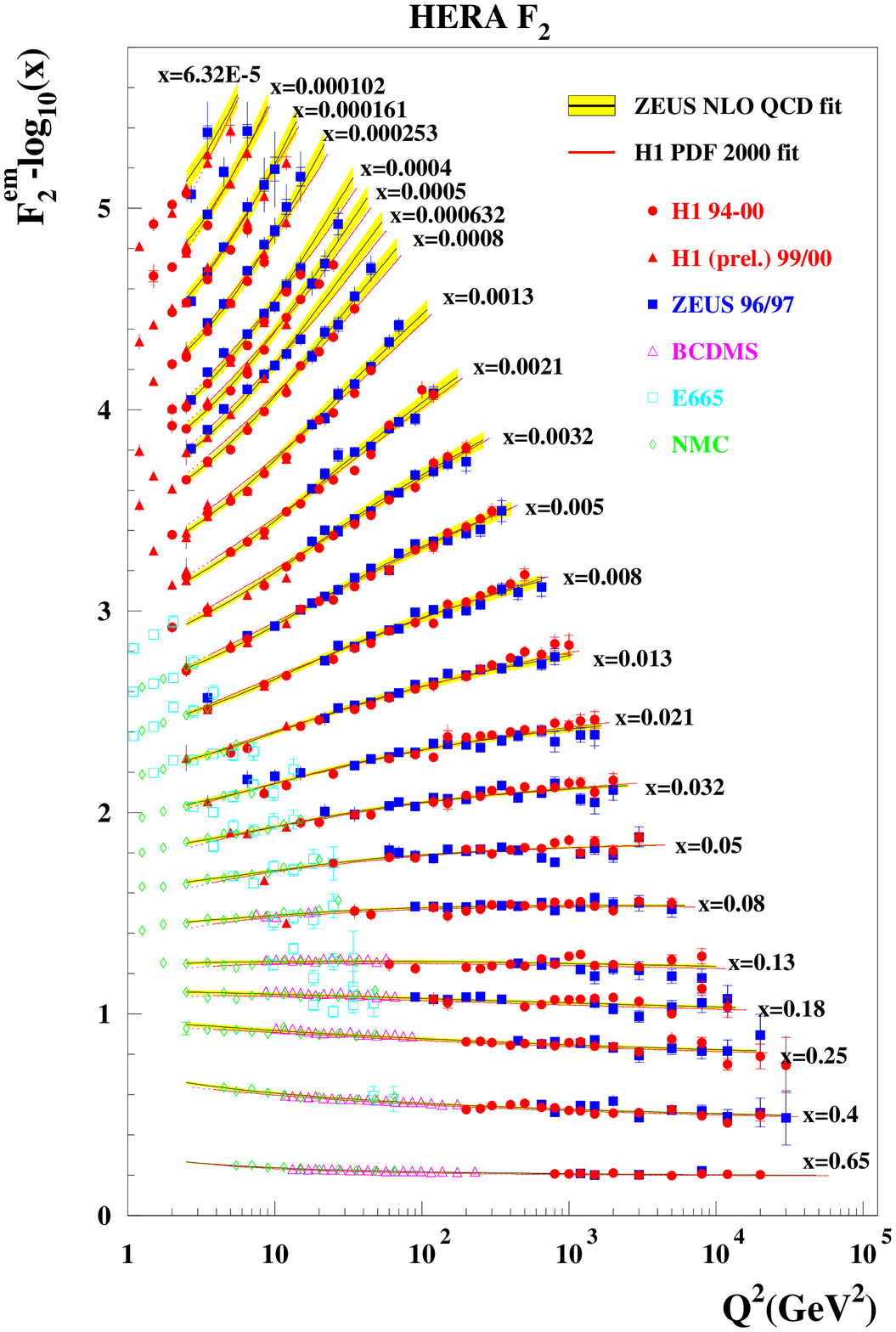}
  \end{minipage}
  \caption{Left: an illustration of the kinematic regions and data
    sets typically used in PDF fits (based on a fit by ZEUS).
    Right: experimental results for $F_2$ as a function of $Q^2$ for
    many different $x$ values, compared to the results of a global
    fit by the ZEUS collaboration. }
  \label{fig:kin-regions-F2}
\end{figure}

Figure~\ref{fig:kin-regions-F2} (left) illustrates the kinematical
regions in the $x$ and $Q^2$ plane covered by the experimental data
sets typically used in global fits.
Everything below the diagonal line corresponds to DIS data, and 
the right-hand plot shows the comparison between a fit (by ZEUS) and
the bulk of the DIS data, illustrating the excellent consistency
between fit and data.
Agreement with such a broad data set is already a non-trivial
achievement.
Figure~\ref{fig:kin-regions-F2}(left) also shows shaded regions that
span the diagonal line. These correspond to hadron-collider jets data,
which provide valuable direct information on the gluon distribution in
global fits, as we will discuss below.
The other topics that we will address here relate to the accuracy of
our determinations of PDFs.

\subsubsection{Factorization and $p\bar p$ jet production}

\begin{figure}
  \centering
  \includegraphics[width=0.3\tw]{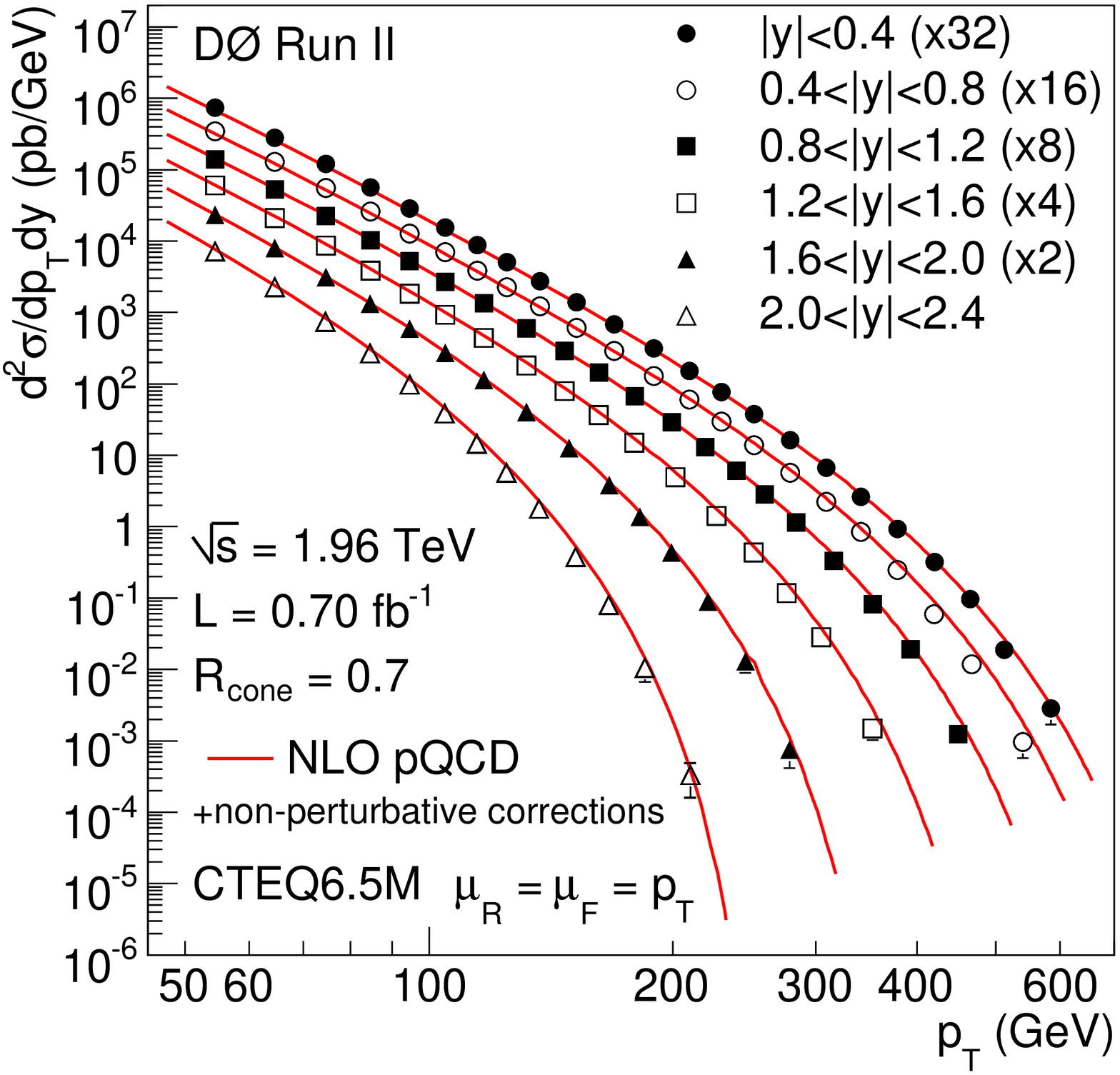}%
  \includegraphics[width=0.3\tw]{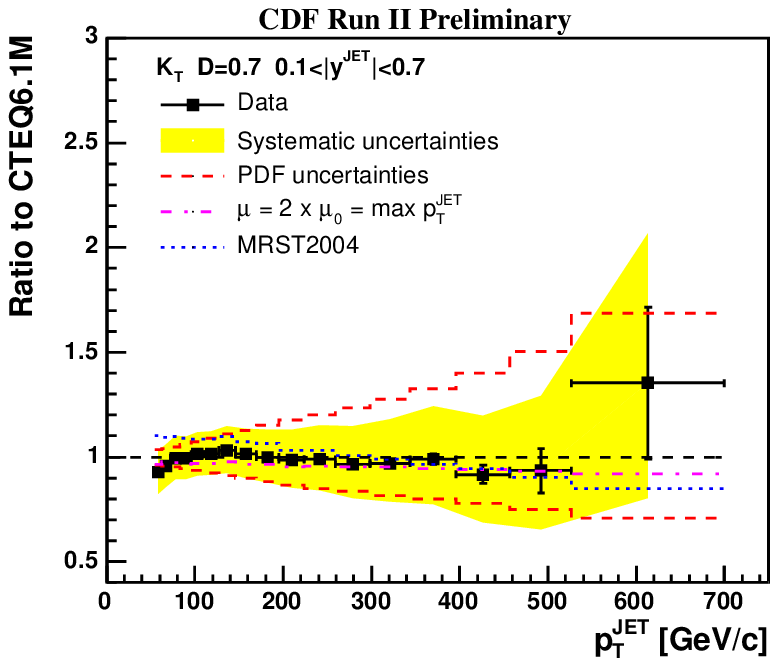}%
  \includegraphics[width=0.4\tw]{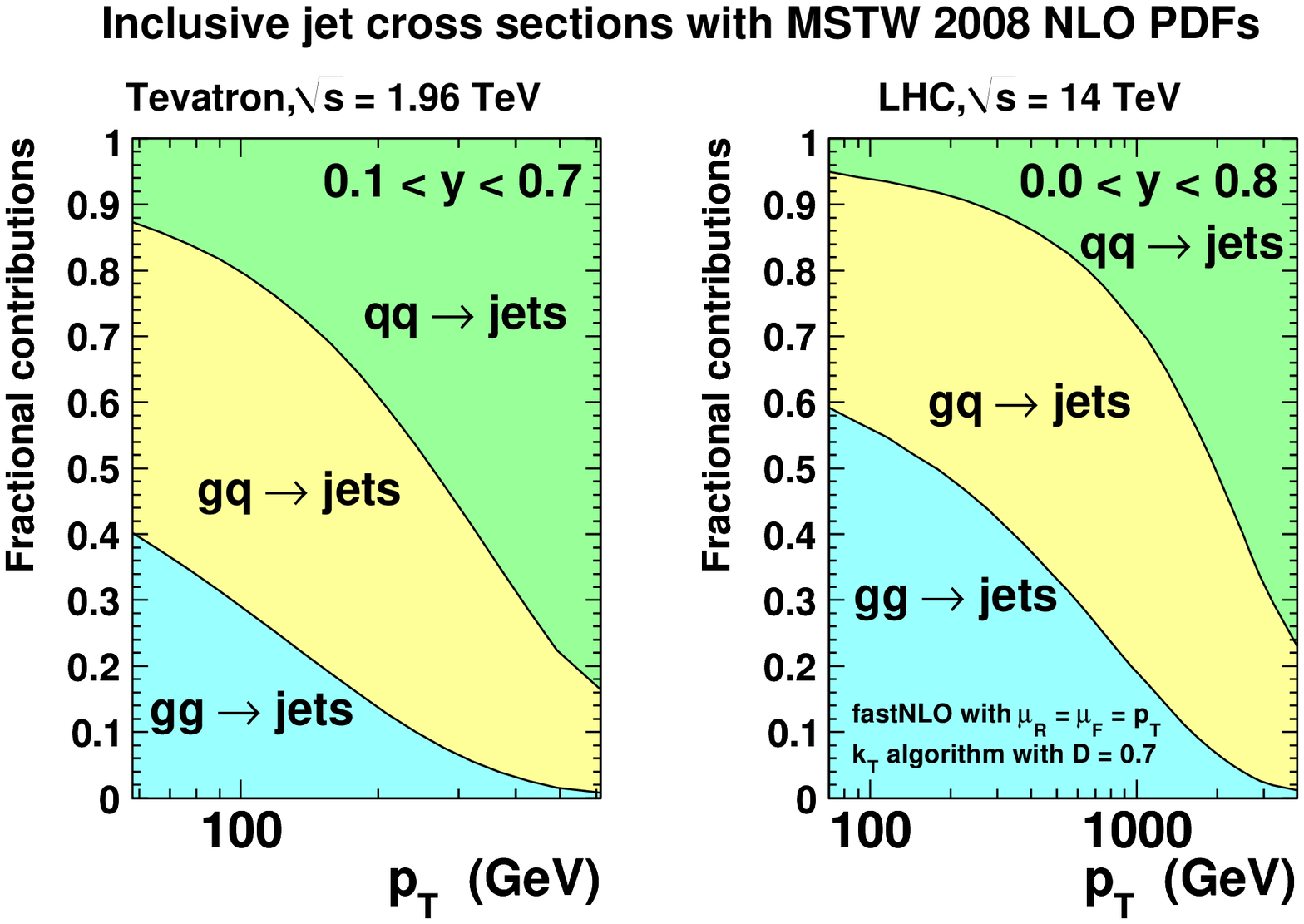}
  \caption{Left: D\O\ inclusive jets data \cite{D0:2008hua} compared
    to predictions with CTEQ6.5M PDFs. Middle: the ratio of CDF data
    from~Ref.~\cite{Abulencia:2007ez} to predictions with MRST2004 PDFs.
    Right: the relative contributions of different scattering channels
    to the Tevatron and LHC jet spectra, as a function of jet
    $p_t$, taken from Ref.~\cite{Martin:2009bu}. }
  \label{fig:inclusive-jets}
\end{figure}

Perhaps the most convincing cross-check of PDF extractions comes from
Tevatron jet data (there are also important jet data from HERA).
The process of factorizing initial-state radiation into the PDF at a
given scale is equally valid in DIS and $p\bar p$ (or $pp$)
collisions, as illustrated in the following pictorial representation
of the two cases:
\begin{equation}
  \begin{minipage}[c]{0.6\linewidth}
    \includegraphics[width=\tw]{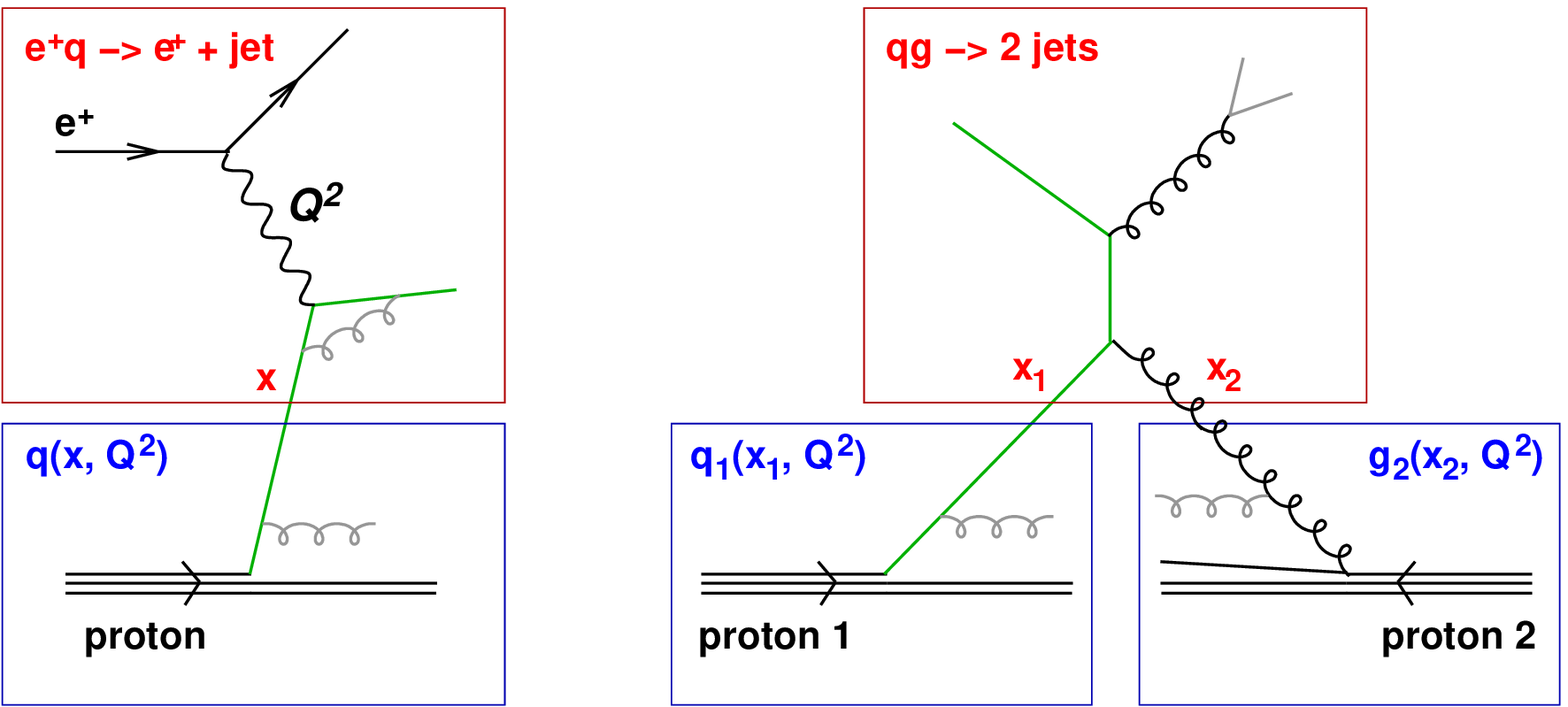}%
  \end{minipage}
\end{equation}
Given factorization and a determination of PDFs in DIS, one can simply
take the expression Eq.~(\ref{eq:parton-model-pp}) for a generic cross
section in hadron--hadron collisions, and rewrite it with
explicit factorization scales:
\begin{equation}
  \label{eq:parton-model-pp-factorized}
  \sigma\sigma_{pp \to ZH} = \int dx_1 { f_{q/p}(x_1,\muF^2)} \int dx_2 {
      f_{\bar q/p}(x_2, \muF^2)}
    \;{ \hat \sigma_{q\bar q \to ZH} (x_1 p_1, x_2 p_2, \muF^2)}\,.
\end{equation}
Such a formula, with $\sigma_{q\bar q \to ZH}$ replaced by
$\sigma_{q\bar q \to q\bar q}$ (summing also over processes with
gluons, etc.)
can be used to obtain predictions for the differential inclusive
jet spectrum at the Tevatron. 
Figure~\ref{fig:inclusive-jets} shows comparisons of data from D\O\ and
from CDF with predictions from CTEQ6.5 \cite{Tung:2006tb} and MRST2004
\cite{Martin:2004ir} PDF parametrizations, illustrating excellent
agreement.
The two right-hand plots show how different incoming partonic
scattering channels contribute to the cross section, highlighting the
significant contribution from gluons.

It is perhaps misleading to use the word `prediction' about
Fig.~\ref{fig:inclusive-jets}: most advanced fully global PDF fits
actually make use of data such as that in
Fig.~\ref{fig:inclusive-jets} as part of the fit.
Still, it is a powerful consistency check that it is possible to
obtain agreement both with the jet data, which is sensitive directly
to quark and to gluon distributions, as well as with DIS data, which is
directly sensitive to the quarks and indirectly to the gluons, through
the scaling violations.

One technical comment is due concerning factorization. While our
discussion has been limited to leading order, many of the figures that
we have shown also involve higher orders (to which we will return).
When using PDFs with predictions beyond LO, it is necessary to specify
the `factorization scheme', i.e., the specific procedure by which
one separates emissions below and above $\muF$. 
The figures in Sections~\ref{sec:DIS} and \ref{sec:dglap} made use of
the `DIS' scheme (hence CTEQ6\emph{D}), defined such that $F_2$ is
given by Eq.~(\ref{eq:F2em}), free of any $\order{\as}$ (or higher)
corrections.
While that scheme has the benefit of pedagogical simplicity, in real
calculations (and all plots in this section) it is usually more
convenient to use the `\MSbar' factorization scheme, based on
dimensional regularization.

\subsubsection{Uncertainties} 
\begin{figure}
  \centering
  \includegraphics[width=0.33\tw]{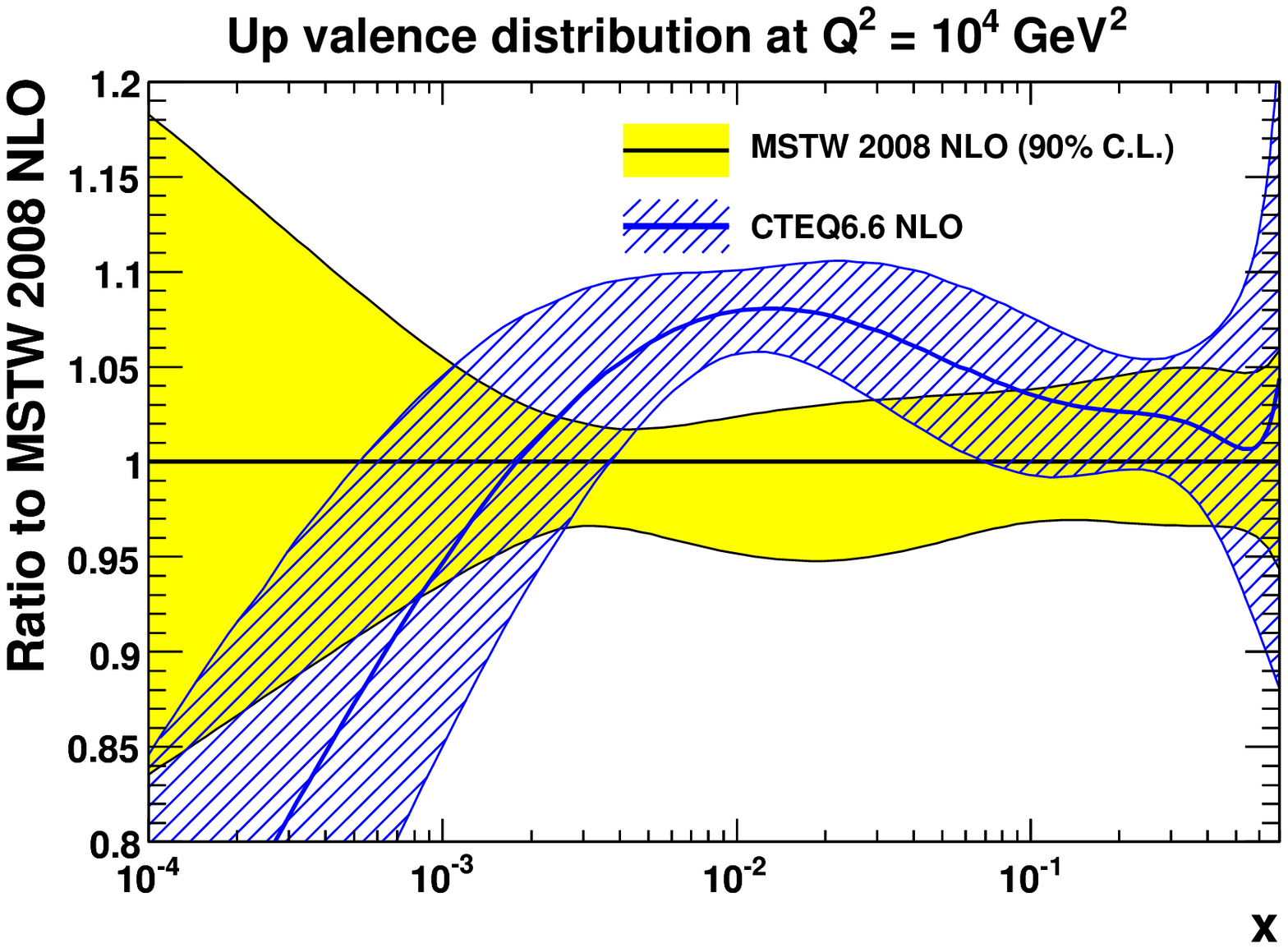}%
  \includegraphics[width=0.33\tw]{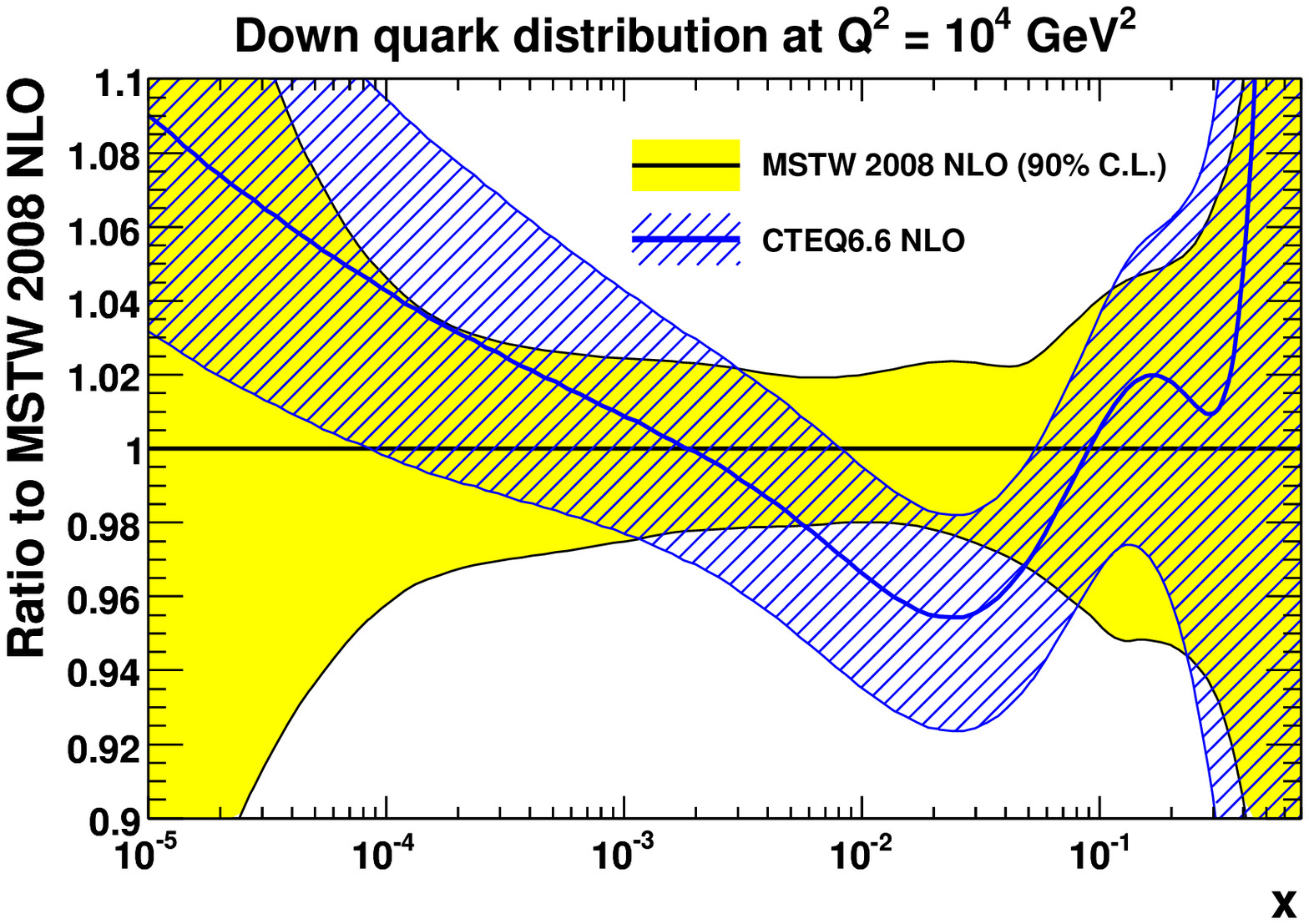}%
  \includegraphics[width=0.33\tw]{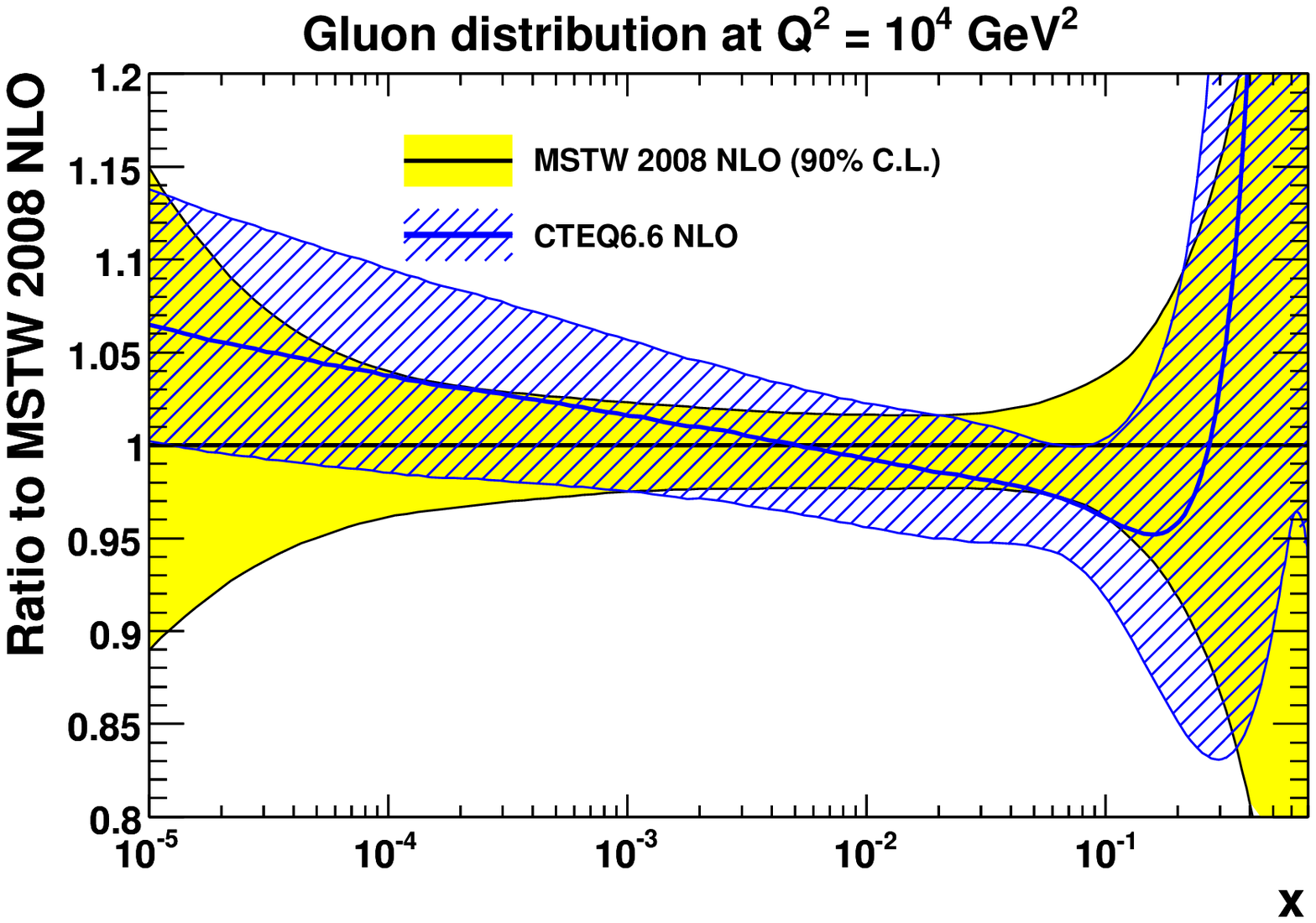}
  \caption{Uncertainties on recent PDFs from the MSTW
    \cite{Martin:2009iq} and CTEQ groups~\cite{Nadolsky:2008zw} at a
    scale of $Q = 100\GeV$ (figure taken from Ref.~\cite{Martin:2009iq})}
  \label{fig:pdf-uncertainties}
\end{figure}

\begin{figure}
  \centering
  \includegraphics[width=0.48\tw]{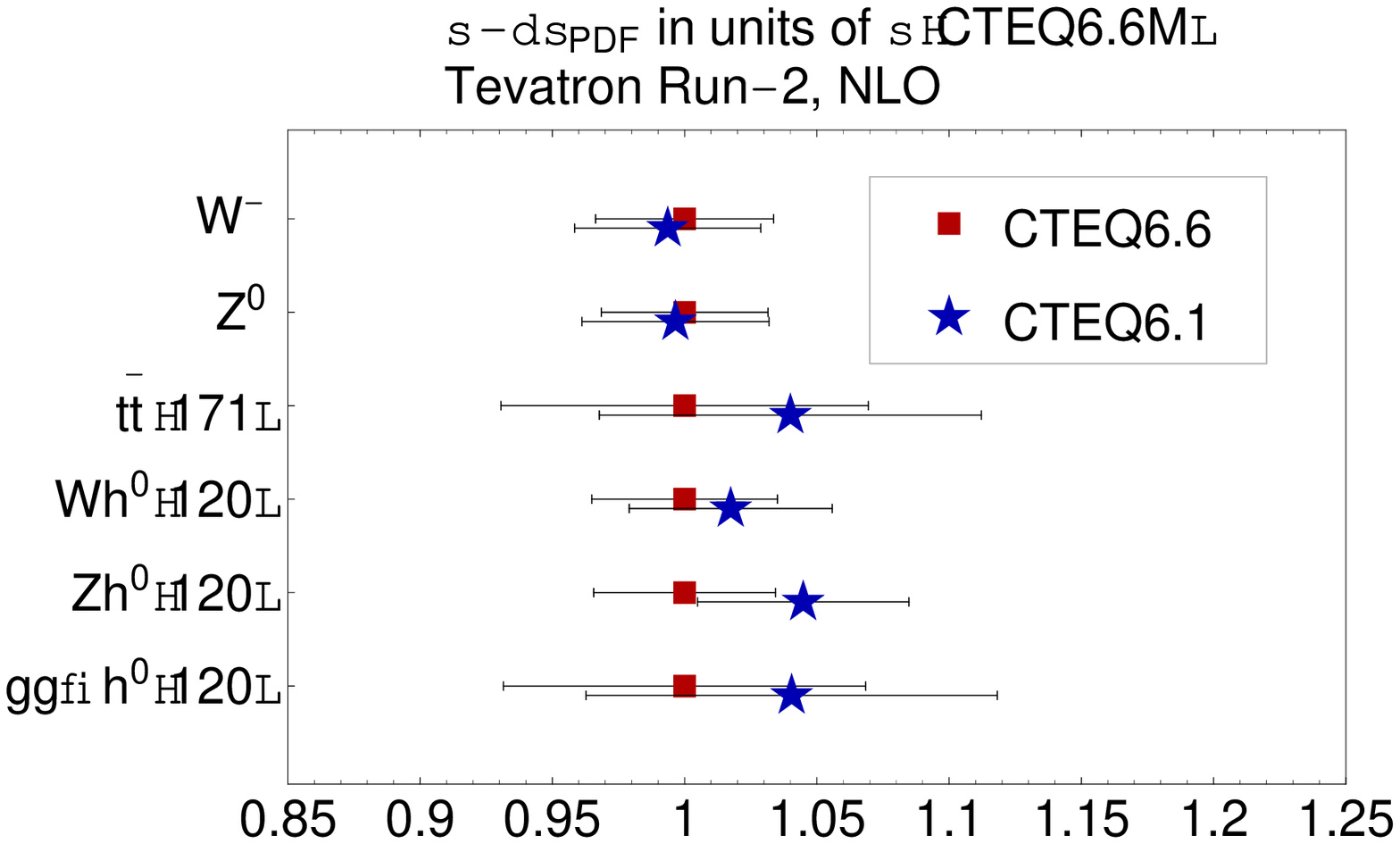}\hfill
  \includegraphics[width=0.48\tw]{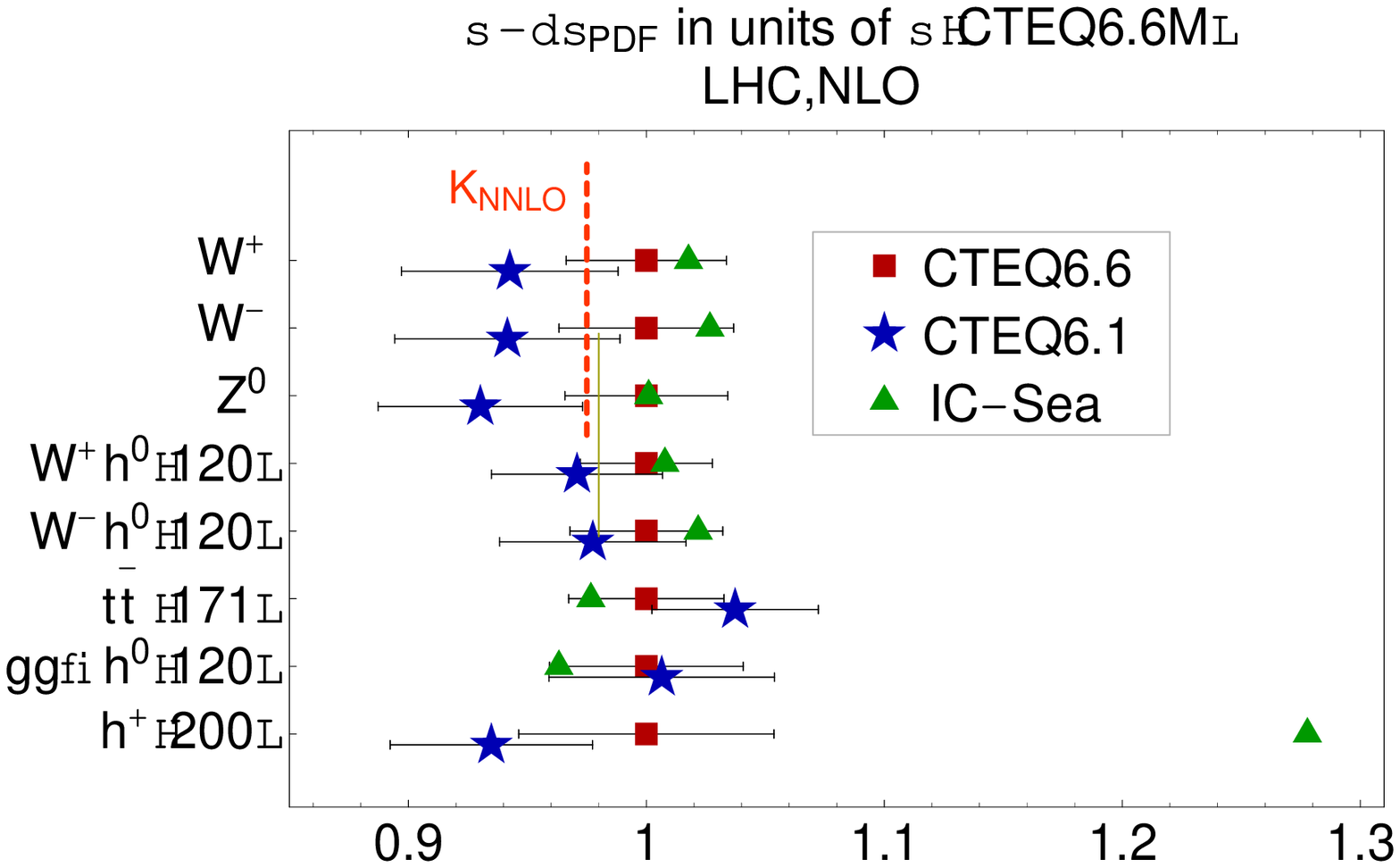}
  \caption{Impact of PDF uncertainties on predictions for standard
    cross sections at the Tevatron and LHC from Ref.~\cite{Nadolsky:2008zw}} 
  \label{fig:pdf-uncertainties-impact}
\end{figure}

An important part of the activity on global
fits in recent years has been geared to the estimation of the
uncertainties on PDFs.
Figure~\ref{fig:pdf-uncertainties} shows the uncertainties on recent PDF
sets from the two most established PDF fitting groups, MSTW and CTEQ,
illustrating uncertainties that are in the couple of per~cent to
ten per~cent range. 
Figure~\ref{fig:pdf-uncertainties-impact} illustrates the impact of the
uncertainties on predictions for a range of cross sections at the
Tevatron and the LHC.

Estimating PDF uncertainties is something of an art: for example, one
must parametrize the PDFs at some input scale and there is freedom in
how flexible a parametrization one uses: too rigid (or with too many
theorist's assumptions) and the global fit may not have the
flexibility to describe the data or may appear to have inappropriately
small uncertainties in regions where there are no data;
with too flexible a parametrization the fits may develop artefacts
that adapt to statistical fluctuations of the data.
Other issues include reconciling barely compatible data sets and
deciding what values of $\chi^2$ variations are reasonable to estimate
errors given the incompatible data sets.
In addition to MSTW and CTEQ (and several other groups), a recent
entrant to PDF fitting is the NNPDF Collaboration, which is developing
procedures that attempt to minimize theoretical bias in questions such
as the parametrizations.
First global fit results including $pp$ data have been given in
Ref.~\cite{Ball:2010de}, though for fully accurate treatment of the HERA
data, one should await their inclusion of heavy-quark effects.
Their results so far tend to be similar to those of MSTW and CTEQ,
except in the regions of small-$x$ near the edge of the available HERA
phasespace and for strange quark distributions, where they find
somewhat larger uncertainties.

\subsubsection{PDFs for LHC and the accuracy of DGLAP evolution} 

\begin{figure}
  \centering
  \begin{minipage}[c]{0.44\linewidth}
    \includegraphics[width=\tw]{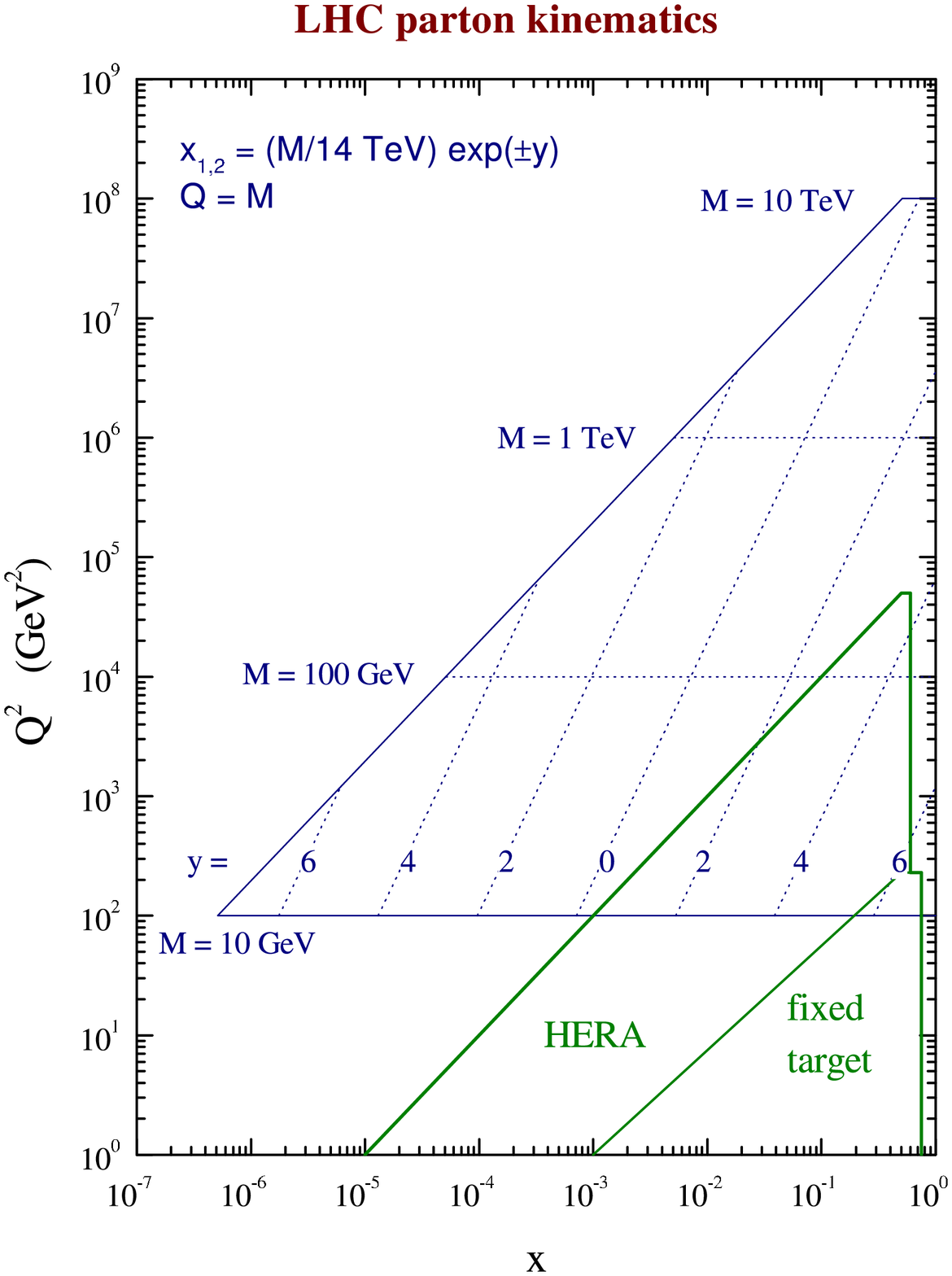}%
    \rput(-2.9,3.5){{\rotatebox{-80}{\Large {$\boldsymbol{\Longleftarrow}$ DGLAP}}}}
    \rput(-1.5,5.0){{\rotatebox{-80}{\Large {$\boldsymbol{\Longleftarrow}$ DGLAP}}}}
  \end{minipage}\hfill
  \begin{minipage}[c]{0.48\linewidth}
    \includegraphics[width=\tw]{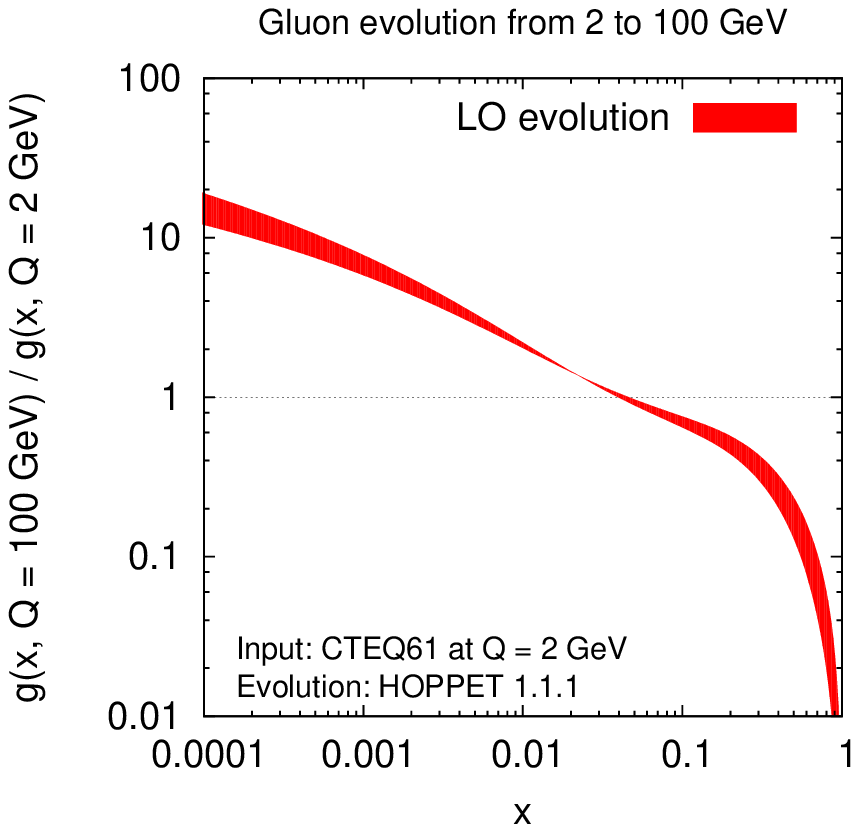}
  \end{minipage}
  \caption{Left: the kinematic regions covered by HERA, fixed-target
    scattering experiments, and by the LHC (adapted from the
    corresponding plot by Stirling). Right: the factor by which the
    gluon distribution evolves in going from a scale of $2\GeV$ to
    $100\GeV$ using CTEQ61 distributions as a fixed input at the low
    scale, and carrying out LO DGLAP evolution with
    HOPPET~\cite{Salam:2008qg} with $x_\mu=\frac12,1,2$ (see text for
    further details).}
  \label{fig:PDFs-to-LHC}
\end{figure}

Figure~\ref{fig:PDFs-to-LHC} (left) illustrates the kinematic region
in the $x$ and $Q^2$ plane that is covered by the LHC (with $\sqrt s =
14\TeV$), compared to that for HERA and fixed-target experiments.
The LHC region is labelled with a grid corresponding to mass ($M$, and
one takes $Q=M$) and rapidity ($y$) of the object that is being
produced. These are related to the incoming momentum fractions $x_1$
and $x_2$ and the $pp$ squared centre-of-mass energy through
\begin{equation}
  \label{eq:lhc-kinematics}
  M = \sqrt{x_1 x_2 s}\,, \qquad\qquad y = \frac12 \ln \frac{x_1}{x_2}\,.
\end{equation}
An object produced at a rapidity $y$ involves $x_{1} = \frac{M}{\sqrt{s}}
e^{+y}$ and $x_{2} =  \frac{M}{\sqrt{s}} e^{-y}$.
One feature that's immediately visible from the plot is that much of
the LHC kinematic plane covers regions where PDFs have not been
measured directly.
We will therefore rely heavily on DGLAP evolution for our
predictions. 
The right-hand plot in Fig.~\ref{fig:PDFs-to-LHC} illustrates just how
much, in that it gives the factor by which the gluon distribution
evolves in going from $Q=2\GeV$ to $Q=100\GeV$.
Depending on the region in $x$, this can be a large factor,
$\order{10}$. 
When compared to the experimental uncertainties on PDFs as shown in
Figs.~\ref{fig:pdf-uncertainties} and
\ref{fig:pdf-uncertainties-impact}, we clearly have to ask how well we
know the evolution.

In Section~\ref{sec:fixed-order} we will discuss in detail how
uncertainties are estimated in theoretical predictions.
For now, essentially, there is freedom in Eq.~(\ref{eq:matrix-dglap})
to choose a renormalization scale $\muR^2$ for $\as$ that differs from
$\muF^2$. A conventional way of estimating the uncertainties is to
choose $\muR^2 = (x_\mu \muF)^2$, varying $x_\mu$ in the range
$\frac12 < x_\mu < 2$ (actually one often just takes three values,
$x_\mu=\frac12,1,2$).
Starting from a fixed input at $2\GeV$ and evolving with different
choices for $x_\mu$ gives the width of the band shown in
Fig.~\ref{fig:PDFs-to-LHC} (right).
That width is much larger than the uncertainties that we see in
Figs.~\ref{fig:pdf-uncertainties} and
\ref{fig:pdf-uncertainties-impact}.

\begin{figure}
  \centering
  \includegraphics[width=0.48\tw]{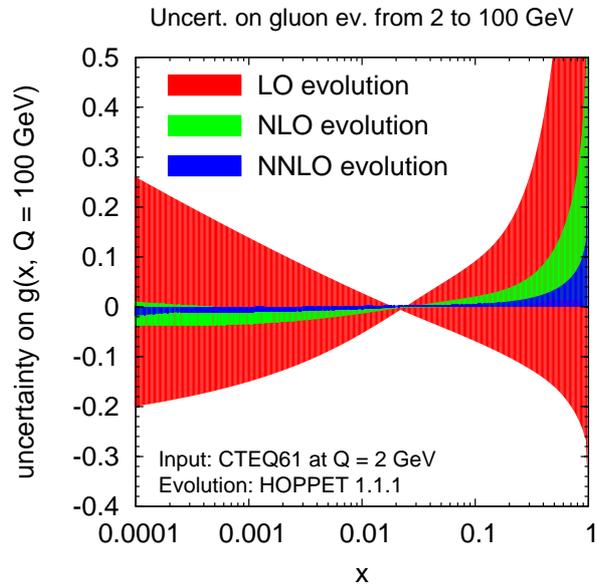}
  \caption{Uncertainties on the evolution of the gluon distribution
    from a fixed (CTEQ61) input at scale $2\GeV$ up to $100\GeV$, for
    LO, NLO and NNLO evolution. The bands correspond to the envelope
    of the results with three scale choices $x_\mu = \frac12,1,2$ and
    what is plotted is the ratio to the result at scale $100\GeV$, as
    obtained with NNLO evolution using $x_\mu=1$. }
  \label{fig:evolution-uncertainties}
\end{figure}

Fortunately we are not limited to leading-order (LO) DGLAP evolution,
i.e., just the $\order{\as}$ term in Eq.~(\ref{eq:matrix-dglap}). The
order $\as^2$ (next-to-leading order --- NLO) corrections to the DGLAP
equation were calculated around
1980~\cite{Furmanski:1980cm,Curci:1980uw} and in 2004 the calculation
of the NNLO corrections was completed~\cite{Vogt:2004mw,Moch:2004pa}.
To give an idea of the complexity of that calculation, rather than
taking a couple of lines as in Eq.~(\ref{eq:splitting-fns}) at LO, the
NNLO results take $\order{10}$ pages to write down!

The impact of including higher orders on the evolution uncertainties
is illustrated in Fig.~\ref{fig:evolution-uncertainties}.
The same CTEQ61 input distribution is evolved from $2\GeV$ to
$100\GeV$ with LO, NLO and NNLO splitting kernels, using three scale
choices, $x_\mu=\frac12,1,2$.
The figure then shows the results for the gluon distribution at scale
$100\GeV$, normalized to gluon distribution obtained with NNLO
evolution and $x_\mu=1$.
One sees how the uncertainty comes down substantially as higher orders
are included: $30\%$ in some regions at LO, $5\%$ over most of the $x$
range at NLO and $2\%$ at NNLO.
In the NNLO case, the uncertainty is usually smaller than the
experimental uncertainties for the PDFs that were shown in
Fig.~\ref{fig:pdf-uncertainties}.%
\footnote{ Figure~\ref{fig:evolution-uncertainties} is to be interpreted
  with some caution. We have taken a fit carried out with $x_\mu=1$,
  and then evolved it to high scales with $x_\mu\neq 1$. However, for
  example, the gluon distribution is partially determined from the
  evolution of $F_2$, so if the fit itself were carried out with
  $x_\mu \neq 1$, the fit result would probably change, introducing
  additional dependence of the $Q=100$ results on the choice of
  $x_\mu$. This could conceivably cancel some of the dependence seen
  in Fig.~\ref{fig:evolution-uncertainties}. }

\subsection{Summary}

Here are some of the points to retain from this section. 
Firstly, the proton really is what we expect it to be, i.e., a $uud$
state; however, fluctuations of the proton introduce many extra $q\bar
q$ pairs (`sea'), as well as a substantial amount of `glue',
carrying $50\%$ of the proton's momentum. The sea and gluon
distributions diverge at small momentum fractions $x$.
 
Determination of the proton's PDFs involves fitting data from a
range of experiments, with direct sensitivity to quarks (e.g., DIS),
indirect sensitivity to the gluon (DIS $Q^2$ evolution), and direct
sensitivity to quarks and gluons (jet data).

One of the keys to being able to measure consistent PDFs from
different experiments, thinking about them in perturbative QCD and
then applying them to predict results at new experiments is
`factorization': initial-state radiation, though collinear
divergent, is process-independent; the divergent part can be absorbed
into the definition of the PDFs, and then a universal set of PDFs,
evolved between different scales with the DGLAP equations, can be used
for any process.

Finally, the accuracy with which we know PDFs is quite remarkable: both
from the experimental side and the theoretical side, in the best cases
we know the PDFs to within a few per~cent.
This will be important in interpreting future signals of new
particles, for example in Higgs-boson production at the LHC when we
want to deduce its electroweak couplings given a measurement of
its cross section.

If you need to use PDFs yourself, the best place to get them is from
the LHAPDF library~\cite{LHAPDF}.

\section{Predictive methods for LHC}
\label{sec:predictive-methods}

In this section we will look at some of the different classes of
technique can be used to make QCD predictions at LHC.
Among the topics that we'll touch on are leading order (LO),
next-to-leading order (NLO) and next-to-next-leading order (NNLO)
calculations, parton-shower Monte Carlos, and then methods to combine
the two types of calculation.

\begin{figure}
  \centering
  \begin{minipage}[t]{0.48\linewidth}
    \centering
    \underline{Signal}
  \end{minipage}\hfill
  \begin{minipage}[t]{0.48\linewidth}
    \centering
    \underline{Background}
  \end{minipage}

  \begin{minipage}[c]{0.24\linewidth}
    \includegraphics[width=\tw]{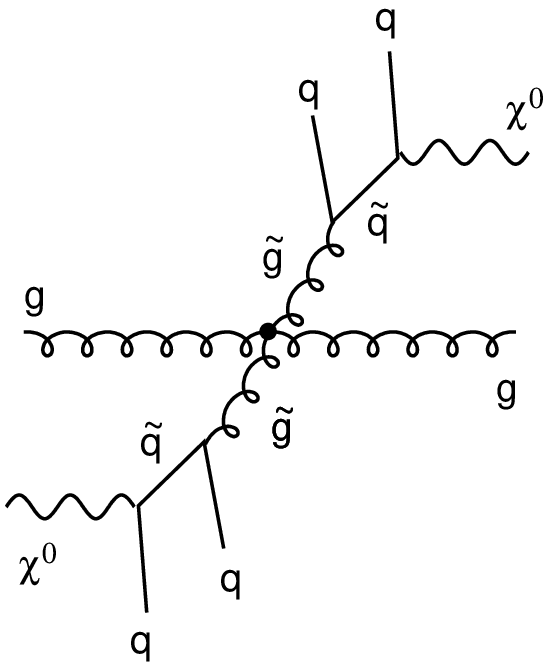}%
  \end{minipage}
  \begin{minipage}[c]{0.24\linewidth}
    \includegraphics[width=\tw]{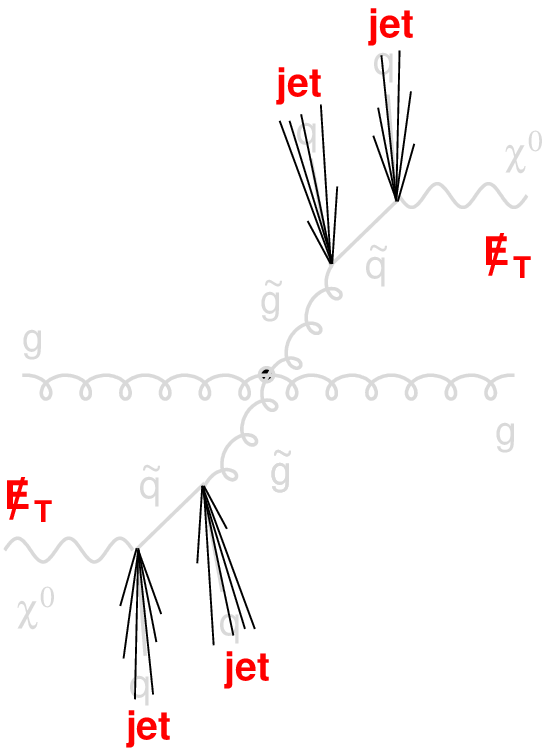}%
  \end{minipage}\hfill
  \begin{minipage}[c]{0.24\linewidth}
    \includegraphics[width=\tw]{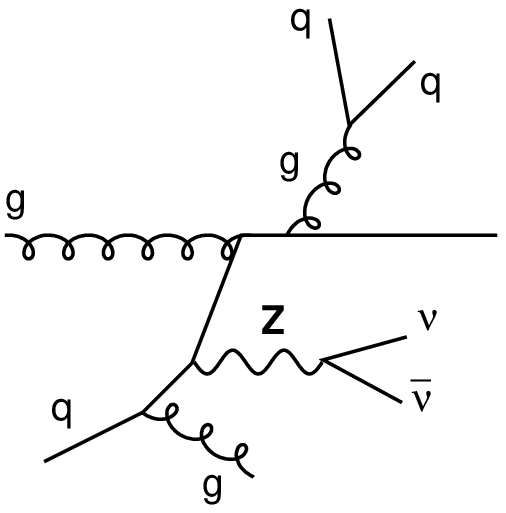}%
  \end{minipage}
  \begin{minipage}[c]{0.24\linewidth}
    \includegraphics[width=\tw]{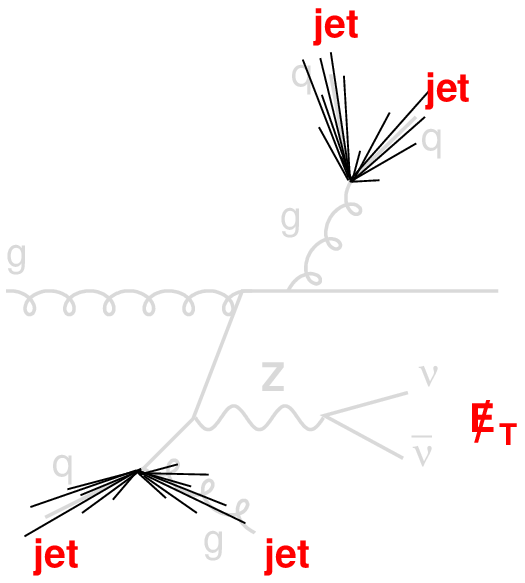}%
  \end{minipage}
  \caption{Left: production of a gluino ($\tilde g$) pair and the
    subsequent decay of each gluino through a squark ($\tilde q$) to
    quarks and a neutralino ($\chi^0$); the experimental signature
    involves four jets and missing transverse energy ($\slE_T$) from
    the unobserved neutralino.
    Right: a background that mimics this signature, with missing
    energy coming from the production of a $Z$-boson that decays to
    neutrinos.}
  \label{fig:signal-v-background-diagrams}
\end{figure}

Many of the examples that we'll use will involve $Z$ (and sometimes
$W$) production at hadron colliders. 
One reason is that $Z$ and $W$ bosons decay to leptons and neutrinos
(missing energy), both of which are easily-taggable handles that are
characteristic of signals in many new-physics scenarios.
An illustration is given in
Fig.~\ref{fig:signal-v-background-diagrams}, which depicts
supersymmetric production of a gluino pair and subsequent decay to four
jets and missing transverse energy from the unobserved neutralinos.
Because of the complexity of the decays and the fact that the missing
energy is the sum of that from two neutralinos, it can be difficult to
extract clear kinematic structures (such as an invariant mass peak)
that make a signal emerge unambiguously over the background.
In such cases the contribution from the signal may just be to give a
cross section that is larger than background expectations over a broad
kinematic range. 
But that will only be a `signal' if we understand what the
backgrounds are. 

The extent to which we will want to (or have to) rely on QCD
predictions of backgrounds in deciding whether there are signals of
new physics at the LHC is a subject that deserves in-depth consideration (for
a nice discussion of it, see Ref.~\cite{Mangano:2008ha}).
But QCD predictions will come into play in many other ways too. 
Monte Carlo parton shower programs, which simulate the full hadronic
final state, are crucial in evaluating detector acceptances and
response.
And knowing QCD predictions (both for backgrounds and possible
signals) is crucial in the design of methods to search for new
physics, as well as for extracting meaning from the data (about
couplings, spins, etc.) when, it is to be hoped, we finally see signals of
something new.

\subsection{Fixed-order predictions}
\label{sec:fixed-order}

Fixed-order predictions, which involve the first couple of terms in
the QCD perturbative expansion for a given cross section, are
conceptually quite simple: it is easy to state which contributions are
included, and as one includes further orders in the expansion one
can reasonably hope to see systematic improvement in the accuracy of
one's predictions.

We'll first look at a couple of examples of fixed-order predictions,
in order to develop a feel for how the perturbative expansion behaves,
and how one estimates its accuracy.
We will then examine more generally what theoretical inputs are needed
for predictions for a given process, and what practical forms the
predictive tools take.

\subsubsection{Example 1: the cross section for $e^+e^- \to $~hadrons
  and its scale dependence}

In Eq.~(\ref{eq:tot-xsct-series}), we wrote the total cross section
for $e^+e^-\to\text{hadrons}$ as a perturbative series expansion in
$\as$ that multiplied the Born cross section $e^+e^-\to q\bar q$.
The expansion was formulated in terms of the coupling evaluated at a
renormalization scale $\muR$ equal to the centre-of-mass energy $Q$,
i.e., $\as(\muR=Q)$.
That choice is, however, arbitrary: for example, the most energetic
gluon that could be produced in $e^+e^-\to q\bar qg$ would be one with
$E=Q/2$, so maybe we should be choosing $\muR = Q/2$.
And in loop diagrams, one integrates over gluon energies that go
beyond $Q$, so maybe $\muR = 2Q$ would be just as reasonable.

\begin{figure}
  \centering
  \includegraphics[width=0.42\tw]{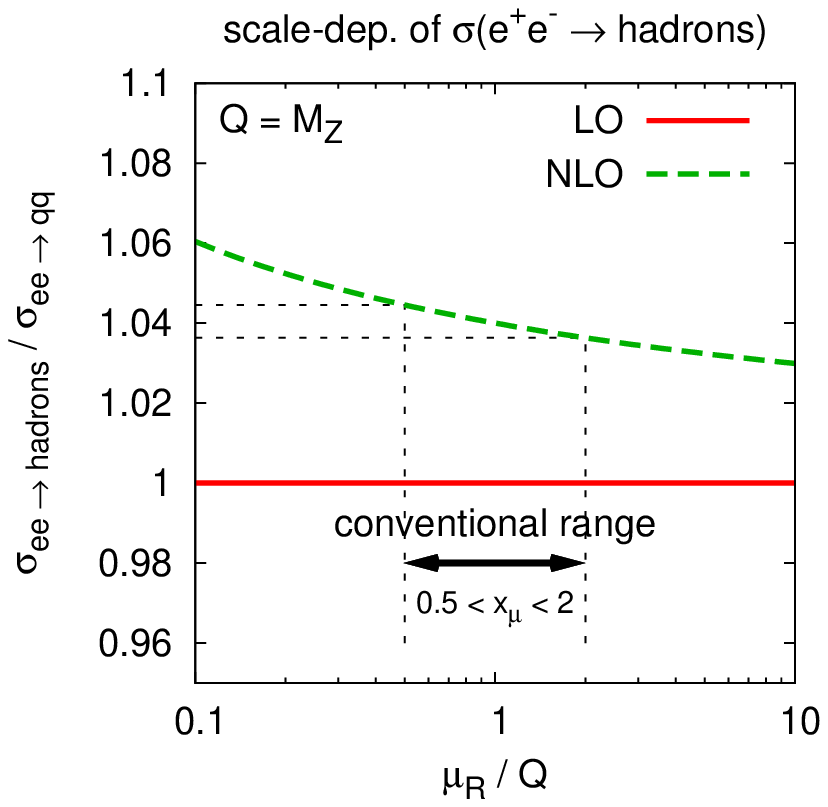}\qquad
  \includegraphics[width=0.42\tw]{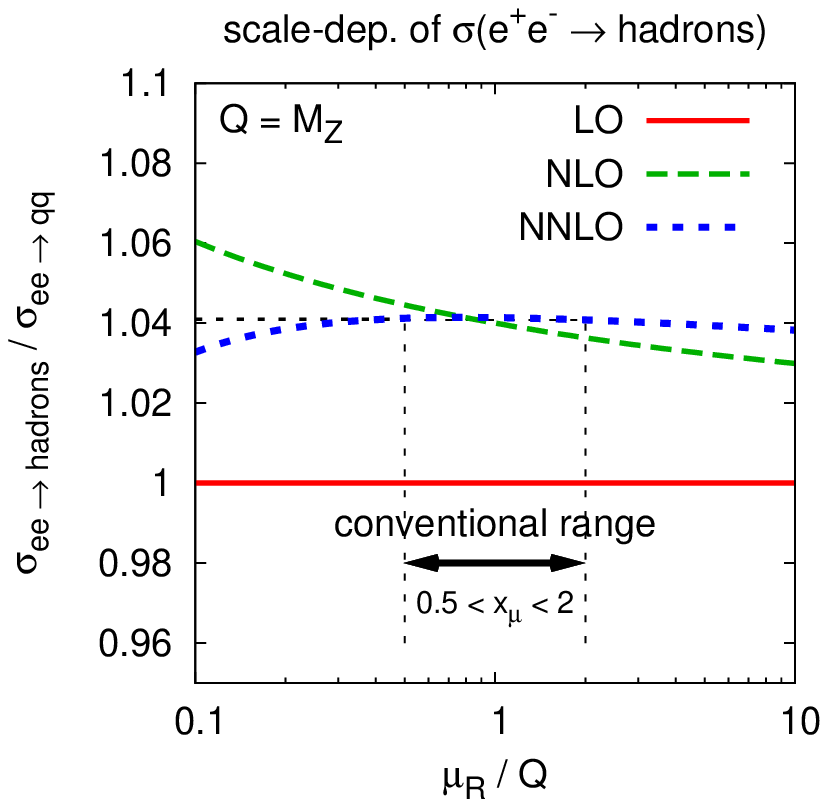}
  \caption{Renormalization-scale dependence of the NLO (left) and NNLO
    (right) predictions for the $e^+e^-\to\text{hadrons}$ total cross
    section, together with an indication of the conventional choice of
    scale-variation range}
  \label{fig:ee2hadrons-scale-dependent}
\end{figure}

Because of this arbitrariness, a convention has emerged whereby one
calculates a `central value' for the prediction by setting the
renormalization scale equal to the main physical scale for the process
(e.g., the centre-of-mass energy at an $e^+e^-$ collider; for
hadron-collider processes the choice may be less obvious). The
uncertainty is then estimated by varying the scale by a factor of two
in either direction from the central value, i.e., taking
$\frac{Q}{2}<\muR<2Q$. 
This is illustrated in Fig.~\ref{fig:ee2hadrons-scale-dependent}
(left), which plots
\begin{equation}
  \label{eq:ee-NLO}
  \sigma^{\text{NLO}} = \sigma_{q\bar q}(1 + c_1 \as(\muR))\,,
\end{equation}
as a function of $\muR$, showing how the $\muR$-dependence translates
into an uncertainty; note that $c_1$ can be read from
Eq.~(\ref{eq:tot-xsct-series}).
Given an expansion of the running coupling (i.e., of the middle result
of Eq.~(\ref{eq:renorm-group-1loop-soln}), $\as(\muR) = \as(Q) - 2
b_0 \as^2(Q) \ln \frac{\muR}{Q} + \order{\as^3}$), we can rewrite
Eq.~(\ref{eq:ee-NLO}) as
\begin{equation}
  \label{eq:ee-NLO-explicit-scale}
  \sigma^{\text{NLO}}(\muR) = \sigma_{q\bar q}\left(1 + c_1 \as(Q) - 2 c_1
  b_0 \as^2(Q) \ln \frac{\muR}{Q} + \order{\as^3}\right)\,.
\end{equation}
This tells us that as we vary the renormalization scale for a
prediction up to $\order{\as}$ (NLO), we effectively introduce
$\order{\as^2}$ (NNLO) pieces into the calculation:
by generating some fake set of NNLO terms, we are probing the
uncertainty of the cross section associated with the missing full NNLO
correction.

If we calculate the actual NNLO cross section for general $\muR$, it
will have a form
\begin{equation}
  \label{eq:ee-NNLO-explicit-scale}
  \sigma^{\text{NNLO}}(\muR) = \sigma_{q\bar q}\left(
    1 + c_1 \as(\muR) + c_2(\muR) \as^2(\muR)\right)\,.
\end{equation}
Observe that the $c_2$ coefficient now depends on $\muR$.  This is
necessary because the second-order coefficient must cancel the
$\order{\as^2}$ ambiguity due to the scale choice in
Eq.~(\ref{eq:ee-NLO-explicit-scale}).
This constrains how $c_2(\muR)$ depends on $\muR$:
\begin{equation}
  \label{eq:ee-NNLO-c2-mu-dep}
  c_2(\muR) = c_2(Q) +  2 c_1 b_0 \as^2(Q) \ln \frac{\muR}{Q}\,,
\end{equation}
where $c_2(Q)$ can again be read from Eq.~(\ref{eq:tot-xsct-series}). 
If we now express $\sigma^\text{NNLO}(\muR)$ in terms of $\as(Q)$, we
will find that the residual dependence on $\muR$ appears entirely at
$\order{\as}$, i.e., one order further than in
Eq.~(\ref{eq:ee-NLO-explicit-scale}).
This is reflected in the right-hand plot of
Fig.~\ref{fig:ee2hadrons-scale-dependent}, which illustrates how the
impact of the scale variation at NNLO is significantly reduced, since
we are now probing the impact of missing $\as^3$ terms, rather than
$\as^2$ terms.

If we had an arbitrarily large number of terms in the $\as$
expansion, the scale dependence would disappear exactly. 
The fact it doesn't in the presence of a fixed number of terms may
initially seem like a drawback, but in some respects it's a blessing
in disguise because it provides a useful handle on the uncertainties. 
This is why scale variation has become a standard procedure.
It's worth bearing in mind that it isn't a failsafe mechanism: a
trivial example comes from the LO curve in
Fig.~\ref{fig:ee2hadrons-scale-dependent}. It doesn't have any scale
variation because they don't depend on $\as$, yet it differs
significantly from the higher-order results.

\subsubsection{Example 2: $pp \to Z$}

At LO the $pp \to Z$ cross section involves a single underlying hard
partonic process, namely $q\bar q \to Z$, which is purely
electroweak. 
To go from the $q\bar q \to Z$ squared matrix element to the $pp \to
Z$ result, one must integrate over the quark distributions
\begin{equation}
  \label{eq:pp2Z-LO}
  \sigma^\text{\sc lo}_{pp\to Z} = \sum_i 
  \int dx_1 dx_2 \, f_{q_i}(x_1,\muF^2)\, f_{\bar q_i}(x_2,\muF^2)
  \,\, \hat \sigma_{0,q_i\bar q_i\to Z} (x_1 p_1, x_2 p_2)\,,
\end{equation}
for which one must choose a factorization scale $\muF$.
A natural choice for this scale is $\muF = M_Z$, but as with the
renormalization scale it is conventional to vary it by a factor of two
either side of the central choice in order to obtain a measure of the
uncertainties in the prediction.

\begin{figure}
  \centering
  \begin{minipage}[c]{0.38\linewidth}
    \includegraphics[width=\tw]{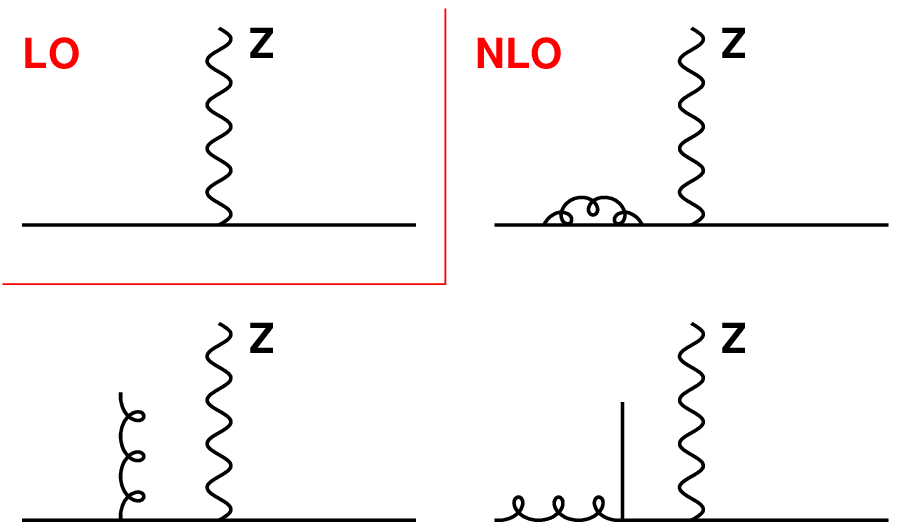}
  \end{minipage}
  \hfill
  \begin{minipage}[c]{0.48\linewidth}
    \includegraphics[width=\tw]{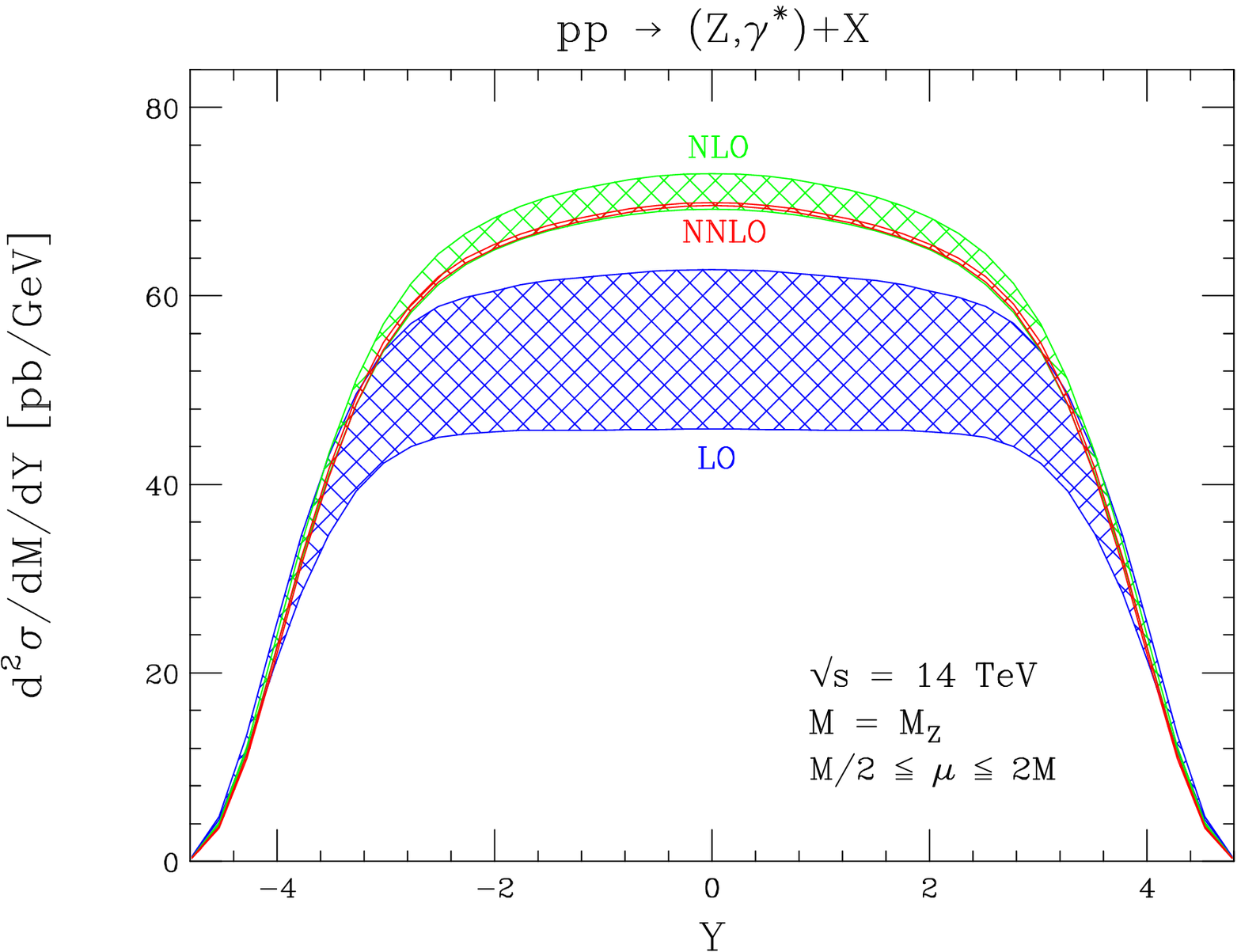}
  \end{minipage}
  \caption{Left: classes of diagram that appear for $pp\to Z$ at LO
    and at NLO. Right: cross section at the LHC for the $Z$-boson,
    differential in rapidity, at LO, NLO and NNLO, as taken from
    Ref.~\cite{Anastasiou:2003ds}.}
  \label{fig:Z-xsct}
\end{figure}

Adding NLO and NNLO terms, the structure becomes
\begin{multline}
  \label{eq:pp2Z-NNLO}
  \sigma^\text{\sc nnlo}_{pp\to Z\!+\!X} = \sum_{i,j} \int dx_1 dx_2 \,
  f_{i}(x_1,\muF^2)\, f_{j}(x_2,\muF^2) \bigg[ \hat 
    \sigma_{0,ij\to Z} (x_1, x_2)\,  
    +\, \as(\muR) \hat\sigma_{1,ij\to Z\!+\!X} (x_1, x_2, \muF)\\[-0.7em]
    +\, \as^2(\muR) \hat\sigma_{2,ij\to Z\!+\!X} (x_1, x_2, \muF, \muR)
    \bigg].
\end{multline}
We now have a sum over the flavours $i$ and $j$ of the initial
partons, because starting from NLO there are contributions from (say)
gluon-quark scattering [cf. Fig.~\ref{fig:Z-xsct}(left)]. 
The cross section is written as being for $Z\!+\!X$, where the $X$
means that we allow anything (e.g., quarks, gluons) to be produced in
addition to the $Z$-boson.
At $\order{\as}$ the $\muF$ dependence of the $\sigma_{1}$ coefficient
partially cancels the dependence present at $\order{\as^0}$ coming
from the $\muF$ dependence of the PDFs.
That dependence is further cancelled at $\order{\as^2}$, as is part of
the $\muR$ dependence that is introduced in the $\order{\as(\muR)}$ term.
The plot on the right of Fig.~\ref{fig:Z-xsct} shows the $Z$-boson cross
section as a function of its rapidity \cite{Anastasiou:2003ds}.
The bands indicate the uncertainty due to scale variation (taking
$\frac12 M_Z < \muR=\muF < 2M_Z$)\footnote{%
  The variation of $\muR$ and $\muF$ simultaneously, though common, is
  not the only possible procedure. An attractive alternative is to vary both
  independently around a central scale, with the additional
  requirement that $\frac12 < \muR / \muF < 2$ \cite{Cacciari:2003fi}.
} and show how this uncertainty
undergoes important reductions going from LO to NLO to
NNLO.

One of the interesting features that comes out of
Fig.~\ref{fig:Z-xsct} is that the LO prediction is only good to within
a factor of $1.5$ to $2$, despite the fact that $\as(M_Z) \simeq
0.118$ would imply $10\%$ accuracy.
This is because the $\order{\as}$ corrections come with large
coefficients. This is not uncommon in hadron-collider cross sections.
Furthermore the LO uncertainty band seems not to provide a faithful
measure of the true uncertainty. 
Other aspects of the perturbative expansion do seem to behave as one
would expect: the size of the uncertainty band decreases significantly
going from LO to NLO (10--20$\%$) to NNLO (a few per~cent). And the
actual shift in the central value in going from NLO to NNLO is
substantially smaller than that from NLO to LO.

Are these characteristics representative of the `typical' situation
for collider observables? We only have predictions up to NNLO in a
handful of cases (see below) and in those it is. 
In cases where we just have NLO predictions, the features of large
`K-factors' (NLO/LO enhancements) with a reduced NLO uncertainty
band are not uncommon, suggesting that beyond NLO corrections should
be small. 
Exceptions are known to arise in two types of case: those where new
enhanced partonic scattering channels open up at NLO (or beyond); and
that involve two disparate physical scales.
For example, if you ask for the $Z$-boson to have a transverse momentum
$p_t$ that is much smaller than $M_Z$, then each power of $\as$ in the
expansion of the cross section will be accompanied by up to two powers
of $\ln^2 M_Z / p_t$, leading to large coefficients at all orders in
the perturbative expansion. These are due to incomplete cancellation
between real and virtual (loop) divergences: loop corrections do not
affect the $Z$-boson $p_t$ and so are fully integrated over, whereas
real emissions do affect the Z $p_t$ and so are only partially
integrated over.

\subsubsection{Predictions for more complex processes}

\begin{figure}
  \centering
  \includegraphics[width=0.6\tw]{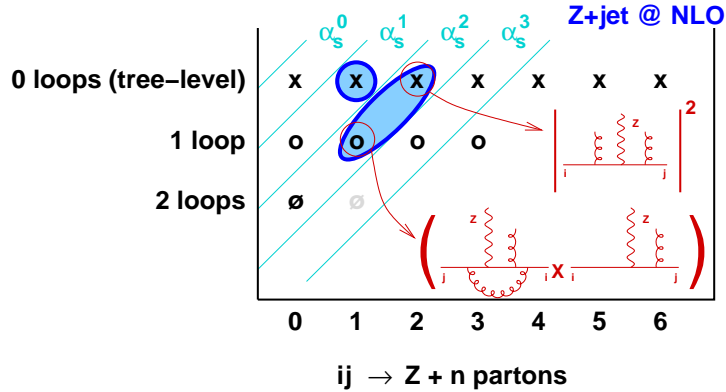}
  \caption{Illustration of the contributions that are known for $ij
    \to Z + \text{n partons}$, where $i$ and $j$ are arbitrary
    incoming partons, according to the number of outgoing partons, the
    number of loops and the number of powers of the coupling.
    An `{\sf x}' represents a squared tree-level diagram, an `{\sf o}'
    represents the interference of a 1-loop diagram with a tree-level
    diagram, and a `{\sf \o}' represents the interference of a
    two-loop diagram with a tree-level diagram or the square of a
    1-loop diagram.  
    Entries in black are known and used; entries in grey are known but
    have not been used.
    The entries in the shaded ellipses are those that are relevant
    for the NLO calculation of the cross section for the production of a
    $Z$-boson with a jet.  }
  \label{fig:loop+legs-static}
\end{figure}

As an example of a more complex process, consider the production of a
$Z$-boson plus a jet.
The leading order cross section requires the calculation of the
$\order{\as}$ squared diagrams for $q\bar q \to Zg$, $qg \to Zq$ and $\bar
qg \to Z\bar q$.
The NLO cross section additionally requires all $\order{\as^2}$
contributions with a $Z$ boson and at least one jet, as illustrated in
Fig.~\ref{fig:loop+legs-static}, i.e., the squared tree-level diagram
for $ij \to Z+\text{2 partons}$ and the interference of the 1-loop and
tree-level diagrams for $ij \to Z+\text{1 parton}$.

More generally, Fig.~\ref{fig:loop+legs-static} allows you to read off
the contributions that you will need for an N$^{p}$LO calculation of
$ij \to Z+\text{n partons}$: just take all entries in
Fig.~\ref{fig:loop+legs-static} with at least $n$ partons, up to order
$\as^{n+p}$.
Entries in black are known and have already been used to obtain
predictions for LHC and Tevatron.
The one entry in grey, the 2-loop $Z+\text{1 parton}$ contribution, is
known but has yet to be used in any hadron-collider prediction, for
reasons that we will discuss below.
Entries that are absent (e.g., $Z+2$~partons at  two loops) have so far
proven to be too complicated to calculate. 

The classes of contributions calculated for $ij \to Z+\text{n
  partons}$ provide a representative view of the situation for other
processes as well, with tree-level diagrams calculated up to quite
high final-state multiplicities, $\sim 10$, 1-loop diagrams having
been used for processes with up to 3 or sometimes 4 final-state
particles, and 2-loop diagrams available and used only for $2\to 1$
type processes, essentially $pp\to W$, $pp\to Z/\gamma^*$ and $pp \to
H$.

It's natural to ask in what form these various calculations are
available. For certain very simple quantities, for example, the total
cross section for $t \bar t$ production, or for $W$, $Z$ or Higgs
production, the result of the perturbative calculation can be written
as in Eq.~(\ref{eq:pp2Z-NNLO}),
\begin{equation}
  \label{eq:incl-xsect-general}
  \sigma^\text{\sc n$^p$lo}_{pp\to A+X} = \sum_{i,j} \int dx_1 dx_2
  \, f_{i}(x_1,\mu_F^2)\, f_{j}(x_2,\mu_F^2)\times
  \sum_{m=0}^{p} \as^{n+m}(\mu_R) \, \hat \sigma_{m,i j \to A+X}
  (x_1 x_2 s, \mu_R, \mu_F)\,,
\end{equation}
where the $\sigma_{m,i j \to A+X} (x_1 x_2 s, \mu_R, \mu_F)$ are
functions whose analytical expressions can be found in the relevant
papers (Refs.~\cite{Hamberg:1990np,Harlander:2002wh} for $W$ and $Z$, and
Refs.~\cite{Harlander:2002wh,Anastasiou:2002yz,Ravindran:2003um,Harlander:2009mq}
for Higgs-boson production).
To obtain a prediction, one just has to type them into a computer
program and then integrate over $x_1$ and $x_2$.

In most cases, however, one wants to calculate a cross section that
incorporates experimental cuts, such as lepton acceptances,
transverse momentum cuts on jets, etc.
In these cases the type of tool that should be used depends on the
order to which you want the answer.

\paragraph{LO predictions}

As long as one is dealing with infrared safe observables, then for a
LO prediction one need only include tree-level diagrams, in kinematic
regions in which their contributions are finite. 
The simplest approach therefore is to carry out Monte Carlo
integration over phase-space points, have a subroutine that determines
whether a given phase-space point passes the cuts, and if it does
calculate the squared matrix elements and PDF factors for each
possible partonic subprocesses.

Quite a number of tools enable you to do this: 
{\sc Alpgen}~\cite{Mangano:2002ea}, 
{\sc Comix/Sherpa}~\cite{Gleisberg:2008fv}, 
{\sc CompHEP}~\cite{Boos:2004kh}, 
{\sc Helac/Phegas}~\cite{Cafarella:2007pc} and 
{\sc MadGraph}~\cite{Alwall:2007st}. 
They allow you to calculate cross sections for a
broad range of $2\to n$ scattering processes with $n$ up to 6--8 (or
in some cases even beyond).
Some of these ({\sc CompHEP, MadGraph}) use formulae obtained from direct
evaluations of Feynman diagrams. This gives them significant
flexibility in terms of the range or processes they support (e.g.,
with easy inclusion of many new physics models), though they suffer at
large $n$ because the number of diagrams to evaluate grows rapidly as
$n$ increases (cf.\ Table~\ref{tab:ndiags}).
Others ({\sc Alpgen, Helac/Phegas} and {\sc Comix/Sherpa}) use methods designed to
be particularly efficient at high multiplicities, such as
Berends--Giele recursion~\cite{Berends:1987me} , which builds up
amplitudes for complex processes by recursively reusing simpler ones
(a nice technical review of the technique is given in
Ref.~\cite{Dixon:1996wi}).

\begin{table}
  \centering
    \caption{\label{tab:ndiags}
The number of Feynman diagrams for tree level $gg\to N
      \text{gluon}$ scattering~\cite{GloverCAPPTalk}}
  \begin{tabular}{lccccccc}\toprule
     {\boldmath $N$}     &  {\bf 2}    &    {\bf  3}  &      {\bf  4}    &     {\bf  5}   &    {\bf   6} &  {\bf 7}  & {\bf 8}\\\midrule
      No. diags  &   4   &      25  &      220    &    2485   &    34300 &
      $5\times 10^5$ & $10^7$\\ \bottomrule
    \end{tabular}
\end{table}

\paragraph{NLO predictions}

When, in Sections \ref{sec:total-xsct} and~\ref{sec:ir-safe-obs}, we looked at the cancellation between divergences in real and
loop diagrams, we wrote the loop diagram with an explicit
integral over phase space  so as to be able to match the divergences
between real and loop diagrams and cancel them easily.

A subtlety that we ignored is that in practical evaluations of loop
diagrams, the integral over loop momenta is carried out in
$4-\epsilon$ dimensions rather than 4 dimensions, in order to
regularize the divergences that appear and obtain an answer whose
finite part is known independently of any related tree-level diagrams.
On the other-hand, experimental cuts are defined in four dimensions, so the real
tree-level diagrams must be integrated in four dimensions, which implies
divergent results if the real diagrams are taken alone.

This mismatch between the ways loop and tree-level diagrams are
handled is one of the main difficulties in carrying out calculations that
include experimental cuts beyond LO.
For the calculation of a process with $n$ partons at LO, the standard
technique nowadays to deal with the problem is to introduce a
$n\!+\!1$-parton \emph{counterterm} where the $n\!+\!1^\text{th}$
parton is always soft and collinear so that it doesn't affect an IR
safe observable.
It is subtracted from the $n\!+\!1$-parton real diagram in four
dimensions and designed so as to cancel all of its soft and collinear
divergences.
It is also designed such that the kinematics of its
$n\!+\!1^\text{th}$ parton can be integrated analytically in
$4-\epsilon$ dimensions so that the result can be easily added to the
loop diagram and cancel \emph{its} divergences.
Since the counterterm is subtracted once and added once, its net
impact is null, and can just be thought of as a way of reshuffling
divergences.
This is known as a `subtraction' procedure and the variant most
widely used in current NLO computer codes is  `dipole'
subtraction~\cite{Catani:1996vz}; other methods that have seen
numerous applications include FKS \cite{Frixione:1995ms} and antenna
\cite{Campbell:1998nn} subtraction.

\begin{figure}
  \centering
    \begin{minipage}{0.42\linewidth}
    \includegraphics[width=\tw]{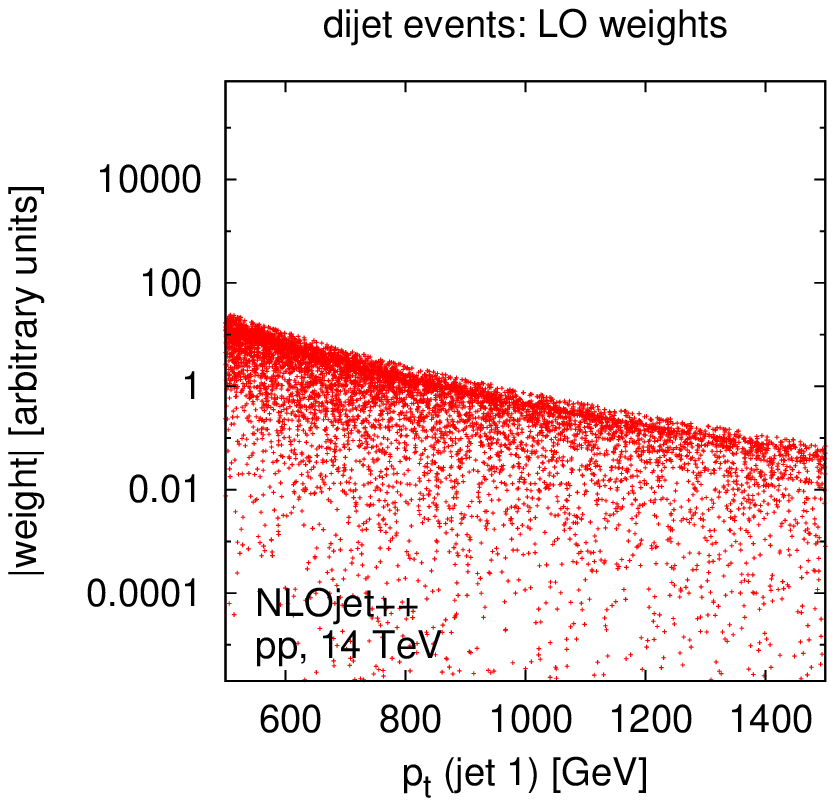}
  \end{minipage}
  \quad
  \begin{minipage}{0.42\linewidth}
    \includegraphics[width=\tw]{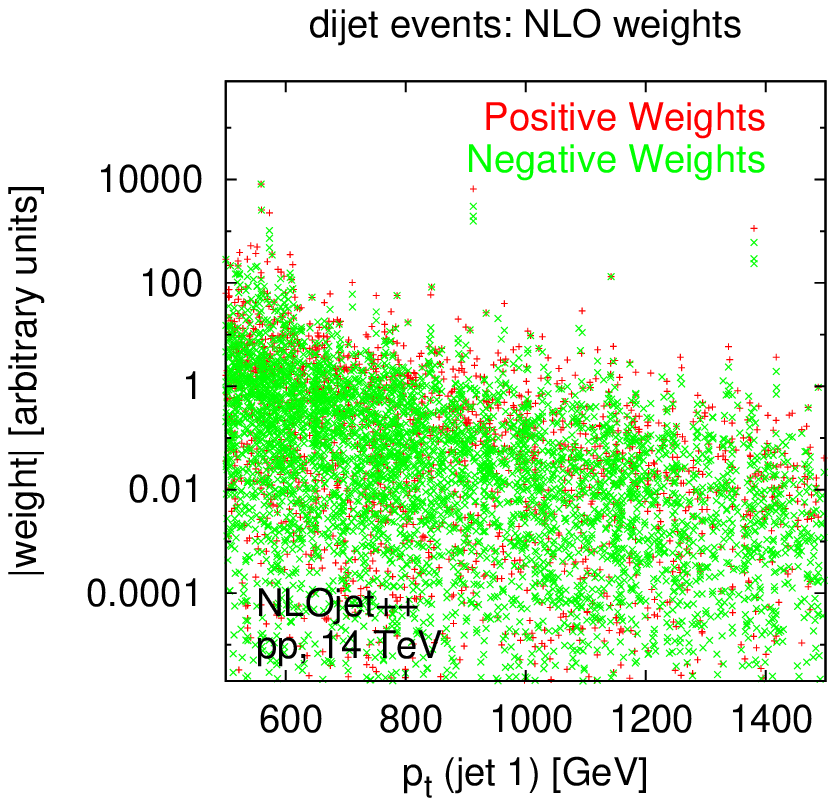}
  \end{minipage}
  \caption{Left: weights associated with tree-level LO $2\to 2$
    scattering events in the calculation of the dijet cross section,
    shown as a function of the transverse momentum of the harder
    jet. Right: weights at NLO from real $2\to 3$ events, subtraction
    counterterm events and loop events (after addition of the
    integrated counterterm), again as a function of $p_t$ of the
    hardest jet. Results obtained with NLOjet\texttt{++}~\cite{Nagy:2003tz}.}
  \label{fig:nlojet-weights}
\end{figure}

An illustration of how subtraction works in practice is given in
Fig.~\ref{fig:nlojet-weights} for dijet production. 
In the left-hand plot we see weights for the LO process as a function
of the jet $p_t$. Though there is some dispersion in the weights,
there is a clear upper bound as a function of $p_t$, reflecting the
finiteness of the matrix-elements.
The right-hand plot shows weights at NLO. Here there is no clear upper
bound to the weights; however we do see that unusually large (nearly
divergent) weights come in clusters: usually one positive (red, `+')
weight, accompanied by one or more negative (green, `$\times$')
weights with identical jet transverse momenta, so that each event in
the cluster contributes to the same bin of the cross section and their
weights sum to a finite result.

Technically, one main consideration has so far limited the range of
processes for which NLO results exist: the availability of the loop
amplitude.
Until recently loop amplitudes were usually calculated semi-manually
for each process.  The complexity of the calculations increased
significantly with the 
number of outgoing legs, limiting available results to those with at
most three outgoing partons.
Many NLO results for $2\to 2$ and $2\to3$ processes are incorporated
into programs such as {\sc NLOJet}\texttt{++} for jet
production~\cite{Nagy:2003tz}, {\sc MCFM} for processes with heavy quarks
and/or heavy electroweak bosons~\cite{Campbell:2000bg}, {\sc VBFNLO} for
vector-boson fusion processes~\cite{Arnold:2008rz}, and the 
{\sc Phox} family~\cite{Binoth:1999qq} for processes with photons in the final state.

In the past couple of years, techniques have come to fruition that
offer the prospect of automated calculation of arbitrary loop
diagrams. 
Though full automation is not here yet, a number of $2\to4$ processes
have now been calculated at NLO thanks to these advances, including
$pp \to t\bar t b \bar b$ \cite{Bredenstein:2009aj,Bevilacqua:2009zn},
$pp \to t\bar t jj$~\cite{Bevilacqua:2010ve} (where $j$ represents a
jet) and $pp \to W\!+\!3j$ \cite{Berger:2009ep,KeithEllis:2009bu} and
$pp \to Z\!+\!3j$~\cite{Berger:2010vm}.%
\footnote{Since the original version of this writeup, first results
  for a $2\to5$ process have also appeared~\cite{Berger:2010zx}.}

It should be said that NLO calculations are very computing intensive:
for some observables it is not unusual to have to devote several years
of CPU time in order to get adequate numerical convergence of the
Monte Carlo integration. 

\paragraph{NNLO predictions}

NNLO predictions suffer from the same problem of cancelling
divergences between real and virtual corrections that is present at
NLO, with the complication that instead of having one soft and one
collinear divergence, there are now two of each, greatly complicating
the task of figuring out counterterms to allow experimental cuts to be
implemented in four dimensions.

As a result, the only general subtraction type approaches that exist
currently are for processes without incoming hadrons, notably $e^+e^-
\to 3j$~\cite{GehrmannDeRidder:2007hr,Weinzierl:2009ms}.
For hadron collider processes it is only $2\to 1$ processes that are
available, specifically vector-boson ({\sc FEWZ}~\cite{Melnikov:2006kv} and
{\sc DYNNLO}~\cite{Catani:2009sm}) and Higgs-boson 
({\sc FeHiP}~\cite{Anastasiou:2005qj} and {\sc HNNLO}~\cite{ Catani:2007vq}) production,
using methods that are not so easily generalizable to more
complicated processes.

\subsection{Monte Carlo parton-shower programs}
\label{sec:MC}

The programs we've discussed so far, known as `Matrix Element Monte
Carlos' provide a powerful combination of accuracy and flexibility as
long as you want to calculate IR and collinear safe observables (jets,
$W$'s, $Z$'s, but not pions, kaons, etc.), don't mind dealing with wildly
fluctuating positive and negative weights, and don't need to study
regions of phase space that involve disparate physical scales. 

All these defects are essentially related to the presence of soft and
collinear divergences. Yet we know that real life does not diverge.
So it is natural to wonder whether we can reinterpret the divergences
of perturbation theory physically.
It turns out that the right kind of question to ask is ``what is the
probability of \emph{not} radiating a gluon above some (transverse
momentum) scale $k_t$''. Starting from a $q\bar q$ system, using the
results of Section~\ref{sec:ee2hadrons}, we know that to $\order{\as}$
the answer in the soft and collinear limit goes as
\begin{equation}
  \label{eq:sudakov-expansion}
  P(\text{no emission above }k_t) \simeq 1 - \frac{2\as\CF}{\pi}
  \int^Q\frac{d{E}}{{E}}
  \int^{\pi/2}\frac{d\theta}{\theta} \Theta(E \theta - k_t)\,.
\end{equation}
It so happens that in the soft and collinear limit, this result is easy to
work out not just at first order, but at all orders, giving simply the
exponential of the first order result
\begin{equation}
  \label{eq:sudakov-exponentiated}
  P(\text{no emission above }k_t) \equiv \Delta(k_t,Q) \simeq \exp\left[ - \frac{2\as\CF}{\pi}
  \int^Q\frac{d{E}}{{E}}
  \int^{\pi/2}\frac{d\theta}{\theta} \Theta(E \theta - k_t) \right]\,.
\end{equation}
Whereas Eq.~(\ref{eq:sudakov-expansion}) had a `bare' infinity if
one took $k_t \to 0$, Eq.~(\ref{eq:sudakov-exponentiated}) is simply
bounded to be between $0$ and $1$. 

The quantity $\Delta(k_t,Q)$ is known as a Sudakov form factor. 
We've been very approximate in the way we've calculated it, neglecting
for example the running of the coupling ($\as$ should be placed inside
the integral and evaluated at the scale $E\theta$) and the treatment
of hard collinear radiation (the $dE/E$ integral should be replaced
with the full collinear splitting function), but these are just
technical details. 
The importance of the Sudakov form factor is that it allows us to
easily calculate the distribution in transverse momentum $k_{t1}$ of
the gluon with largest transverse momentum in an event:
\begin{equation}
  \label{eq:dP_dkt1}
  \frac{dP}{dk_{t1}} = \frac{d}{d k_{t1}} \Delta(k_{t1},Q)\,.
\end{equation}
This distribution is easy to generate by Monte Carlo methods: take a random 
number $r$ from a distribution that's uniform in the range $0<r<1$ and
find the $k_{t1}$ that solves $\Delta(k_{t1},Q) = r$. Given  $k_{t1}$
we also need to 
generate the energy for the gluon, but that's trivial.
If we started from a $q\bar q$ system (with some randomly
generated orientation), then this gives us a $q\bar q g$ system.
As a next step one can work out the Sudakov form factor in the
soft/collinear limit for there to be no emission from the $q\bar q g$
system as a whole above some scale $k_{t2}$ ($<k_{t1}$) and use this
to generate a second gluon.
The procedure is then repeated over and over again until you find that
the next gluon you would generate is below some non-perturbative
cutoff scale $Q_0$, at which point you stop.
This gives you one `parton shower' event.

This is essentially the procedure that's present in the shower of
{\sc Pythia} 8~\cite{Sjostrand:2007gs} and the $p_t$ ordered option of
{\sc Pythia} 6.4~\cite{Sjostrand:2006za}, as well as
{\sc Ariadne}~\cite{Lonnblad:1992tz} and {\sc Sherpa}~1.2 (the {\sc Sherpa}
reference~\cite{Gleisberg:2008ta} describes an earlier version).
It is also possible to choose other ordering variables: the original
{\sc Pythia} shower\cite{Sjostrand:2000wi,Sjostrand:2006za} is based on
virtuality ordering (plus an angular veto). 
This is still the most widely used shower which works well on a range
of data (though there are theoretical issues in some formulations of
virtuality ordering).
In the {\sc Herwig} family of programs~\cite{Corcella:2000bw,Bahr:2008pv} it
is angular ordering that is used.

The above shower descriptions hold for final-state branching. With
initial-state hadrons, one also needs to be careful with the treatment
of the PDFs, since the collinear splitting that is accounted for in
the parton shower is also connected with the way the PDF is built up
at the scale of the hard scattering.

\begin{figure}
  \centering
  \begin{minipage}[t]{0.45\linewidth}
    a)\\
    \includegraphics[width=\tw]{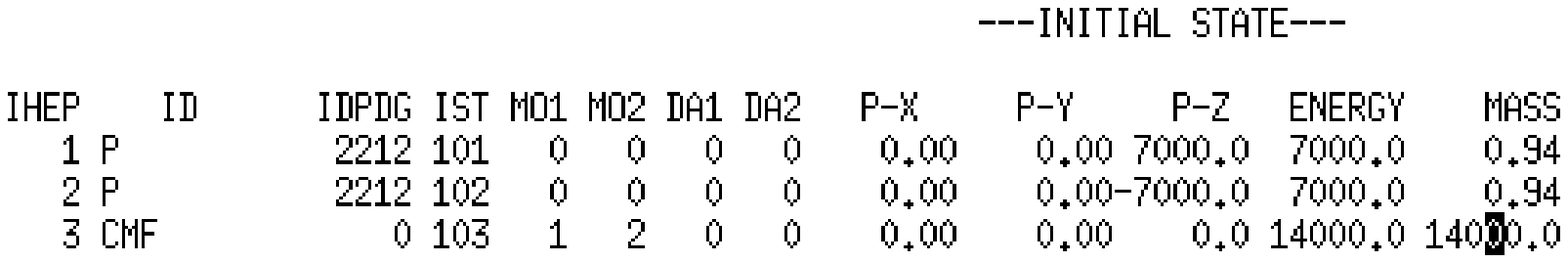}\\
    b)\\
    \includegraphics[width=\tw]{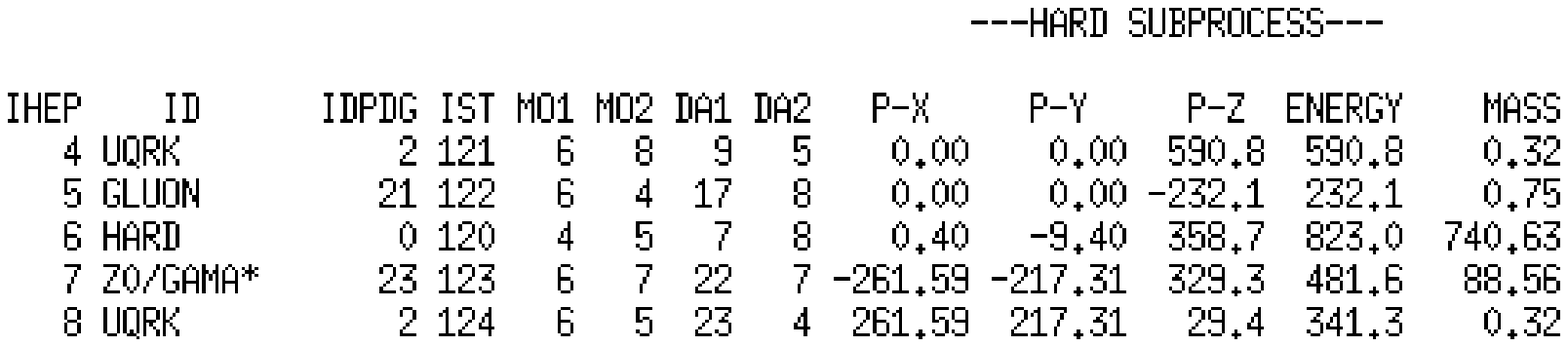}
  \end{minipage}
  \hfill
  \begin{minipage}[t]{0.52\linewidth}
    c)\\
    \includegraphics[width=\tw]{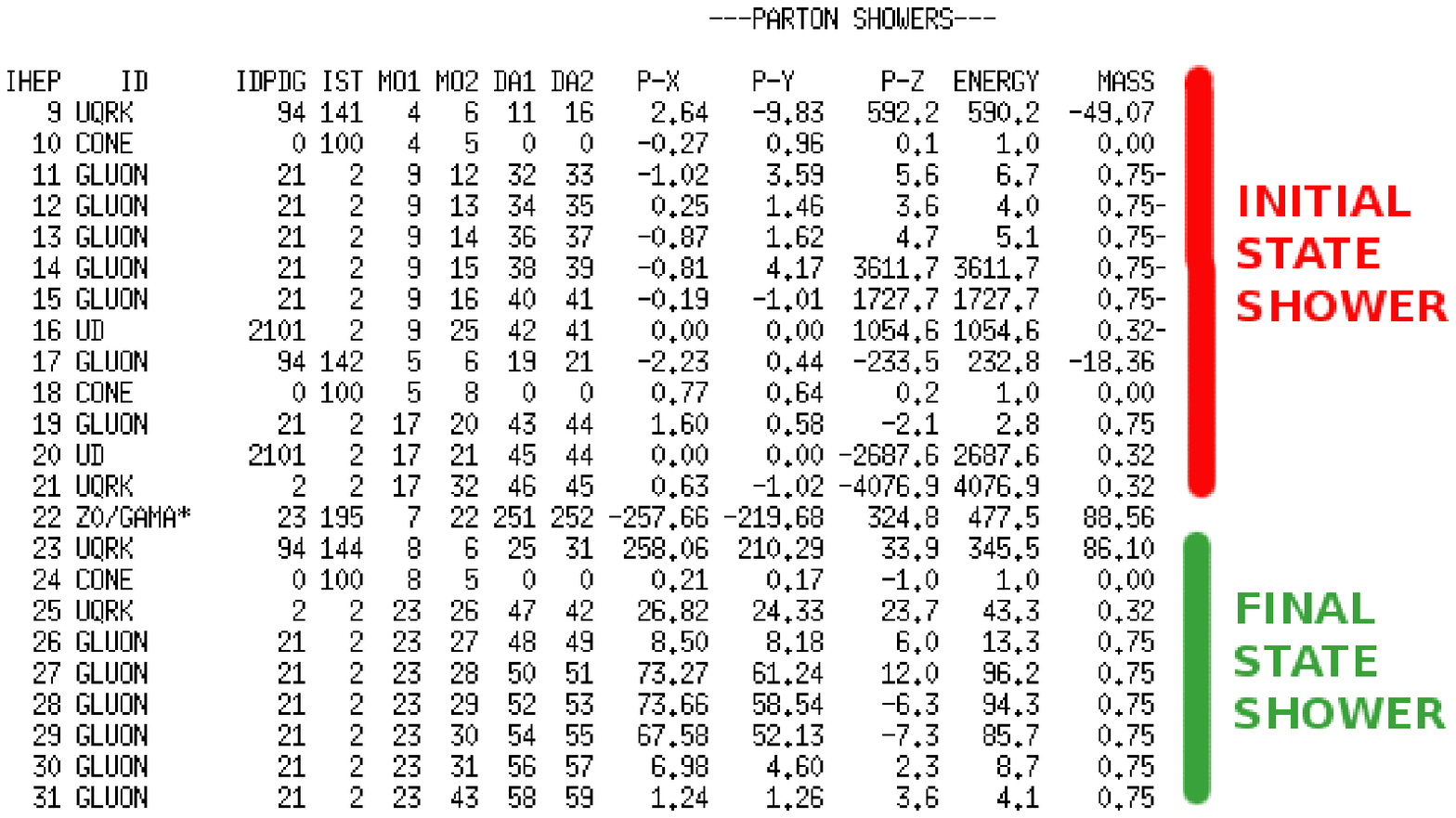}
  \end{minipage}
  \caption{Sequence of steps in the generation of of a $pp\to Z+j$
    event in {\sc Herwig}: (a) specification of the colliding beams and their
    energy, (b) generation of the kinematics and partonic flavour of
    the hard subprocess, $ug \to Zu$, and (c) generation of the
    initial- and final-state parton showers}
  \label{fig:showering}
\end{figure}

The sequence of steps for the generation of a parton-shower event in
$pp$ collisions is illustrated in Fig.~\ref{fig:showering}.

\begin{figure}
  \centering
  \begin{minipage}{0.36\linewidth}
    \includegraphics[width=\tw]{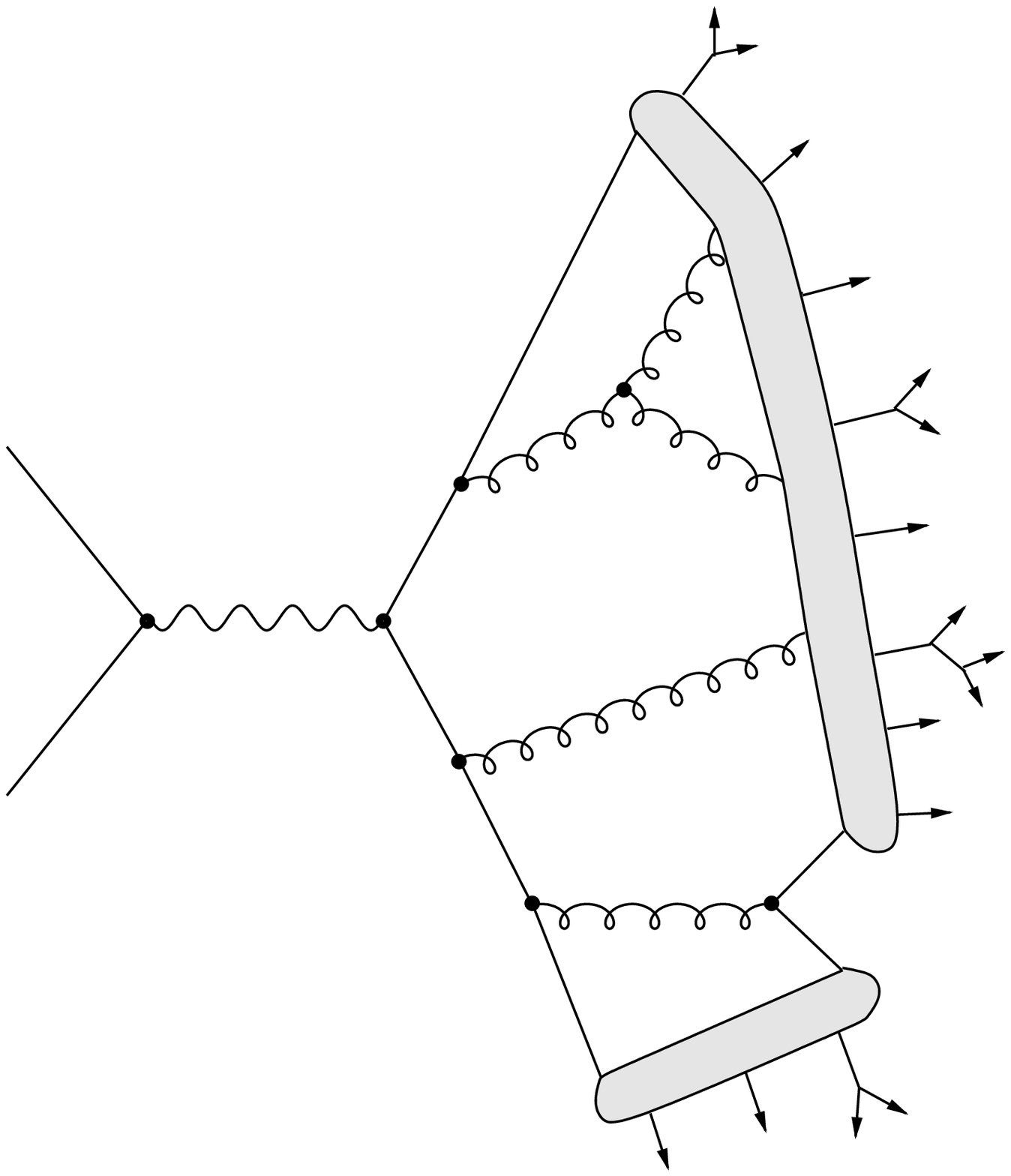}
  \end{minipage}  \qquad
  \begin{minipage}{0.36\linewidth}
    \includegraphics[width=\tw]{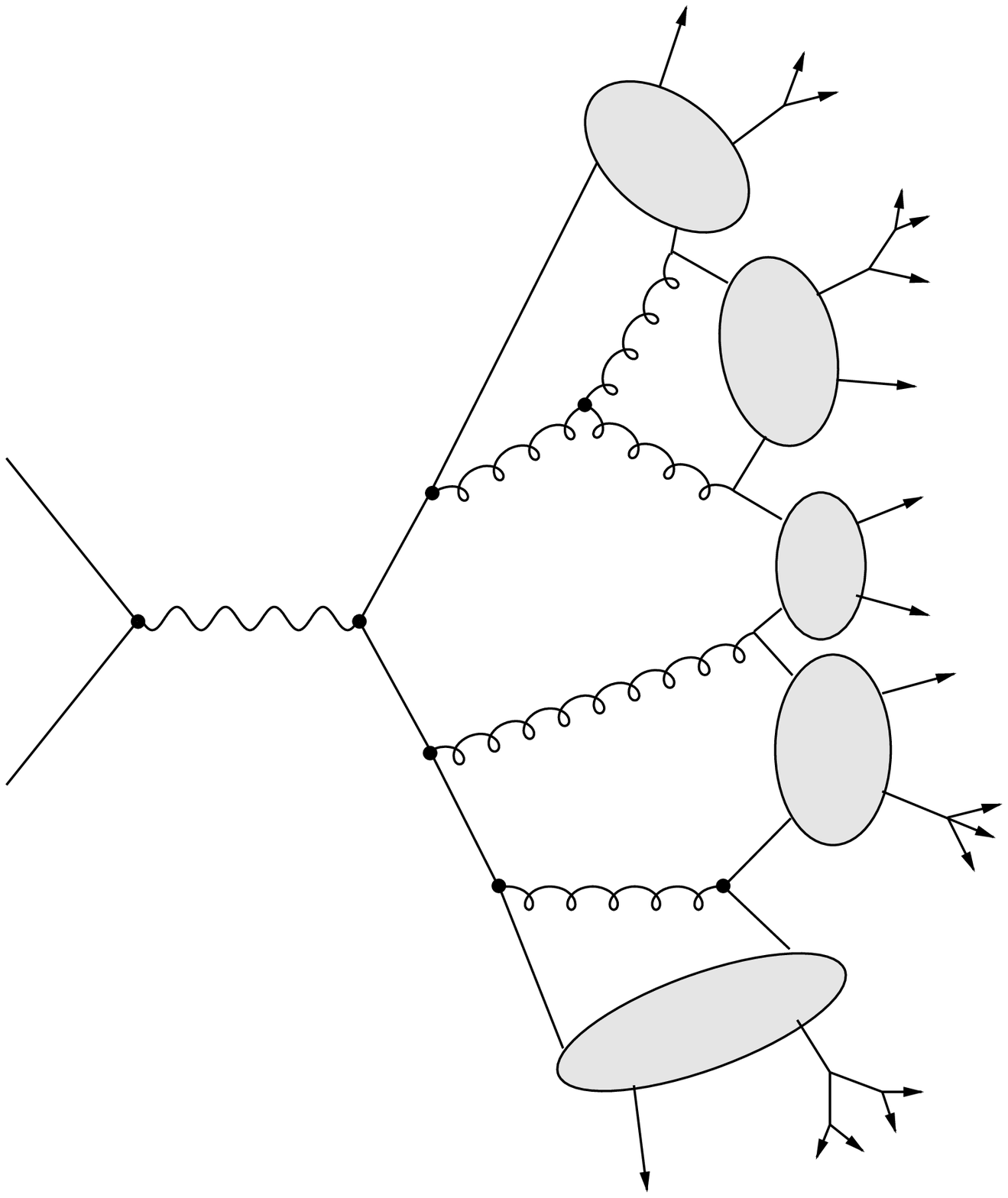}
  \end{minipage}  
  \caption{Illustration of string (left) and cluster (right) fragmentation, taken
    from Ref.~\cite{Ellis:1991qj}}
  \label{fig:hadronization}
\end{figure}

Real events consist not of partons but of hadrons. Since we have no
idea how to calculate the transition between partons and hadrons,
Monte Carlo event generators resort to `hadronization' models. 
One widely-used model involves stretching a colour `string' across
quarks and gluons, and breaking it up into
hadrons~\cite{Andersson:1983ia,Sjostrand:1984ic}. For a discussion of
the implementation of this `Lund' model in the MC program {\sc Pythia},
with further improvements and extensions, Ref.~\cite{Sjostrand:2000wi}
and references therein provide many details.
Another model breaks each gluon into a $q\bar q$ pair and then groups
quarks and anti-quarks into colourless `clusters', which then give
the hadrons. This cluster type hadronization is implemented in the
{\sc Herwig} event generator
\cite{Webber:1983if,Corcella:2000bw,Bahr:2008pv} and recent versions
of {\sc Sherpa}.
Both approaches are illustrated in Fig.~\ref{fig:hadronization}

\begin{figure}
  \centering
  \includegraphics[width=0.4\tw]{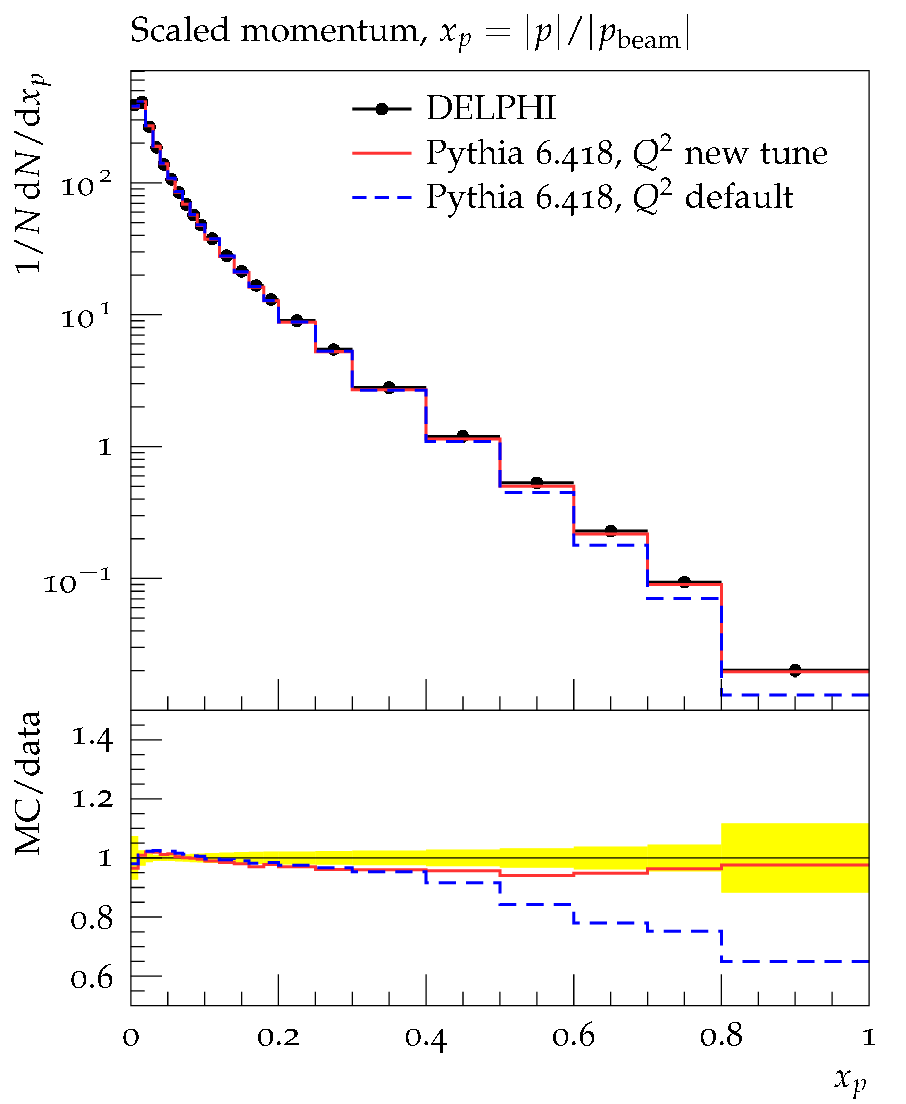}
  \quad
  \includegraphics[width=0.4\tw]{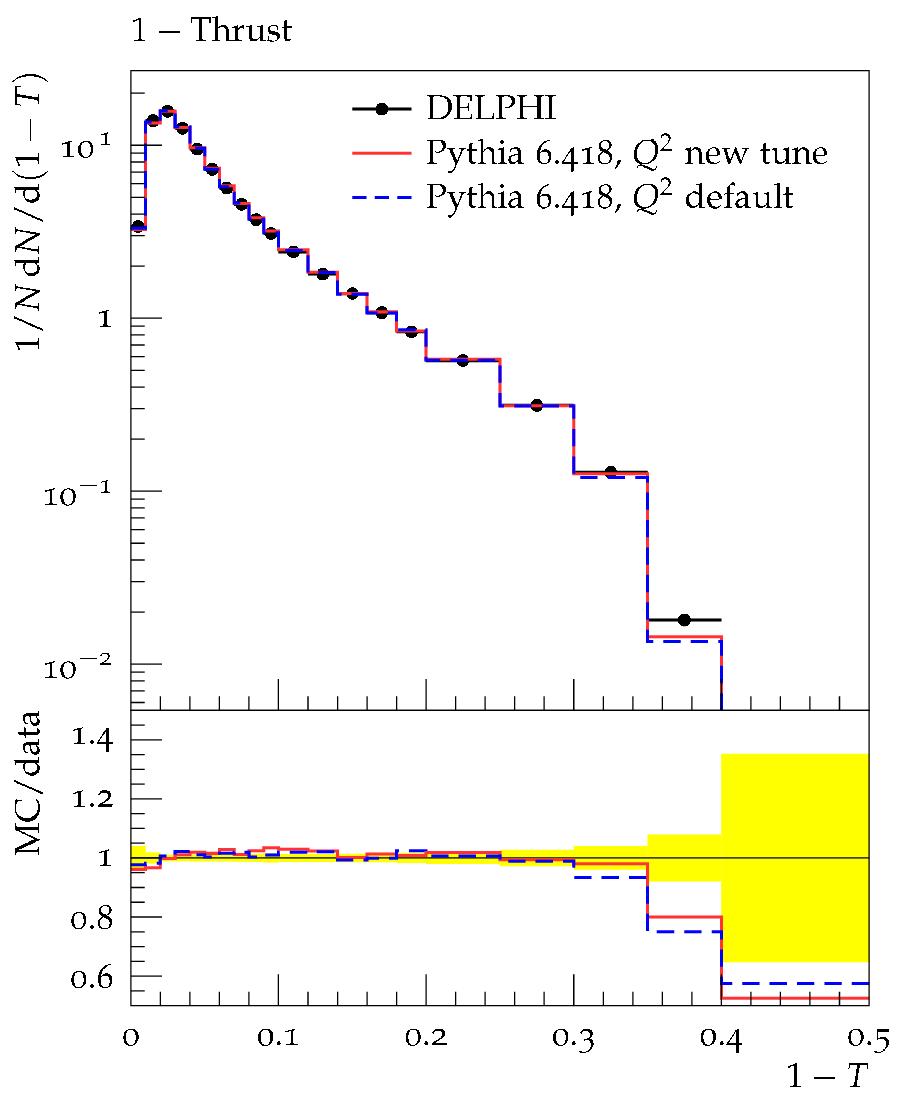}
  \caption{A comparison of  DELPHI $e^+e^-$ data for the particle
    (scaled) momentum distribution (left) and the thrust event shape
    distribution (right) with two tunes of the {\sc Pythia} event
    generator~\cite{Buckley:2009bj}.}
  \label{fig:tuning}
\end{figure}

Hadronization models involve a number of `non-perturbative'
parameters. 
The parton-shower itself involves the non-perturbative cutoff $Q_0$.
These different parameters are usually tuned to data from the LEP
experiments.
The quality of the description of the data that results is illustrated
in Fig.~\ref{fig:tuning}. 

A final point in our brief description of Monte Carlo event generators
concerns the `underlying event' in $pp$ and $\gamma p$ collisions. 
In addition to the hard interaction that is generated by the Monte
Carlo simulation, it is also necessary to account for the interactions
between the incoming proton (or photon) remnants.
This is usually modelled through multiple extra $2\to 2$ scattering,
occurring at a scale of a few GeV, and known as multiple parton
interactions.
This modelling of the underlying event is crucial in order to give an
accurate reproduction of the (quite noisy) energy flow that
accompanies hard scatterings in hadron-collider events.

Our description here of Monte Carlo event generators has been fairly
brief. 
For a more complete discussion, a good starting point is the lectures
notes by Sj\"ostrand \cite{Sjostrand:2006su} from the 2006 School.

\subsection{Comparing fixed-order and parton-shower programs}
\label{sec:MC-v-fixed-order}

Parton-shower Monte Carlo programs do a good job of describing most of
the features of common events, including the hadron-level detail that is
essential for the correct simulation of detector effects on
event reconstruction.
Another nice feature of theirs is that events have equal weight, just
as with real data.

A drawback of parton-shower Monte Carlos is that, because they rely on
the soft and collinear approximation, they do not necessarily generate
the correct pattern of hard large-angle radiation.
This can be important, e.g., if you're simulating backgrounds to
new-physics processes, for which often the rare, hard multi-jet
configurations are of most interest.
In contrast, fixed-order programs do predict these configurations
correctly.

The purpose of this section is to give two examples of comparisons
between parton-shower predictions and fixed-order predictions, in
order to help illustrate their relative strengths. 

\subsubsection{Jet production}

\begin{figure}
  \centering
  \begin{minipage}{0.2\linewidth}
    \includegraphics[width=\tw]{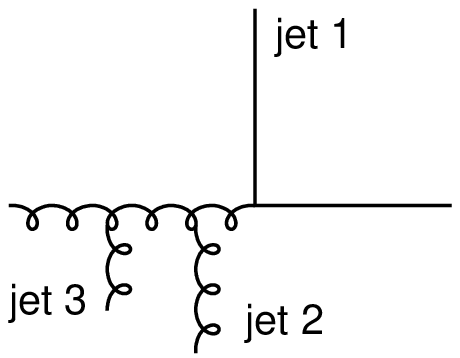}
  \end{minipage}\qquad
  \begin{minipage}{0.45\linewidth}
    \includegraphics[width=\tw]{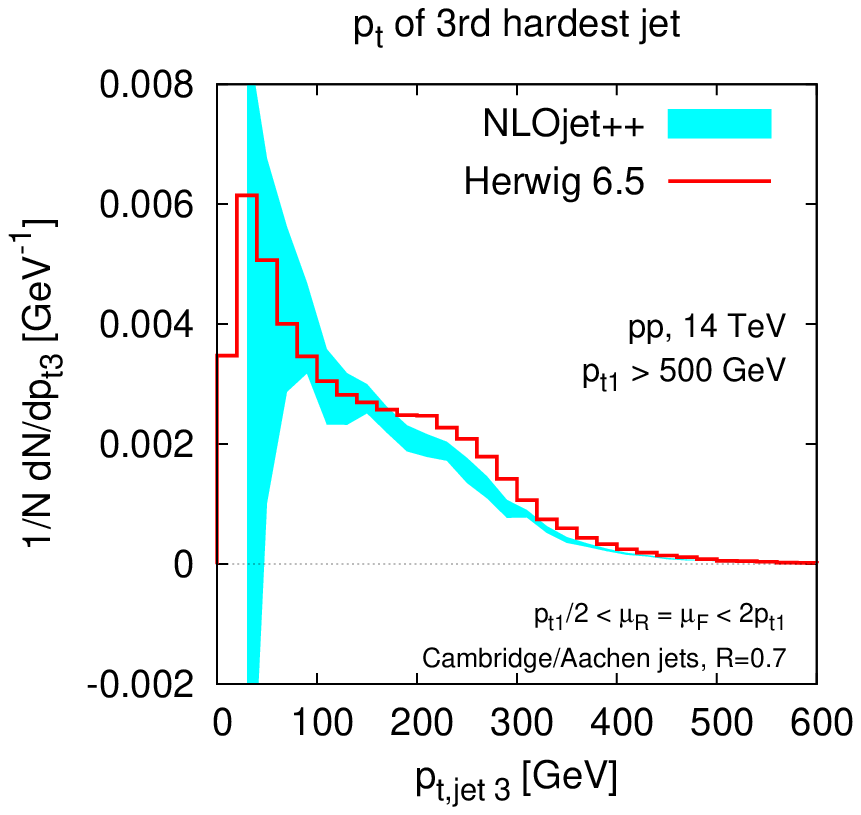}
  \end{minipage}
  \caption{The predicted $p_t$ distribution for the jet with the third largest
    transverse momentum, $p_{t3}$, in $14\TeV$ pp events where the
    hardest jet has transverse 
    momentum $p_{t1} > 500\GeV$. The solid histogram is the result
    from {\sc Herwig}~6.5~\cite{Corcella:2000bw}. The band corresponds to
    the ratio of the NLO 3-jet cross section, differential in $p_{t3}$
    (with the $p_{t1}$ cut), to the NLO cross section for events to
    pass the $p_{t1}$ cut. 
    The width of the band corresponds to the scale uncertainty and the
    results have been obtained with {\sc NLOJet}\texttt{++}
    \cite{Nagy:2003tz}.}
  \label{fig:3rd-jet}
\end{figure}

In plain jet production, parton shower Monte Carlos start from a hard event
that consists of parton--parton scattering diagrams like $qq\to qq$,
$qg \to qg$, etc., and then rely on the showering to generate extra
radiation.
While the showering is only correct in the soft and/or collinear
limit, it does sometimes generate extra hard radiation. 
In Fig.~\ref{fig:3rd-jet} we can see distribution in transverse
momentum of the 3rd hardest jet, in {\sc Herwig} events in which we have
imposed a cut on the hardest jet, $p_{t1} > 500\GeV$.
That is compared to the NLO ({\sc NLOJet}\texttt{++}) prediction for the
same distribution, together with its uncertainty band from scale
variation.

In much of the range, one observes relatively good agreement between
the two distributions: the 20--30$\%$ differences that are visible
around $p_{t3} \equiv p_{t,\text{jet 3}} \sim 250\GeV$ are no larger
than the uncertainties that would be obtained from a LO calculation,
despite the fact that in this region {\sc Herwig} does not even include the
exact LO $2\to 3$ matrix element.
Of course, it is hard to be sure whether the good agreement is
meaningful generally, or instead just a coincidence specific to our
particular choice of observable --- and the only way to be sure is,
for each new observable, to also generate the NLO prediction.

The NLO prediction is not without its limitations though: at low
$p_{t3}$, the uncertainty band on the NLO prediction blows up.
This is a manifestation of large higher-order corrections, which
compromise the convergence of the perturbative series.
They arise because we have a large ratio of scales between $p_{t1}
(\gtrsim 500\GeV$) and $p_{t3}$ (a few tens of GeV). 
Such large scale ratios translate into NLO corrections whose size
relative to the LO contribution go as $\as \ln^2 p_{t1}/p_{t3} \sim 1$
and $\as \ln p_{t1}/p_{t3}$.

\subsubsection{Vector-boson plus production}

\begin{figure}
  \centering
  \begin{minipage}{0.55\linewidth}
    \includegraphics[width=\tw]{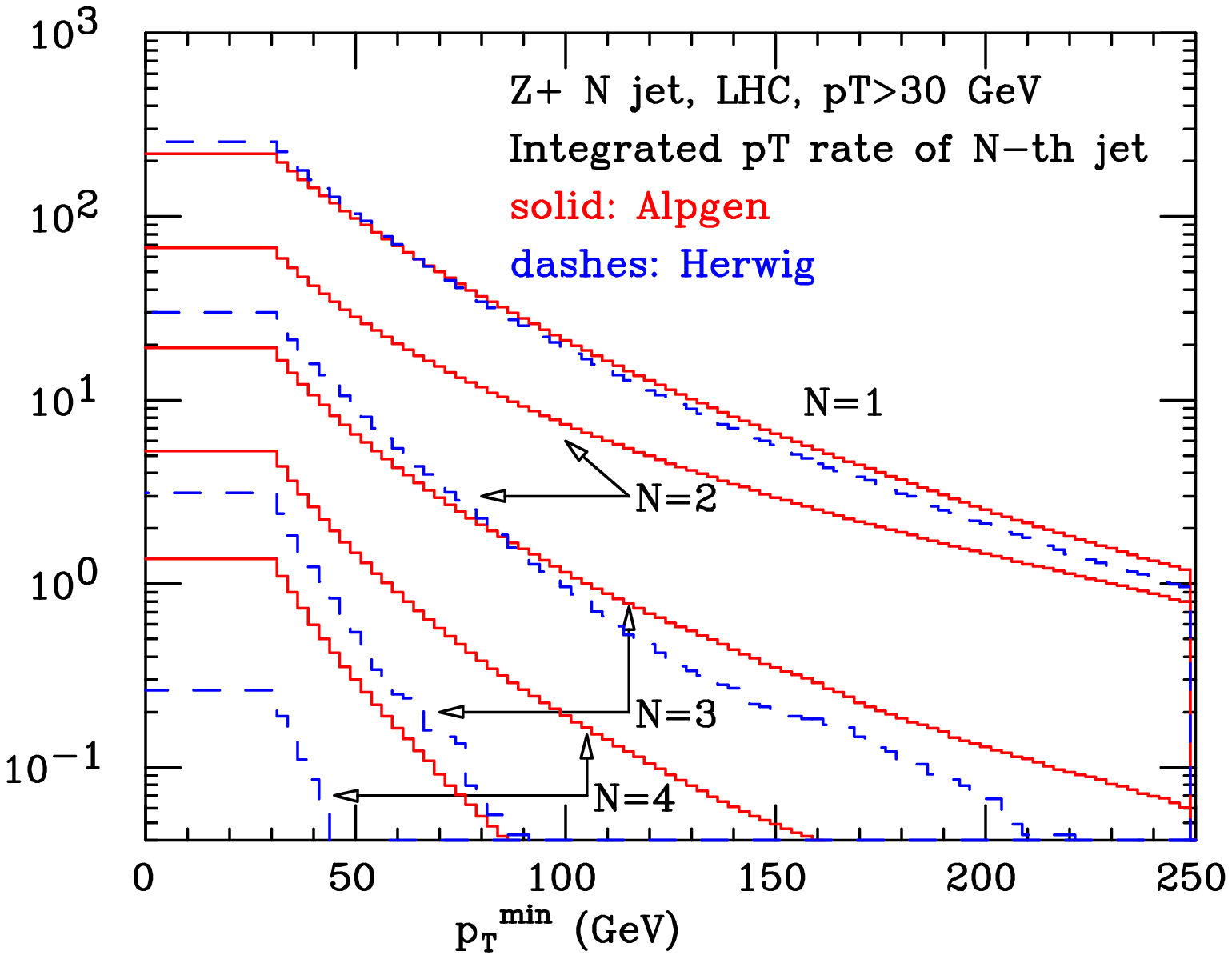}
  \end{minipage}
  \begin{minipage}{0.43\linewidth}
    $Z+j$: 
    \begin{minipage}{0.9\linewidth}
      \centering
      \includegraphics[width=0.48\tw]{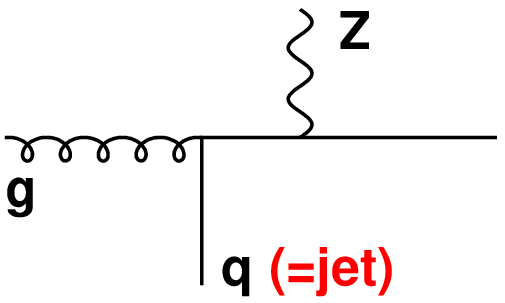}
    \end{minipage}\vspace{2em}

    $Z+2j$: 
    \begin{minipage}{0.9\linewidth}
      \includegraphics[width=0.48\tw]{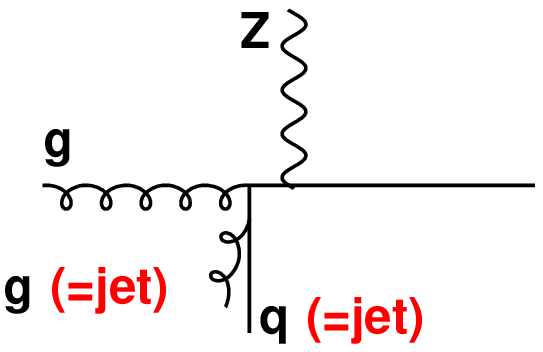}
      \includegraphics[width=0.48\tw]{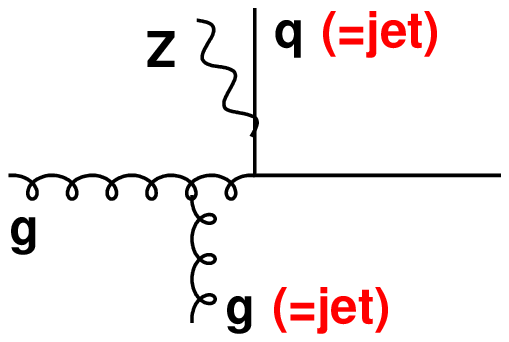}
    \end{minipage}
\end{minipage}
  \caption{Left: the cross section for the $N^\text{th}$ jet to have a
    transverse energy above a given $E_T$, in (14~TeV) LHC events with
    a $Z$-boson, as calculated with {\sc Herwig} and a tree-level (LO)
    prediction from {\sc Alpgen}. Figure taken from Ref.~\cite{Mangano:2008ha}.
    Right: kinematic configurations contributing to $Z+$jet and $Z+2$\,jet
    events.
  }
  \label{fig:mangano-Z-multijet}
\end{figure}

The picture seen above of good agreement between parton shower Monte
Carlo and fixed-order predictions does not always hold.
Events with vector bosons are among those that parton shower programs
have the greatest difficulty reproducing. 
This is illustrated in Fig.~\ref{fig:mangano-Z-multijet} (left), which
shows the `integrated $E_T$ spectrum for the $N^\text{th}$ jet' in
events with a $Z$-boson, i.e., the cross section for the $N^\text{th}$
jet to have a transverse energy above $E_T$.
Results are given both from {\sc Herwig} and from {\sc Alpgen}, which provides an
exact LO (tree-level) prediction for each jet's spectrum.

The distribution for the first  jet is fine: this is by construction,
since {\sc Herwig} (like {\sc Pythia}) includes the full matrix element for
$Z\!+\!\text{parton}$ production.
What is shocking is the result for the second (and higher) jets, for
which the $E_T$ spectra are in complete disagreement between {\sc Herwig}
and {\sc Alpgen}.

At first sight it is mysterious how {\sc Herwig} could be doing such a good
job for pure jet production, Fig.~\ref{fig:3rd-jet}, yet such a poor
job when there's also a $Z$-boson in the event,
Fig.~\ref{fig:mangano-Z-multijet}.
The diagrams in the right-hand part of
Fig.~\ref{fig:mangano-Z-multijet} help explain what's going on. 
{\sc Herwig} generates hard configurations like those in the upper line, labelled
Z+j.
Events with two jets are generated in {\sc Herwig} by emission of a gluon off
(say) the high-$p_t$ quark.
However, there are also events (bottom right) which look like a dijet
event with a $Z$-boson radiated off a quark line. Since at high $p_t$
the $Z$-boson's mass becomes irrelevant in its production, such diagrams
acquire soft and collinear enhancements (just like gluon radiation).
However, today's parton-shower Monte Carlos only include QCD
showering, not electroweak showering and therefore they are never in a
position to start from a dijet event and radiate a $Z$-boson from it.
Therefore they miss a very large part of the cross section. 

This example helps illustrate a general feature of the use of Monte
Carlos: if you are to trust the results, it is crucial that you know
what you have asked the Monte Carlo to generate and whether the observable you
are interested in is truly likely to be dominated by what the Monte Carlo can
generate.

\subsection{Combining fixed-order and parton-shower methods}
\label{sec:MC-plus-fixed-order}

In the above subsections we saw various strengths and weaknesses of
different predictive techniques: NLO codes give predictions with well
controlled normalizations, for a reasonable range of processes, as
long as one isn't faced with observables that involve disparate scales.
Tree-level (LO) predictions can be generated up to quite high
multiplicities for a broad range of processes, though without good
control of the normalization (i.e., often no better than a factor of
two).
And parton shower Monte Carlos provide reliable behaviour in
soft-collinear regions, giving a fully exclusive final state, though
they have normalizations which at best are no better than LO
normalizations and sometimes they do dramatically badly in reproducing
multi-jet structure.

It is natural to ask whether one can develop tools that combine (or
merge) the
advantages of all three. This is an active research topic, and here we
will just outline the ideas behind two well-established merging
tasks: combination of different multiplicity LO matrix elements with
parton showers; and combination of NLO and parton shower predictions.

\subsubsection{Matrix elements with parton showers (MEPS)}
\label{sec:matrix-elements-with-parton-showers}

Suppose you ask for $Z$+jet production as the initial hard process in
{\sc Pythia} or {\sc Herwig}. 
As we saw above, these programs contain the correct matrix element
(ME) for $Z$+parton production, but do a very bad job of estimating
$Z$+2\,jet production.

\begin{figure}
  \centering
  \includegraphics[width=0.7\tw]{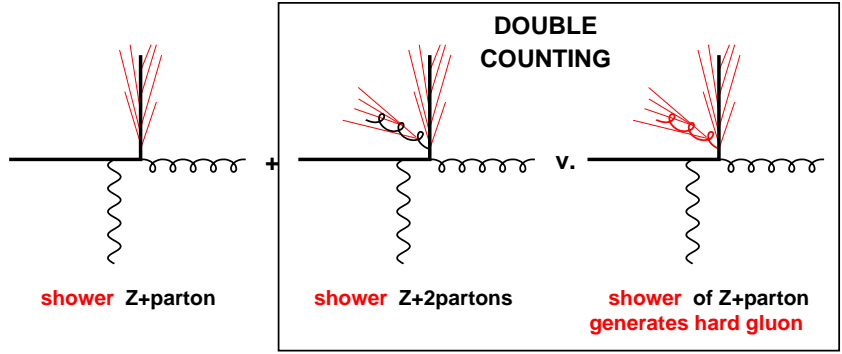}
  \caption{Illustration of the double-counting issues that can arise
    if one attempts to shower $Z$+parton and $Z$+2-parton events}
  \label{fig:me+ps-double-counting}
\end{figure}

One naive solution to this problem would be to generate $Z$+2-parton events
with {\sc Alpgen, Madgraph}, or some other preferred LO ME tool and then
ask {\sc Herwig} or {\sc Pythia} to shower those configurations. 
However, if one showers both $Z$+parton and $Z$+2-parton events, then one
is faced with a double-counting issue, as illustrated in
Fig.~\ref{fig:me+ps-double-counting}. 
In some events (left) the showering off the $Z$+parton configuration
just leads to soft and collinear emissions. 
Similarly off the $Z$+2-parton events (middle). 
However, sometimes (right) the showering off the $Z$+parton configuration leads
to the production of a relatively hard, large-angle gluon that is in
the same phase-space region that is already covered in the $Z$+2-parton
event sample. 

Two main methods exist to avoid this double counting: CKKW
matching~\cite{Catani:2001cc} and MLM
matching~\cite{Alwall:2007fs}. The latter, named after its inventor,
M.~L.~Mangano, is the one we will describe here (it is the
simpler of the two).
Let's examine the basics of how it proceeds:
\begin{itemize}
\item Introduce a (dimensionful) transverse momentum cutoff scale
  $Q_{ME}$ and a (dimensionless) angular cutoff scale $R_{ME}$ for
  matrix element generation.

\item Generate tree-level events for $Z$+1-parton, $Z$+2-partons, \ldots up
  to $Z$+$N$-partons, where all partons must have $p_t > Q_{ME}$ and be
  separated from other partons by an angle greater than $R_{ME}$ (we
  will discuss the definition of this `angle' later in Section~\ref{sec:jets}).
  The numbers of events that one generates in the different samples
  are in proportion to their cross sections with these cuts.

\item For each tree-level event from these samples, say one with $n$
  partons, shower it with your favourite parton-shower program.

\item Apply a jet algorithm to the showered event (choose the
  algorithm's angular reach $R$ to be $\gtrsim R_{ME}$) and identify
  all jets with $p_t > Q_{merge}$, where the merging scale $Q_{merge}$
  is to be taken $\gtrsim Q_{ME}$. 
  
\item If each jet  corresponds to one of the partons (i.e., is nearby
  in angle) and there are no extra jets above scale $Q_{merge}$, then
  accept the event. (For the sample with $n=N$, the condition is that
  there should be no extra jets with $p_{t} > p_{tN}$.)

\item Otherwise, reject the event. 
\end{itemize}
\begin{figure}
  \centering
  \includegraphics[width=0.7\tw]{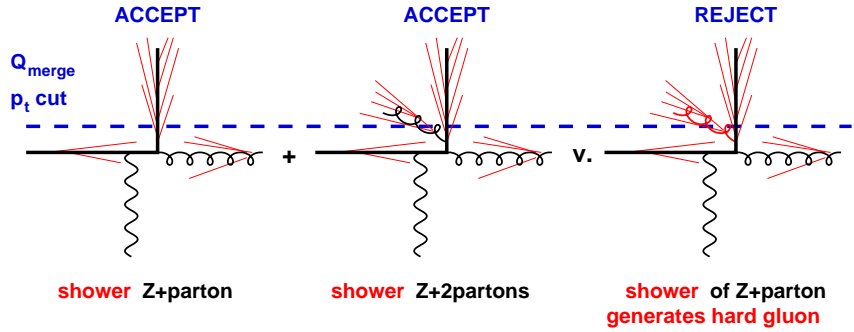}
  \caption{Illustration of the application of the MLM matching
    procedure to the events of Fig.~\ref{fig:me+ps-double-counting},
    with the $Q_{merge}$ $p_t$ cutoff represented by the dashed
    line}
  \label{fig:mlm-in-action}
\end{figure}%
The action of the MLM procedure on the events of
Fig.~\ref{fig:me+ps-double-counting} is illustrated in
Fig.~\ref{fig:mlm-in-action}, showing which events would be accepted
and which ones rejected. 
One immediately sees how the double-counting issue disappears: in the
rightmost event, the showering of a $Z$+parton event leads to an extra jet
above $Q_{merge}$; since this event now has more jets above
$Q_{merge}$ than it had partons, it is rejected. 
In contrast the middle event, which also has two jets above
$Q_{merge}$, was generated from a $Z$+2-parton event and is accepted. 
So is the leftmost event with only one jet, starting from a
$Z$+1-parton event.

By providing a remedy for the double-counting issue, ensuring that the
hard jets always come just from the matrix element, the MLM procedure
also ensures that hard jets above $Q_{merge}$ have distributions given
by the tree-level matrix-element calculations.

The rejection of extra jets also plays another important role:
when there are big differences in scales between the jets and
$Q_{merge}$ (or between different jets), the Monte Carlo showering
would want to `fill up' that phase space with emissions.  However,
whenever it does so, the event gets rejected by the matching
procedure.
As long as the Monte Carlo is carrying out a reasonable showering of
these multi-parton events,\footnote{This can depend on subtleties of
  how the Monte Carlo showers multi-parton events and the
  communication of information on colour flows between the fixed-order
  program and the Monte Carlo.} then the procedure is equivalent to the
introduction of a Sudakov form factor encoding the probability of not
emitting radiation.

The above `MLM' merging of matrix-elements and parton showers (MEPS)
is the main procedure available with {\sc Alpgen}, for use with both {\sc Herwig}
and {\sc Pythia}.
It is also provided (in a variant form) within {\sc Madgraph}.
The {\sc Sherpa} Monte Carlo also has its own matrix-element generator(s)
and provides `CKKW' MEPS matching~\cite{Catani:2001cc}, which
instead of the veto steps of MLM matching, uses an analytical
calculation of the Sudakov form factors.
These and other matrix-element/parton-shower merging schemes are
discussed in detail in Ref.~\cite{Alwall:2007fs}. They all share the
feature of a matching scale to separate the region under the
`responsibility' of matrix elements and that delegated to the parton
shower.
In all cases physics predictions well above the matching scale should
be independent of the scale. Additionally, distributions at scales
around the matching scale should be reasonably smooth, as long as the
matching scale has been chosen in a region where the parton shower can be
expected to give a reasonable description of emission (for caveats,
see Ref.~\cite{Lavesson:2007uu}; for a method that avoids the need for
a matching scale, see Ref.~\cite{Giele:2007di}).

\begin{figure}
  \centering
  \includegraphics[width=0.7\tw]{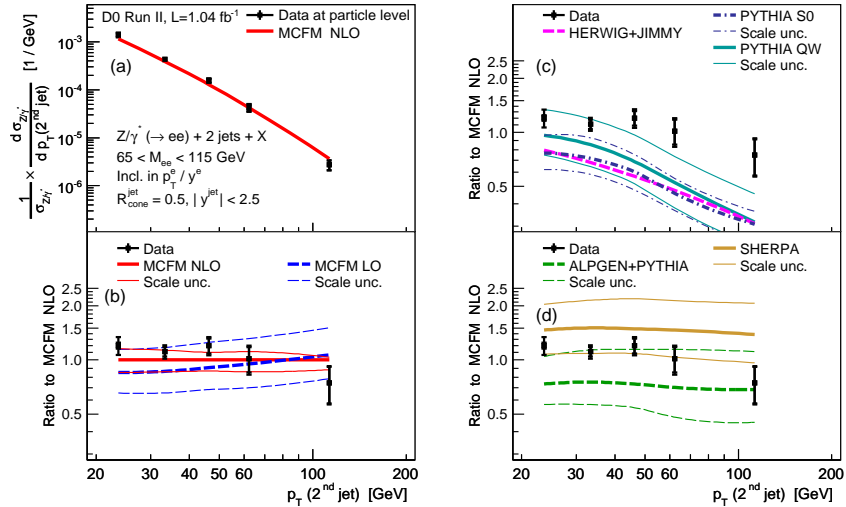}
  \caption{Cross section for $Z$+2-jet events at the Tevatron,
    differential in the transverse momentum of the second jet, as
    measured by D\O~\cite{Abazov:2009av}. Comparisons are shown to LO
    and NLO predictions (from MCFM), parton-shower predictions
    ({\sc Pythia, Herwig}) and merged matrix-element + parton-shower
    predictions ({\sc Alpgen+Pythia, Sherpa}). }
  \label{fig:meps-in-action}
\end{figure}

MEPS predictions (as well as other predictive methods) are compared to
experimental results for $Z$+2-jet production in
Fig.~\ref{fig:meps-in-action} (bottom right).
The MEPS results show good agreement for the shape of the observable
in question, the $p_t$ distribution of the second jet, and they are much
more successful than plain parton-shower predictions.
Since their normalizations are based on LO calculations, they do,
however, suffer from substantial scale dependence, which is much
larger than the scale uncertainty one would obtain at NLO (bottom
left).

As a result of their considerable success in describing the shapes of
experimental distributions, MEPS predictions have become one of the
main tools for a broad range of Tevatron and LHC analyses, especially
those involving complex final states.

\subsubsection{Parton showers and NLO accuracy}
\label{sec:parton-showers-nlo}

We've seen how to obtain a parton shower structure with LO accuracy
for a range of different jet multiplicities.
However, given the large uncertainty on the normalizations of the
predictions, there would be considerable advantages to obtaining NLO
accuracy.
One might think that since we've had NLO predictions and parton shower
predictions for a couple of decades, such a task should not be too
hard. 
The main difficulty comes from the fact that NLO predictions involve
divergent event weights (cf. Fig.~\ref{fig:nlojet-weights}), which
aren't even positive definite.
Two approaches are in use to get around this problem, the
MC@NLO \cite{Frixione:2002ik} and {\sc Powheg}~\cite{Nason:2004rx} methods.

The idea behind the MC@NLO approach is to `expand' the Monte Carlo
parton shower to first order in $\as$.
I.e., the Monte Carlo's parton showers already contain some (partially
wrong) estimate of the true NLO corrections and the aim is to figure
out what that estimate is.
This requires a deep understanding of the Monte Carlo program.
As a next step, one calculates the difference between the true NLO
contributions and the partial ones included in the Monte Carlo. One of the keys
to the MC@NLO method is that as long as the Monte Carlo gives the correct
pattern of soft and collinear splitting (which it is supposed to),
then the differences between true NLO and Monte Carlo partial NLO should be
\emph{finite}.
Then one can generate partonic configurations with phase-space
distributions proportional to those finite differences and shower
them. 

Symbolically, if we imagine a problem with one phase-space variable,
say energy $E$, then we can write the `expansion' of the Monte Carlo
cross section as
\begin{equation}
  \label{eq:mc-expansion}
  \sigma^{MC} = 1\times \delta(E) + \as \sigma_{1R}^{MC}(E)
  + \as \sigma_{1V}^{MC} \delta(E) + \order{\as^2}
\end{equation}
where $\sigma_{1R}^{MC}(E)$ is the coefficient of $\as$ for real
emission of a gluon with energy $E$ in the Monte Carlo and
$\sigma_{1V}^{MC}$ is the (divergent) coefficient of the virtual
corrections. 
The MC@NLO prediction is then given by 
\begin{equation}
  \label{eq:mc@nlo}
  \text{MC@NLO} = \text{MC} \times \left(1 + \as(\sigma_{1V} -
    \sigma_{1V}^{MC}) + \as\int dE (\sigma_{1R}(E) -
    \sigma_{1R}^{MC}(E)) \right)\,,
\end{equation}
where $\sigma_{1R}(E)$ and $ \sigma_{1V}$ are the true NLO real and
virtual coefficients.
Each term in (small) brackets in Eq.~(\ref{eq:mc@nlo}) should
separately be finite and corresponds to a given topology of event to
be showered: a LO topology for the ``$1$''  and $(\sigma_{1V} -
\sigma_{1V}^{MC})$ terms, and a NLO real topology for the
$(\sigma_{1R}(E) - \sigma_{1R}^{MC}(E))$ term.
(A more complete and clearer `toy' explanation is given in the
original MC@NLO paper\cite{Frixione:2002ik}, which makes for very
instructive reading.)

The MC@NLO approach has the practical characteristic that all event weights are
$\pm 1$. 
Quite a range of processes are available in the MC@NLO approach for
{\sc Herwig}, including the production of a Higgs boson, single and double
vector-bosons, a heavy-quark pair, various single-top processes, and
$H+W$ and $H+Z$.
The characteristic of these processes is that they are nearly all free
of light jets at LO (except for one of the single-top processes),
because this simplifies the set of divergences that need to be dealt
with. 
Very recently one {\sc Pythia}~\cite{Torrielli:2010aw} and a couple of
{\sc Herwig}{\tt ++} processes (Ref.~\cite{LatundeDada:2009rr} and
references therein) were also interfaced within the MC@NLO
approach.\footnote{In the time since the original version of these
  lectures were written up, many more processes have been implemented
  for {\sc Herwig}{\tt ++}~\cite{Frixione:2010ra}.}

An alternative to MC@NLO is {\sc Powheg}~\cite{Nason:2004rx}.
It aims to avoid the (small fraction of) negative weights that are
present in MC@NLO and also seeks to be less tightly interconnected
with a specific Monte Carlo program. 
The principle of the {\sc Powheg} approach is to write a simplified Monte
Carlo that generates just one emission beyond LO. This single-emission
Monte Carlo is designed in such a way as to be sufficient to give the correct NLO
result. 
It essentially works by introducing a Sudakov form factor such as
Eq.~(\ref{eq:sudakov-exponentiated}) in which the contents of the
square bracket are replaced by the integral over the exact real
radiation probability above $k_t$ (plus a constant term that accounts
for the finite part of the 1-loop correction).
Then emissions with transverse momenta below the scale of the first
emission are left to the default Monte Carlo program, for example
{\sc Herwig} or {\sc Pythia} (implementing a transverse momentum veto to ensure
that nothing is generated above the scale of the first, {\sc Powheg},
emission).
The range of processes available within {\sc Powheg} is gradually catching
up with that for MC@NLO, and it is hoped that this will be helped by the
availability of systematic tools to help with the development of new
processes~\cite{Alioli:2010xd}.

\begin{figure}
  \centering
  \includegraphics[width=0.6\tw]{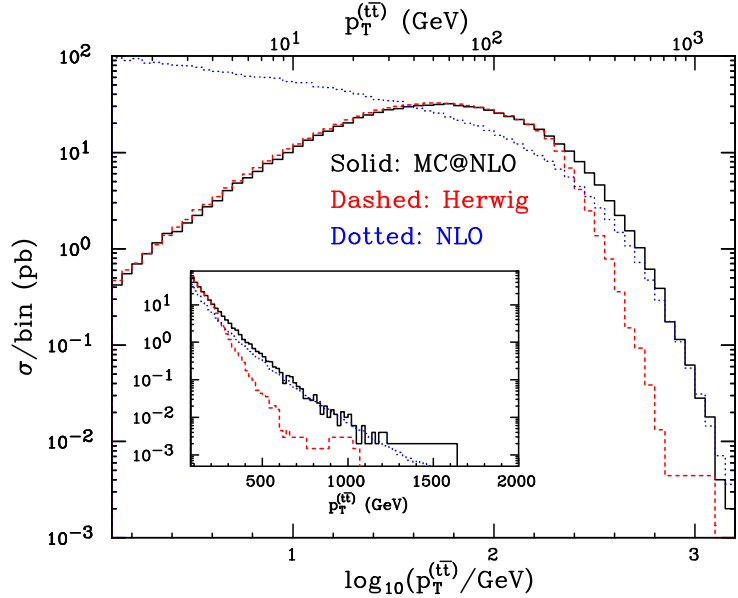}
  \caption{The transverse-momentum distribution of $t\bar t$ pairs in
    the MC@NLO approach, compared to the plain
    {\sc Herwig} result (rescaled by the
    $\sigma_\text{NLO}/\sigma_\text{LO}$ $K$-factor) and to the NLO
    calculation.  Shown for a $14\TeV$ LHC. Figure taken from
    Ref.~\cite{Frixione:2003ei}. }
  \label{fig:mcatnlo-result}
\end{figure}

An illustration of a result with MC@NLO is given in
Fig.~\ref{fig:mcatnlo-result} for the transverse-momentum distribution
of a $t\bar t$ pair. It illustrates how MC@NLO reproduces the {\sc Herwig}
shape in the low-$p_t$ region (with NLO normalization), the NLO
distribution at high $p_t$, and that neither {\sc Herwig} nor plain NLO are
able to properly describe both regions.

\subsection{Summary}
\label{sec:summary-of-predictions}

In this section, we have seen quite a range of different predictive
methods for QCD at hadron colliders.
%
%
Two big classes of predictive methods exist: partonic fixed-order
calculations, which have well controlled accuracy, but wildly
fluctuating positive and negative event weights; and Monte Carlo
parton shower tools, which give a much more complete description of
events (and uniform event weights).

Development is active on both sets of tools individually.
On one hand we've mentioned the challenge of having a broader range of
processes known at NLO and a handful at NNLO.
And though we have not really touched on it, there is also a very
active programme to develop parton shower Monte Carlos in C++ as
replacements for venerable but ageing Fortran codes like {\sc Pythia} 6.4
and {\sc Herwig~6.5}.

In addition there are methods with the advantages of both fixed-order
and parton-shower programs.
It is widespread nowadays to merge different LO `tree-level'
predictions (e.g., $Z$+parton, $Z$+2-partons, $Z$+3-partons, etc.) together
with parton showers.
And for simple processes, those with at most one light parton in the
final state, it is possible to combine NLO accuracy with the full
parton-shower description.
Ultimately the hope is to be able to combine $Z$+jet, $Z$+2-jets, $Z$+3-jets all
at NLO accuracy also merging them with parton showers, so as to obtain
accurate descriptions for the whole range of processes that are
relevant at the LHC, both as backgrounds and as signals of new particles
and new physics.

\section{Jets}
\label{sec:jets}

The concept of a jet has already arisen in various contexts, so in
this final section we will examine jet-related ideas in more detail. 

\begin{figure}
  \centering
  \includegraphics[width=0.31\tw]{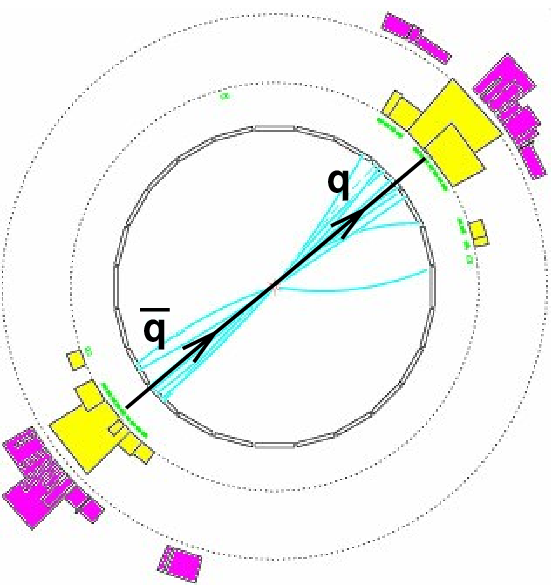}\;
  \includegraphics[width=0.32\tw]{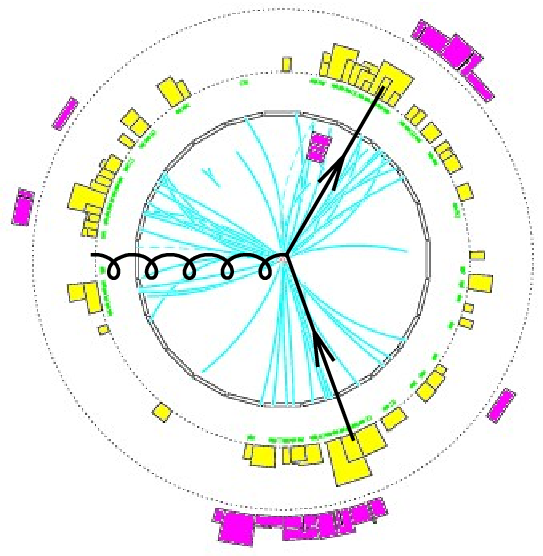}\;
  \includegraphics[width=0.32\tw]{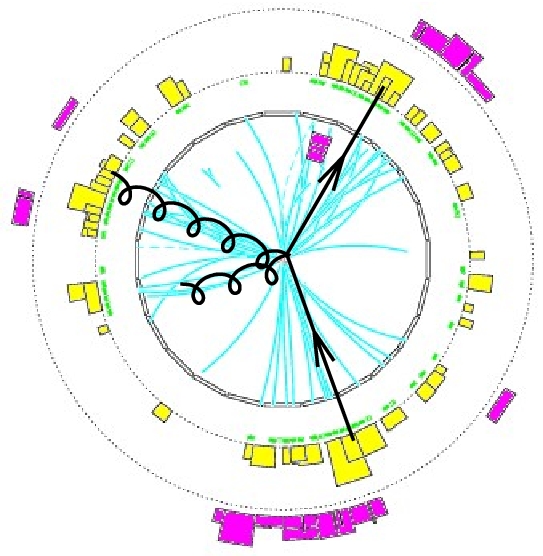}
  \caption{Left: an $e^+e^-$ event that can be interpreted as having a
    2-jet, $q\bar q$-like structure; middle: an event that can be
    interpreted as having a 3-jet, $q\bar q g$, structure; right: the
    same event reinterpreted as having a 4-jet structure, $q\bar q
    gg$.}
  \label{fig:jet-ambiguities}
\end{figure}

Consider the three events of Fig.~\ref{fig:jet-ambiguities}. In the
left-hand one, one interpretation is that we're seeing an $e^+e^- \to
q\bar q$ event, in which there has been soft and collinear showering
followed by a transition to hadrons. 
This is a classic picture of a `2-jet' event. 
The middle event is more complex: energy flow is not limited to two
cones. 
One interpretation of the event is that a $q\bar q$ pair has emitted a
hard gluon $g$, and all three have undergone soft and collinear
showering. 
However, the same event can also be interpreted (right) as a $q\bar q g
g$ event, with further soft and collinear showering.
Deciding between these two interpretations means choosing just how
hard and separated in angle an emission has to be in order for it to
be considered a separate jet (cf.\ the angular and energy parameters,
$\delta$ and $\epsilon$, in our discussion of the 2-jet cross section
in Section~\ref{sec:ir-safe-obs}).

\begin{figure}
  \centering
  \includegraphics[width=0.65\tw]{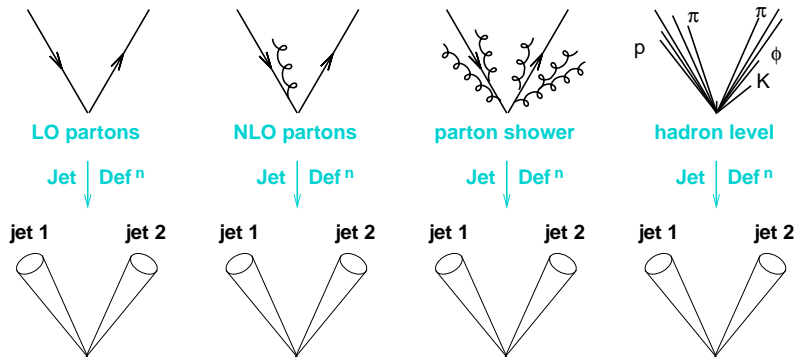}
  \caption{The application of a jet definition to a variety of events
    that differ just through soft/collinear branching (and
    hadronization), should give identical jets in all cases.}
  \label{fig:jet-projections}
\end{figure}

In making this choice, it would be highly painful to visually inspect
each of the $\order{10^9}$ events written to disk every year at the
LHC.
Instead one uses a set of rules, a `jet definition', by which a
computer can take a list of particle momenta for an event (be they
quark and gluons, or hadrons, or even calorimeter deposits), and
return a list of jets.
If one modifies an event just through soft and collinear emission,
then the set of jets should not change, i.e., the result of applying
the jet definition should be insensitive to the most common effects of
showering and hadronization, as illustrated in
Fig.~\ref{fig:jet-projections}. 

Jets are central to collider physics: both theory and experimental
results are often presented in terms of jet cross sections, and thus
jets provide the meeting point between the two. 
As we saw in Section~\ref{sec:matrix-elements-with-parton-showers},
jets are also used to assemble together different kinds of theory
predictions.
And jets are an input to almost all physics analyses: to new physics
searches (since new particles may decay to quarks or gluons, giving
jets), in Higgs searches, top physics, Monte Carlo validation, fits of
PDFs, etc. 

\subsection{Jet definitions}
\label{sec:jet-definitions}

The construction of a jet involves different considerations:
\begin{itemize}
\item Which particles are grouped together into a common jet? The set
  of rules that one follows for deciding this is usually known as a
  jet algorithm, and it comes with parameters that govern its exact
  behaviour. A common parameter is $R$ which determines the angular
  reach of the jet algorithm.

\item How do you combine the momenta of particles inside a jet? One
  needs to specify a `recombination scheme'. The most common is to
  simply add the 4-vectors of the particles (the `E-scheme'). 
  This gives jets that are massive (so jets cannot be thought of as a
  direct stand-in for partons, which are massless).
\end{itemize}
Taken together, the algorithm, its parameters and the recombination
scheme specify a `jet definition'. 

Two broad classes of jet definition are in common use: cone
algorithms, which take a top-down approach, and sequential
recombination algorithms, based on a bottom-up approach.
Below we'll give a brief discussion of each of kind of algorithm,
referring the reader to Ref.~\cite{Salam:2009jx} for a more complete
description of all the variants that exist.

\subsection{Cone algorithms}
\label{sec:cone-algorithms}

There are many types of cone algorithm, but all rely on the idea that
soft and collinear branching doesn't modify the basic direction of
energy flow. 

One of the simpler cone algorithms (we'll call it IC-PR, for iterative
cone with progressive removal of particles) is that used by the CMS
experiment during much of their preparation for LHC running.
One first sorts all particles according to their transverse momentum,
and identifies the one with largest transverse momentum.
This is referred to as a seed particle, $s$. 
One draws a cone of radius $R$ around the seed --- in hadron-collider
variables this means identifying all particles with $\Delta R_{is}^2 =
(y_i - y_s)^2 + (\phi_i - \phi_s)^2 < R^2$, where $y_i = \frac12 \ln
\frac{E_i + p_{zi}}{E_i + p_{zi}}$ is the rapidity of particle $i$,
$\phi_i$ its azimuth, and $y_s$ and $\phi_s$ the rapidity and azimuth
of the seed.
One then identifies the direction of the sum of the momenta of those
particles.
If it doesn't coincides with the seed direction then one uses that sum
as a new seed direction, and iterates until the sum of the cone
contents coincides with the previous seed.
This is referred to as finding a stable cone.
The particles inside that stable cone make a jet, and they're removed
from the list of particles in the event.
The procedure is then repeated on the particles that remain, removing
particles each time one finds  a stable cone ($\to$ jet), until no
particles remain and one has the complete set of jets.
Of these jets one retains only those above some transverse-momentum
threshold $p_{t,\text{min}}$.

\begin{figure}
  \centering
  \includegraphics[width=0.65\tw]{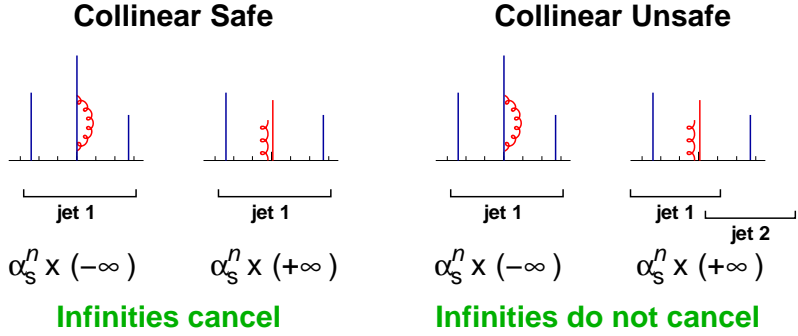}
  \caption{Sample events illustrating the result of applying collinear safe and
    unsafe jet algorithms. The height of a given line corresponds to the particle's
    transverse momentum, its horizontal position to its rapidity
    ($\phi=0$ for all particles here).
    Left: expectations for the behaviour of a collinear-safe
    jet algorithm, where the jets found by the algorithm should be
    independent on the collinear splitting of the hardest particles. 
    Right: in a collinear unsafe algorithm such as the IC-PR type CMS
    cone, the splitting of the central hard particle causes the
    leftmost particle to become the hardest in the event, leading to a
    two-jet rather than a one-jet event. }
  \label{fig:coll-unsafety}
\end{figure}

There is one major drawback to the above procedure: the use of the
particles' $p_t$'s to decide which one to take as the first seed.
This is problematic, because particle $p_t$'s are not collinear safe
quantities.
As illustrated in the two right-hand events of
Fig.~\ref{fig:coll-unsafety}, in an IC-PR algorithm, if the hardest
particle undergoes a collinear splitting then this can result in
another particle in the event becoming the `new' hardest particle,
giving a different set of final jets as compared to events without the
splitting.
Thus in the example of Fig.~\ref{fig:coll-unsafety} there is a
divergent (real, positive) contribution to the 2-jet cross section and
a separate divergent (1-loop virtual, negative) contribution to the
1-jet cross section.
In contrast, for a collinear-safe algorithm (two leftmost events), the
collinear-splitting of the hardest particle does not change the set of
final jets. Then the real and virtual divergences both contribute to
the 1-jet cross section and so cancel each other.

Collinear unsafety means that certain cross sections cannot be
calculated at NLO (or sometimes NNLO) --- one will only obtain
nonsensical infinite answers.
Furthermore, even if one is working at some low perturbative order
which is not divergent (e.g., LO), the fact that higher orders diverge
means that the convergence of the whole perturbative series becomes
questionable, compromising the usefulness even of the low orders.

Over the past two decades there has been significant discussion of
such problems. 
There are many other variants of cone algorithm, and nearly all suffer
from problems either of collinear safety, or infrared safety. 
One class that has been widely used at the Tevatron avoids the
ordering of initial seeds, and instead obtains stable cones using all
possible particles as seeds:
stable cones are not immediately converted into jets, but instead,
once one has the list of the stable cones found by iterating from all
possible seeds one then uses a `split--merge' procedure to decide
how particles that appear in multiple stable cones should be
unambiguously assigned to a single jet.

\begin{figure}
  \centering
  \includegraphics[width=0.7\tw]{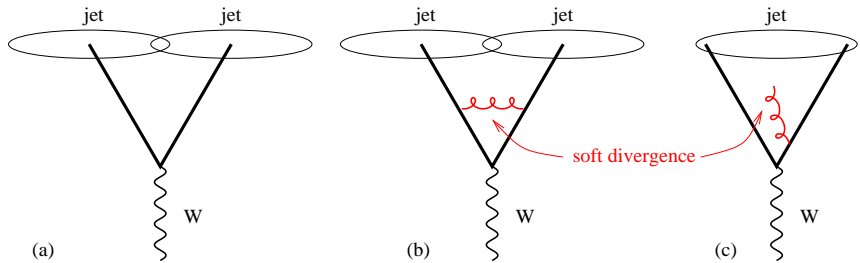}
  \caption{Configurations illustrating IR unsafety of iterative cone
    algorithms with a split--merge procedure,
    in events with a $W$ and two hard partons. The addition
    of a soft gluon converts the event from having two jets to just
    one jet. 
    In contrast to Fig.~\ref{fig:coll-unsafety}, here the explicit
    angular structure is shown (rather than $p_t$ as a function of
    rapidity).}
  \label{fig:IR-unsafety}
\end{figure}

This procedure avoids the collinear-safety issue (the order of particles'
$p_t$'s no longer matters), however, it turns out that it instead
introduces an infrared-safety issue: adding an extra soft particle
creates a new seed, which can lead to an extra stable cone being
found, which feeds through the split--merge procedure, altering the
final set of (hard) jets.
This tends to happen when two hard particles are separated by a
distance $\Delta R$ that satisfies $R < \Delta R < 2R$ (so that a cone
centred on either fails to capture the other and each hard particle
leads to its own jet) and one adds a soft particle in between the two
(so that a cone centred on the soft particle includes both hard ones,
which then end up in a single jet), as illustrated in
Fig.~\ref{fig:IR-unsafety}.
A partial fix for the problem was given in Ref.~\cite{midpoint} (adopted by
the Tevatron in Ref.~\cite{RunII-jet-physics}), which involves adding extra
seeds at the midpoints between all pairs of stable cones and uses
those as new starting points for finding additional stable cones
before starting the split--merge step.
A full fix involves a non-seed-based approach to exhaustively finding
all possible stable cones in an event, in an algorithm known as the
Seedless Infrared Safe Cone (SISCone)~\cite{Salam:2007xv}.

\subsection{Sequential-recombination algorithms}
\label{sec:seq-rec-algorithms}

Sequential-recombination jet algorithms take a bottom-up approach to
constructing jets, as if they were inverting the sequence of
splittings of the parton shower. 
Of course that sequence doesn't really exist unambiguously, since
gluon emission is 
a quantum-mechanical process involving coherent emission from all
colour sources in an event.
However, for collinear emissions the picture that there is a single
`emitter' is not a poor approximation.

\subsubsection{The $e^+e^-$ $k_t$ algorithm}
The most widely used sequential recombination algorithm to date is the
$k_t$ algorithm, originally formulated for $e^+e^-$ events~\cite{Kt}.
Recall, from Eq.~(\ref{eq:fs-universal-splitting}), that the soft and
collinear limit of the gluon-emission probability for $a \to i j$ is
\begin{equation}
  \label{eq:splitting-again}
  dS \simeq \frac{2\as C_i}{\pi} \frac{dE_{i}}{\min(E_i,E_j)}
  \frac{d\theta_{ij}}{\theta_{ij}}\,,
\end{equation}
where $C_i$ is $C_A$ ($C_F$) if $a$ is a gluon (quark), and where
we've written $\min(E_i,E_j)$ in the denominator to avoid specifying
which of $i$ and $j$ is the soft particle.

The essential idea of the $k_t$ algorithm is to define a `distance
measure' $y_{ij}$ between every pair of particles $i,j$,
\begin{equation}
  \label{eq:yij-kt-alg}
  y_{ij} = \frac{2\min(E_i^2,E_j^2)(1-\cos\theta)}{Q^2}\,.
\end{equation}
In the collinear limit, $y_{ij}$ reduces to $\min(E_i^2,E_j^2)
\theta_{ij}^2$, which is the relative transverse momentum between
particles $i$ and $j$ (hence the name $k_t$ algorithm), normalized to
the total visible (or sometimes centre-of-mass) energy $Q$.
Apart from the normalization, this is just what appears in the
denominator of the splitting probability,
Eq.~(\ref{eq:splitting-again}),
so that pairs of particles that would arise from a splitting with a
strong divergence are considered very close together.

The algorithm works by identifying the pair of particles that has the
smallest $y_{ij}$ and recombining them into a single particle (also
called a `pseudojet').
It then recalculates all $y_{ij}$ distances taking into account the
new particle, and again recombines the pair that's closest.
The procedure is repeated until all $y_{ij}$ distances are above some
threshold $y_{\text{cut}}$, at which point the pseudojets that remain
become the event's jets.

\subsubsection{The $k_t$ algorithm for hadron collisions}

For hadron collisions and deep-inelastic scattering, the version of
the $k_t$ algorithm that is most commonly used reads as
follows~\cite{KtHH,Kt-EllisSoper}. For every pair of particles define
a (dimensionful) inter-particle distance $d_{ij}$,
\begin{equation}
  \label{eq:dij-kt}
  d_{ij} = \min(p_{ti}^2, p_{tj}^2) \frac{\Delta R_{ij}^2}{R^2}\,,
\end{equation}
where $R$ is a parameter whose role is similar to that of $R$ in cone
algorithms.
Also define a beam distance for every particle,
\begin{equation}
  \label{eq:diB-kt}
  d_{iB} = p_{ti}^2\,.
\end{equation}
The algorithm proceeds by searching for the smallest of the $d_{ij}$
and the $d_{iB}$. If it is a $d_{ij}$ then particles $i$ and $j$ are
recombined into a single new particle. 
If it is a $d_{iB}$ then $i$ is removed from the list of particles,
and called a jet.
This is repeated until no particles remain.

Note that the distance in Eq.~(\ref{eq:dij-kt}) just reduces to that
of Eq.~(\ref{eq:yij-kt-alg}) in the collinear limit (modulo $Q^2$
normalization). So one is still dealing with the relative transverse
momentum between pairs of particles.
As with cone algorithms, in this `inclusive longitudinally invariant
$k_t$ algorithm,' arbitrarily soft particles can form jets. 
It is therefore standard to place a $p_{t,\min}$ cutoff on the jets
one uses for `hard' physics.

One can verify that $R$ in the $k_t$ algorithm plays a similar role to
$R$ in cone
algorithms, using the following observations: if two particles $i$ and
$j$ are within $ R$ of each other, i.e., $\Delta R_{ij} < R$, then
$d_{ij} < d_{iB}, d_{jB}$ and so $i$ and $j$ will prefer to recombine
rather than forming separate jets.
If a particle $i$ is separated by more than $R$ from all other
particles in the event then it will have $d_{iB} < d_{ij}$ for all $j$
and so it will form a jet on its own.

\begin{figure}
  \centering
  \includegraphics[width=0.48\tw]{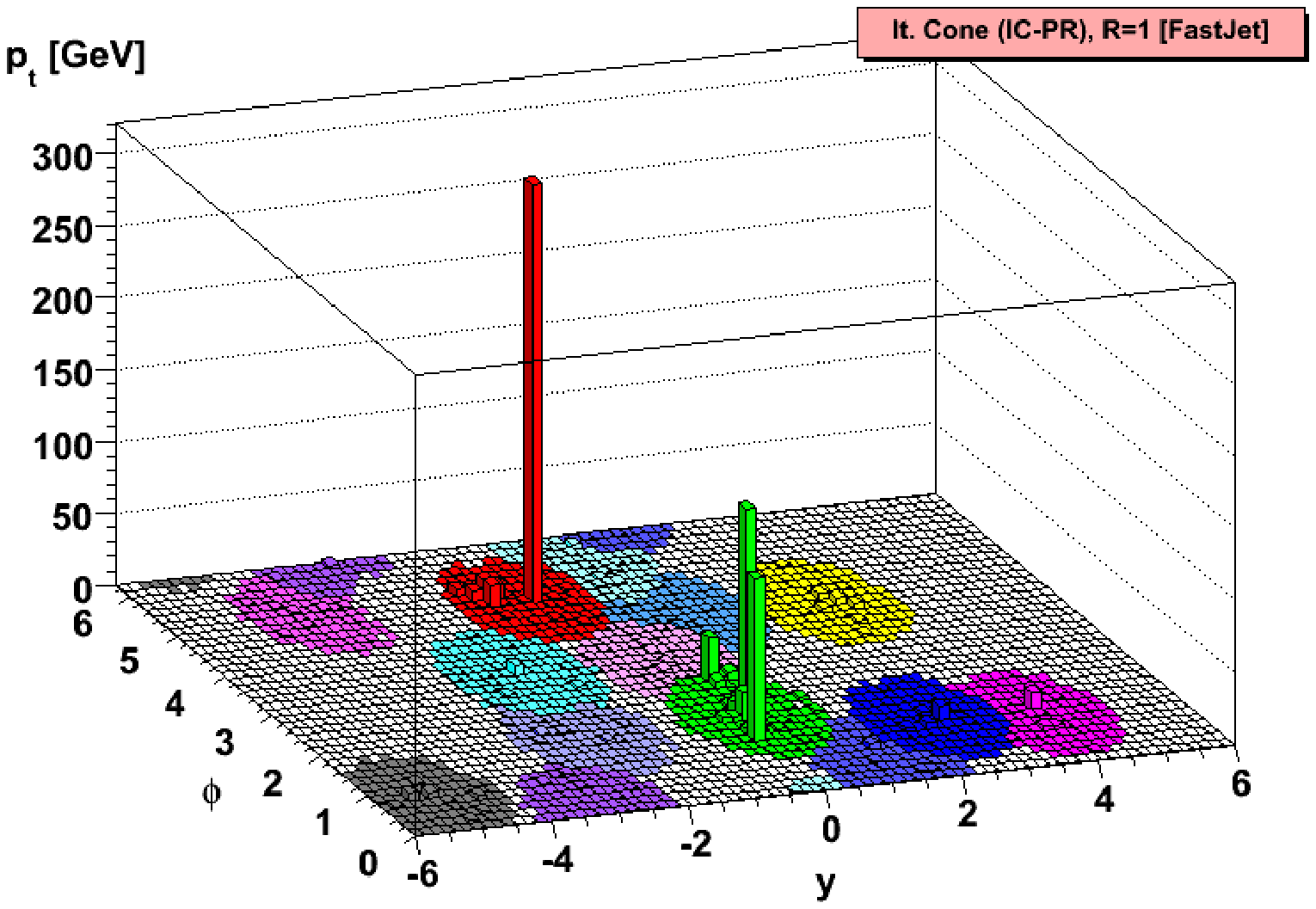}\hfill
  \includegraphics[width=0.48\tw]{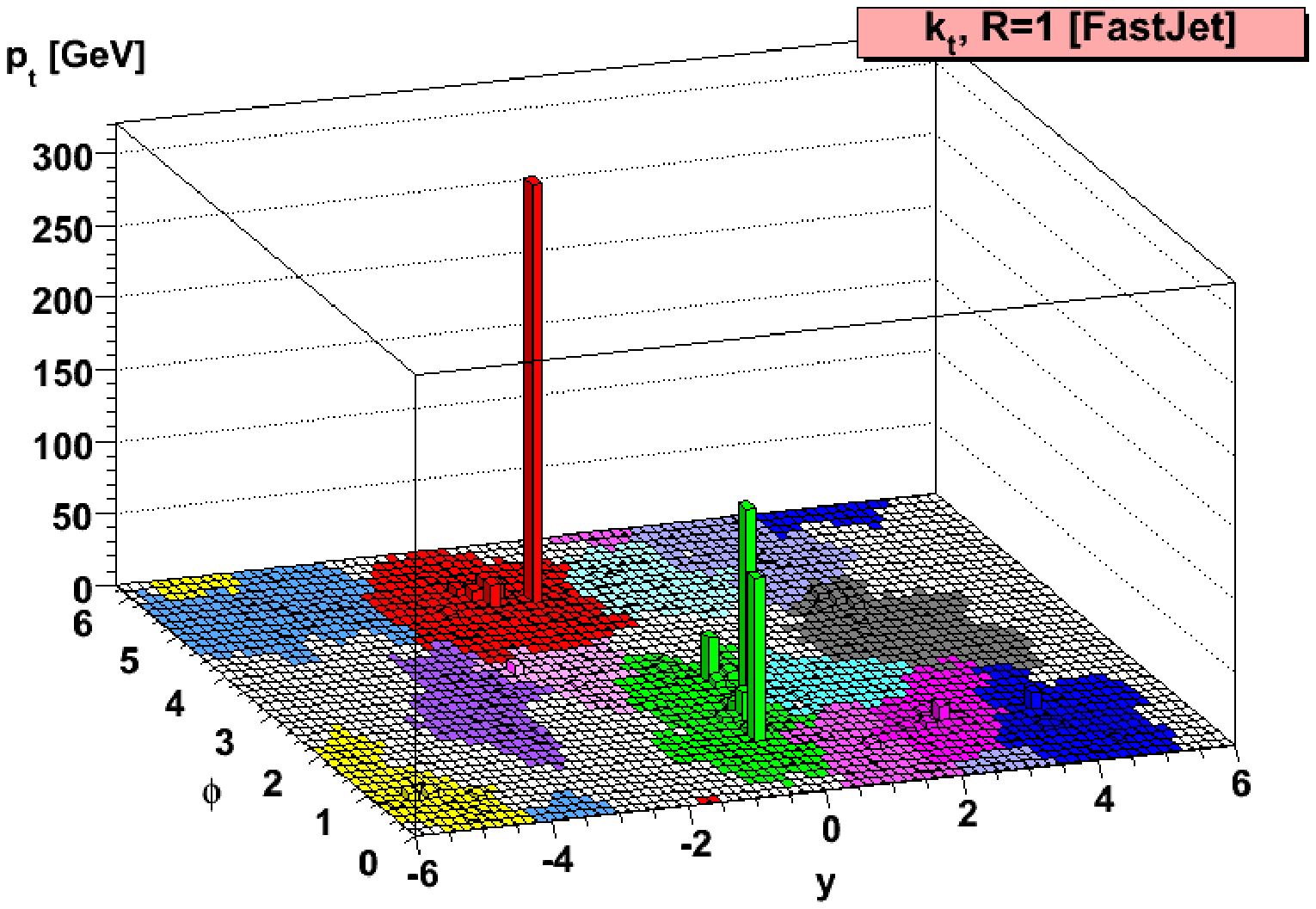}
  \caption{Regions of the $y$--$\phi$ plane covered by jets in an
    illustrative (simulated) hadron-collider event with an IC-PR type
    cone algorithm (left) and the inclusive longitudinally-invariant
    $k_t$ algorithm (right).
    The jet finding was carried out with the {\sc FastJet}
    package~\cite{Cacciari:2005hq,FastJet}, with the addition of very
    soft ghost particles to trace the extent of the
    jets~\cite{Cacciari:2008gn}.  }
  \label{fig:shapes}
\end{figure}

Despite this similarity to the behaviour of cone algorithms for pairs
of particles, the $k_t$ algorithm gives jets that `look' somewhat
different. 
Figure~\ref{fig:shapes} illustrates what happens when one clusters a
simulated hadron-collider event with an IC-PR type cone algorithm and
with the $k_t$ algorithm.
In both cases the figure shows (in a given colour) the calorimeter cells that
are included in each jet. 
For the IC-PR algorithm the hardest jets all appear circular, as
expected given the use of cones in the definition of the algorithm (in
cone algorithms with split--merge steps, the jets are often not
circular, because of the less trivial mapping from stable cones to
jets).
For the $k_t$ algorithm, the jets have quite irregular (or jagged) edges, because
many of the soft particles cluster together early in the recombination
sequence (owing to their small $p_t$ and hence $d_{ij}$) in patterns
that are determined by the random distributions of those particles in
$p_t$ and rapidity and azimuth.

The irregularity of the jet boundaries has often been
held against the $k_t$ algorithm.
One reason is that it makes it harder to calculate acceptance
corrections: for example, if you know that some part of a detector is
misbehaving, it is not obvious how far a $k_t$ jet must be from that
part of the detector in order not to be affected by it.
Another reason relates to the linearity of the dependence of the jet
momentum on soft-particle momenta: in the IC-PR algorithm, the hard
core of the jet essentially determines the jet boundary, and the
algorithm depends linearly on the momenta of any soft particles within
the boundary, and is independent of particles outside it. 
In the $k_t$ algorithm, varying the momentum of one soft particle in the
vicinity of the jet core can affect whether it and other soft
particles get clustered with that core or not. 
This can complicate energy calibrations for the jet algorithm, though
techniques exist to correct for this to some extent (jet-area-based
subtraction~\cite{Cacciari:2007fd}, which leaves just a residual term
known as back-reaction~\cite{Cacciari:2008gn}).

A feature of the $k_t$ algorithm that is attractive is that it not
only gives you jets, but also assigns a clustering sequence to the
particles within the jet. 
One can therefore undo the clustering and look inside the jet.
This has been exploited in a range of QCD studies (e.g., Ref.~\cite{D0KtSubJets}), 
and also in discussions of searches of hadronic
decays of boosted massive particles such as $W$, $H$, or $Z$ bosons, top
quarks, or new particles (early examples include
Refs.~\cite{Seymour:1993mx,Butterworth:2002tt}; for more recent examples,
see the reviews in Refs.~\cite{Salam:2009jx,Butterworth:2010ym}).
Jet substructure studies are also often carried out with the
Cambridge/Aachen (C/A) algorithm~\cite{Cam,Aachen}, which essentially
replaces $p_{ti}\to 1, p_{tj} \to 1$ in
Eqs.~(\ref{eq:dij-kt},\ref{eq:diB-kt}) but is otherwise like the $k_t$
algorithm.

\subsubsection{The anti-$k_t$ algorithm}
\label{sec:antikt}

\begin{figure}
  \centering
  \includegraphics[width=0.48\tw]{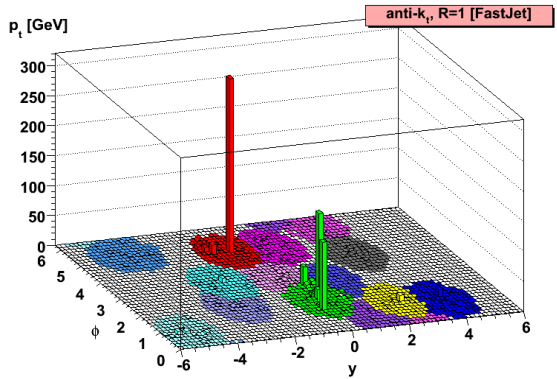}
  \caption{As in Fig.~\ref{fig:shapes}, but shown for the anti-$k_t$
    algorithm}
\label{fig:antikt}
\end{figure}

It turns out that it is possible to design a sequential-recombination
algorithm with many of the nice properties of cone algorithms via a
simple modification of the $k_t$ algorithm's distance
measures~\cite{Cacciari:2008gp}:
\begin{subequations}
  \begin{align}
    \label{eq:dij-antikt}
    d_{ij} &= \frac{1}{\max(p_{ti}^2, p_{tj}^2)} \frac{\Delta
      R_{ij}^2}{R^2}\,,
    \\
    \label{eq:diB-antikt}
    d_{iB} &= \frac1{p_{ti}^2}\,.
  \end{align}
\end{subequations}
The correspondence with the divergences of
Eq.~(\ref{eq:splitting-again}) is partially lost: objects that are
close in angle still prefer to cluster early, but that clustering
tends to occur with a hard particle (rather than necessarily involving
soft particles).
This means that jets `grow' in concentric circles out from a hard
core, until they reach a radius $R$, giving circular jets just as with
the IC-PR cone, as shown in Fig.~\ref{fig:antikt}.
However, unlike the IC-PR cone, this `anti-$k_t$' algorithm is
collinear (and infrared) safe, meaning that it is safe to use with
fixed-order QCD predictions.
This, combined with the fact that it has been implemented efficiently
in the {\sc FastJet} jet-finding code~\cite{Cacciari:2005hq,FastJet}, has led to it
being adopted as the default jet algorithm by both the ATLAS and CMS
collaborations.

Note that the anti-$k_t$ algorithm does not provide useful information
on jet substructure: if a jet contains two hard cores, then the $k_t$
(or C/A) algorithms first reconstruct those hard cores and merge the
resulting two subjets. 
The anti-$k_t$ algorithm will often first cluster the harder of the two cores
and then gradually agglomerate the contents of the second hard core.

\subsection{Using jets}
\label{sec:using-jets}

Because of the intricacies of calibrating jets, past experiments have
tended to concentrate their efforts on just one or two jet
definitions, which are used across the board for QCD studies,
top-quark studies, Higgs and new physics searches.
Typically the choices have been for cone-type algorithms with $R$
values in the range 0.4--0.7.

At the LHC there are prospects of experimental methods (for example
topoclusters at ATLAS and particle flow at ATLAS) that make it easier
to use a broad range of jet definitions.
We've already mentioned, above, the use of jet substructure in new
physics searches. 
These methods will definitely benefit from experiments' flexibility in
choosing different jet definitions (for example, many of the proposed
searches use quite a large jet radius, $R\sim$~ 1--1.5).
More generally, when using jets, it is important to establish what
you are trying to get out of the jet algorithm, what it is that you're
trying to optimize.

Different studies will want to optimize different things. For example
in QCD studies, such as the inclusive jet-spectrum measurement that
goes into PDF fits, one criterion might be to choose a jet definition
that minimizes the non-perturbative corrections that need to be
applied when comparing perturbative predictions with data.

\begin{figure}
  \centering
  \includegraphics[width=0.65\tw]{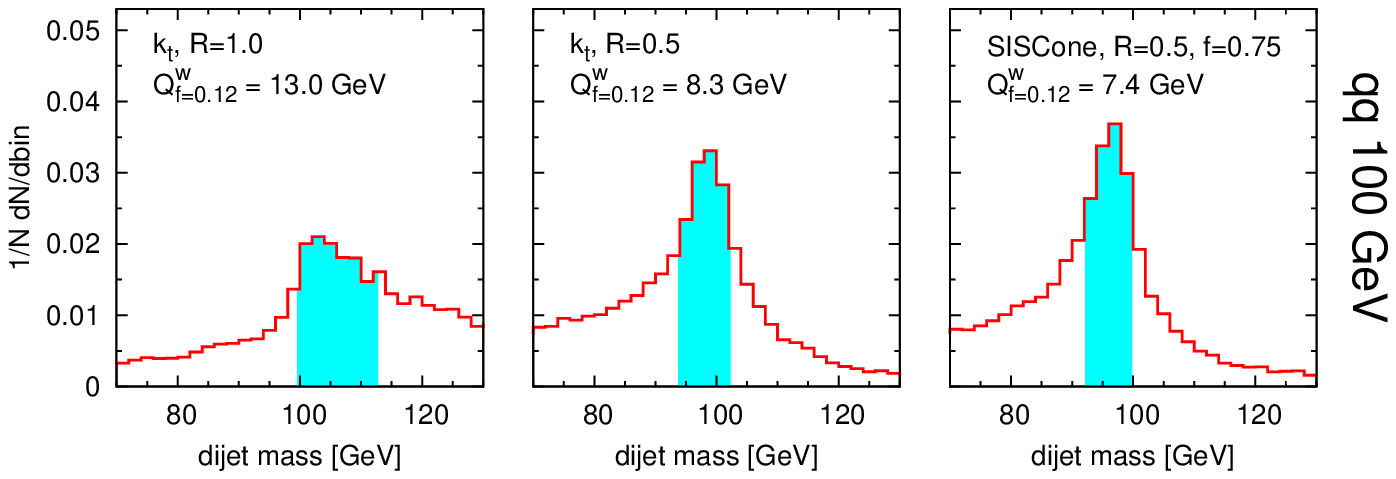}\\
  \includegraphics[width=0.65\tw]{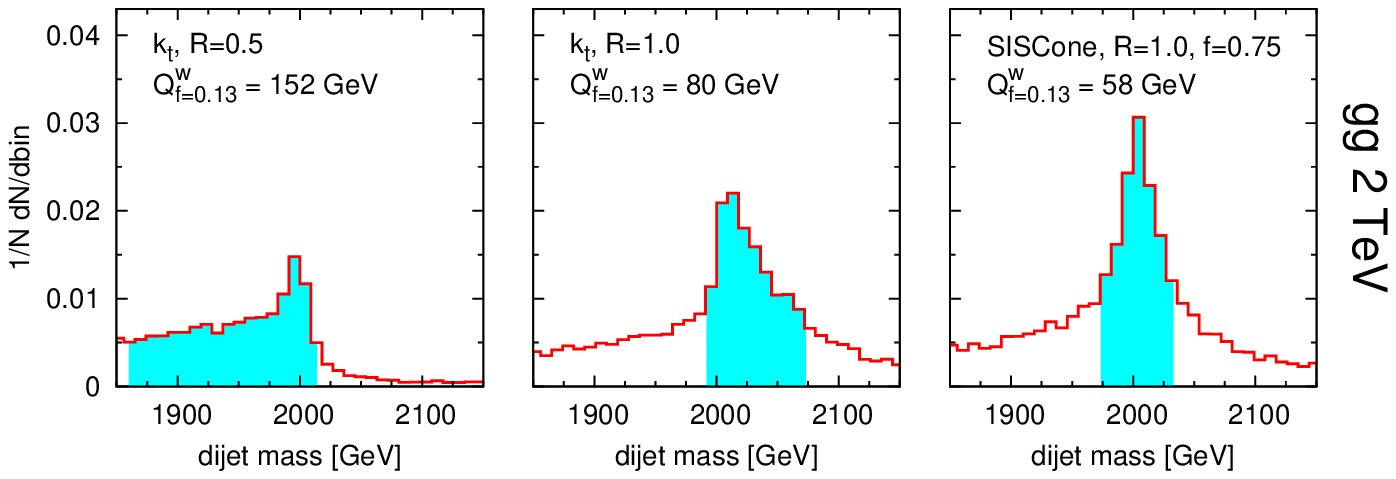}
  \caption{Illustration of the impact of different jet definition on
    the quality of reconstruction of dijet invariant mass peaks for a
    $100\GeV$ resonance decaying to $q\bar q$ (top) and a $2\TeV$
    resonance decaying to $gg$ (bottom).
    The $Q^w_{f=0.12}$ indicates the width of the region that contains
    $12\%$ of the resonance events (corresponding to about $25\%$ of
    the resonance events that pass basic selection cuts).
    Smaller numbers imply better reconstruction.
    Figure taken from Ref.~\cite{Cacciari:2008gd}. Many further plots
    available from Ref.~\cite{quality.fastjet.fr}.
  }
  \label{fig:quality}
\end{figure}

In searches for new physics, for example when looking for a resonance
in the dijet invariant mass spectrum, the criterion might be the
following: given a narrow resonance, what jet definition will lead to
the narrowest peak in the dijet mass spectrum?
It turns out that the answer depends significantly on properties of
the resonance you're trying to reconstruct. 
Figure~\ref{fig:quality} illustrates how SISCone with $R=0.5$ does well
for reconstructing a $100\GeV$ resonance that decays to $q\bar q$.
While for a $2\TeV$ resonance decaying to $gg$, you're better off with
an algorithm (such as SISCone, or others supplemented with the trick
of filtering and/or variants~\cite{Butterworth:2008iy,Krohn:2009th})
with a substantially larger radius, $R\sim
1$~\cite{Buttar:2008jx,Cacciari:2008gd,quality.fastjet.fr,VariableR}.
And if you find a resonance, you might then want to know which jet
definition will allow you to measure its mass most accurately, which
may not be the same choice that allowed you to find the resonance most
easily.

Understanding analytically which jet choices work well is the subject
of ongoing theoretical
work~\cite{Dasgupta:2007wa,Cacciari:2008gn,Rubin:2010fc}, which
involves figuring out how perturbative radiation, hadronization, the
underlying event and pileup all come together to affect jet momenta.

\section{Conclusions}
\label{sec:conclusions}

In these lectures we have covered material that ranges from the QCD
Lagrangian through to $e^+e^-$ scattering, soft and collinear
divergences, PDFs, fixed-order calculations, parton showers and jets.
It is my hope that the material here will have given you enough
background information to at least understand the basics of these
different concepts.

The aspects of QCD that we have examined will play a role in nearly
all analyses at the LHC. Whether because Monte Carlo parton-shower
programs are used for estimating detector effects on almost every
measurement; or because in searches for new physics one might look for
an excess with respect to QCD predictions of backgrounds; or perhaps
because an understanding of QCD will make it possible to more clearly
pull out a signal of new physics that involves QCD final states.

If you want to find out more about the topics discussed here (or those
left out due to time constraints, notably heavy quarks), good places
to look include the
textbooks~\cite{Ellis:1991qj,Brock:1993sz,Dissertori:2003pj} mentioned
at the beginning of these lectures, the references given
throughout, 
and transparencies from the various summer schools dedicated to QCD,
for example, the CTEQ (and MCNET) schools~\cite{cteq-mcnet,cteq09}.


\section*{Acknowledgements}

I wish to thank the organizers of the 2009 European School
of High-Energy Physics for their warm hospitality during the school
itself and for their kind patience in waiting for these lecture notes to
be completed.
I am also grateful to the Physics Department of Princeton University
for hospitality while finishing these lecture notes.
This work was supported by the French Agence Nationale de la
Recherche, under grant ANR-09-BLAN-0060.


\end{document}